\documentclass[pdflatex,sn-mathphys-num]{sn-jnl}

\usepackage{graphicx}%
\usepackage{multirow}%
\usepackage{amsmath,amssymb,amsfonts}%
\usepackage{amsthm}%
\usepackage{mathrsfs}%
\usepackage[title]{appendix}%
\usepackage{xcolor}%
\usepackage{textcomp}%
\usepackage{manyfoot}%
\usepackage{booktabs}%
\usepackage{algorithm}%
\usepackage{algorithmicx}%
\usepackage{algpseudocode}%
\usepackage{listings}%
\usepackage{tabularx}
\usepackage{caption}
\usepackage[many]{tcolorbox}
\theoremstyle{thmstyleone}%
\theoremstyle{thmstyletwo}%

\theoremstyle{thmstylethree}%

\newtcolorbox{boxK}{
    sharpish corners, 
    boxrule = 0pt,
    toprule = 0.5pt, 
    enhanced,
    fuzzy shadow = {0pt}{-2pt}{-0.5pt}{0.5pt}{black!35} 
}

\begin{document}

\title[Article Title]{An Empirical Investigation on the Challenges in Scientific Workflow Systems Development}


\author*{\fnm{Khairul} \sur{Alam}}\email{kha060@usask.ca}
\author{\fnm{Banani} \sur{Roy}}\email{banani.roy@usask.ca}
\author{\fnm{Chanchal K.} \sur{Roy}}\email{chanchal.roy@usask.ca}

\author{\fnm{Kartik} \sur{Mittal}}\email{wte705@usask.ca}

\affil{\orgdiv{Department of Computer Science}, \orgname{University of Saskatchewan}, \orgaddress{\city{Saskatoon}, \postcode{S7N 5A2}, \state{Saskatchewan}, \country{Canada}}}




\abstract{Scientific Workflow Systems (SWSs) are advanced software frameworks that drive modern research by orchestrating complex computational tasks and managing extensive data pipelines. These systems offer a range of essential features, including modularity, abstraction, customization, interoperability, workflow composition tools, resource management, error handling, and comprehensive documentation. Utilizing these frameworks accelerates the development of scientific computing, resulting in more efficient and reproducible research outcomes. Despite their significance, developing a user-friendly, efficient, and adaptable SWS poses several challenges that are not always well-documented or understood. This study explores these challenges through an in-depth analysis of interactions on Stack Overflow (SO) and GitHub, key platforms where developers and researchers discuss and resolve issues. In particular, we leveraged topic modeling (BERTopic) to understand the topics SWSs developers discuss on these platforms. Then, we examined the popularity and difficulty of those topics. We identified 10 topics developers discuss on SO (e.g., Workflow Creation and Scheduling, Data Structures and Operations, Workflow Execution) and found that workflow execution is the most challenging among them. By analyzing GitHub issues, we identified 13 topics (e.g., Errors and Bug Fixing, Documentation, Dependencies) and discovered that errors and bug fixing is the most dominant topics in this context. We found system redesign and API migration to be the most challenging topics utilizing GitHub data. A cross-platform comparison revealed overlapping concerns such as task management and data operations. Additionally, we categorized each topic by type (How, Why, What, and Others) and observed that the How type consistently dominates across all topics, indicating a need for procedural guidance among developers. This dominance of the How type is also prevalent in other domains, such as Chatbots and Mobile development. We believe that our study will guide future research in proposing tools and techniques to help the community overcome the challenges developers face when developing SWSs.}



\maketitle

\section{Introduction}\label{introduction}
Scientific Workflow Systems (SWSs) are software frameworks designed to streamline the design, execution, and management of complex, data-intensive scientific experiments and research \citep{10479414}. They allow users to create, visualize, and automate workflows composed of interconnected steps, each carrying out specific data operations. These systems orchestrate tasks, manage data dependencies, and ensure reproducibility by consistently capturing and executing workflows \citep{liu2015survey}. By enhancing the efficiency of running and analyzing experiments, they facilitate collaboration and meet the rigorous demands of scientific research. SWSs serve as the bedrock upon which efficient, transparent, and impactful research is built, propelling knowledge and innovation across diverse fields. They accomplish mundane yet essential tasks intrinsic to scientific inquiry—ranging from data acquisition to analysis and reporting. By liberating researchers from time-consuming tasks, SWSs allow them to focus on more complex and creative work, which helps speed up and improve research results \citep{almeida2018modular}.
They provide a platform to create and distribute the necessary computational steps for tasks involving data analysis and simulations, all while concealing the intricate infrastructural complexities \citep{olabarriaga2014scientific}. More importantly, they capture the entire scientific experimentation process, offering valuable insights for reproducing, reusing, or adapting these procedures. SWSs such as Galaxy \citep{giardine2005galaxy}, KNIME \citep{berthold2009knime}, Snakemake  \citep{koster2012snakemake}, and Nextflow \cite{di2017nextflow} support the design, execution, and management of complex scientific workflows across various domains (e.g., Genomics and Bioinformatics, Microbiology and Virology, Computational Biology,  Epidemiology and Public Health). These systems offer diverse interfaces ranging from user-friendly graphical tools (e.g., Galaxy, KNIME) to script- and DSL-based platforms (e.g., Airflow, Snakemake, Nextflow), enabling reproducible, scalable, and efficient computational research.

Traditionally, various types of scientific analyses are implemented by practitioners using either scripting languages (Python, Perl, or Bash) requiring fairly sophisticated computational skills or by manually calling and storing intermediate results, which requires a considerable amount of time organizing data files and managing the overall analysis process \citep{mcphillips2009scientific}. Alternatively, SWSs can help practitioners by providing higher-level modeling approaches for explicitly specifying analyses as well as by providing several generic services for optimizing workflow execution, visualizing workflows and workflow results, managing intermediate data products, and storing the details of past workflow executed results \citep{bowers2012scientific}. Lin et al. \cite{lin2009reference} defined scientific workflow as a computerized facilitation or automation of a scientific process, in whole or part, which usually streamlines a collection of scientific tasks with data channels and dataflow constructs to automate data computation and analysis to enable and accelerate scientific discovery. The ability of SWSs to handle large volumes of data and integrate various computational tools makes them essential for advancing research across diverse scientific domains. Furthermore, SWSs afford researchers, developers, and project managers real-time insights into the progress of research endeavors, empowering them to identify bottlenecks, allocate resources judiciously, and optimize workflow execution. The customizable nature of SWSs allows practitioners to tailor workflows to their specific needs and preferences, enhancing their utility and versatility. In essence, SWSs emerge not merely as tools of convenience but as indispensable allies in the relentless pursuit of scientific excellence.
\begin{figure}[htbp]
    \centering
    \includegraphics[width=\linewidth]{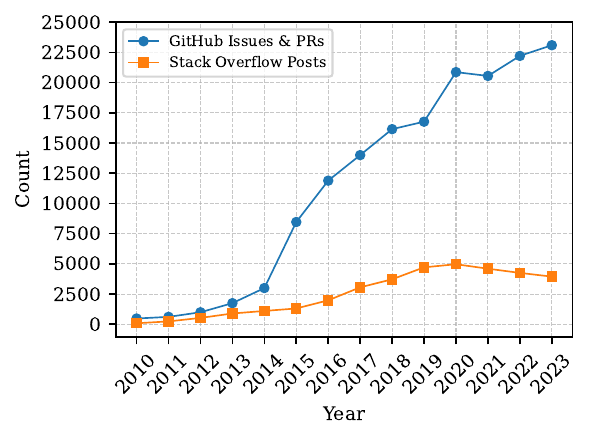}
    \caption{Year-wise distribution of posts on Stack Overflow, and issues \& pull requests on GitHub related to selected SWSs, reflecting the growing developer engagement and activity in SWS development over time.}
    \label{fig:swsdatainqaforums}
    \vspace{-1em}
\end{figure}
The extensive adoption of SWSs in diverse fields (e.g., machine learning \citep{deelman2019role}, data mining \citep{marozzo2016workflow}, bioinformatics \citep{oinn2004taverna}, chemistry \citep{fernandez2013chemistry}, geosciences \citep{li2015enabling}, high-performance computing \citep{hendrix2016tigres}) has led the software engineering community to develop numerous software tools tailored for SWSs. Thousands of developers are contributing to SWSs in various capacities, undertaking tasks such as system architecture design (\emph{determining the overall structure, components, and interactions between different modules to ensure flexibility, scalability, and performance}), algorithm development (\emph{developing algorithms for data processing, analysis, and automation, leveraging techniques from machine learning, natural language processing, and data mining to extract insights from scientific data}), UI design and development (\emph{user-friendly interfaces, i.e., intuitive dashboards, visualization tools, and workflow editors}), backend development (\emph{backend logic of SWSs, including data storage, processing pipelines, and workflow execution engines, writing codes in different languages and integrating with databases and external services}), integration with external tools and services (\emph{interoperability and data exchange}). 
They are also involved in testing and quality assurance, documentation, deployment and maintenance, security and compliance, and performance optimization by identifying bottlenecks, tuning algorithms and configurations, and leveraging parallel processing and distributed computing techniques to efficiently handle large-scale data analysis. Managing these vast amounts of tasks comes with numerous challenges. Developers usually turn to SO to ask questions, find solutions, learn and understand new technologies or programming languages, validate best practices, and engage with the community \citep{mamykina2011design}. GitHub also plays a crucial role in collaborative development \citep{dabbish2012social}. Through GitHub issues and pull requests, developers can report problems, suggest improvements, engage in discussions, contribute to projects, and participate in the open-source community. Such technical Q\&As, issue reports, and pull requests may communicate developers' challenges when developing SWSs.

Practitioners encountered a significant deficiency in usability, effectiveness, and reliability when dealing with these SWSs \cite{mork2015contemporary}. Furthermore, the consistent rise in the number of posts, issues, and pull requests (for selected SWSs) shown in Figure \ref{fig:swsdatainqaforums} implies a growing demand for increased support for developers engaged in SWSs. Given the widespread use of SWSs across various domains and the interest they have garnered, including from the software analytics community, it becomes crucial to comprehend these challenges for advancing the efficiency and effectiveness of these systems, which are integral to data-intensive research and complex scientific computations. Practitioners often encounter unique difficulties in SWSs, such as managing heterogeneous data sources, ensuring reproducibility, and optimizing workflows for high-performance computing environments. By systematically identifying these challenges, researchers and practitioners can tailor tools and methodologies to better support the development of robust, scalable, and user-friendly workflows. Addressing these challenges not only enhances the usability and reliability of SWSs but also accelerates scientific discovery by enabling researchers to focus on their domain-specific tasks rather than the intricacies of the workflow system itself. Moreover, understanding these challenges provides valuable insights that can drive innovation in SWSs design, leading to more efficient computational solutions that meet the evolving needs of the scientific community \citep{deelman2019role}. Therefore, we analyzed SO posts, GitHub issues, and pull requests related to SWSs to identify developer challenges in our research.

In this paper, we conducted an empirical investigation to rigorously understand the needs and desiderata of developers working with SWSs. Our objective is to gain a precise understanding of the challenges developers face and identify potential areas for improvement, aiming to pinpoint specific issues and obstacles encountered in their work. To achieve this, we compiled a dataset of 35,619 SO posts and 163,118 GitHub issues and pull requests related to SWSs. Using a combination of manual analysis and automated topic modeling, we examined these SO posts and GitHub data to uncover the specific challenges developers encounter. In particular, our study aims to answer the following three research questions (RQs):
\begin{itemize}
    \item \textbf{RQ1: What topics are SWSs developers asking about on Stack Overflow, and which topics are raised in the GitHub projects?} The SWSs-related posts, issues, and pull requests may reflect developers' challenges when learning or developing an SWS. To understand these challenges, we utilized topic modeling to extract semantic topics from SWSs-related data on SO and GitHub. From SO posts, we identified 10 topics, which include traditional software engineering topics (e.g., data structures and operations, dependencies management) and SWSs-specific challenges (e.g., workflow creation and scheduling, workflow execution). Similarly, analysis of GitHub data yielded 13 topics, encompassing traditional software engineering topics (e.g., errors and bug fixing, dependencies) and SWSs-specific topics (e.g.,data shuffling and partitioning strategies, task coordination). By comparing the challenges developers face on SO and GitHub, we aimed to identify any similarities. By answering \emph{RQ1}, we aimed to pinpoint the areas in SWSs development where developers frequently seek assistance, thereby highlighting the aspects that commonly pose challenges for them.
    
    \item \textbf{RQ2: What types of questions are SWSs developers asking?} Developers ask different (i.e., How, Why) types of questions to address different challenges. To answer \emph{RQ2}, we adopted a methodology similar to previous research on identifying types of posts on Stack Overflow \cite{rosen2016mobile, abdellatif2020challenges, treude2011programmers}. We selected a statistically significant number of posts for each topic and classified them into types such as \emph{How, What, Why}, and \emph{Others}. This approach enabled us to systematically analyze the nature of the questions developers ask and gain deeper insights into their needs and challenges. Our analysis showed that SWSs developers primarily seek guidance on specific implementation routines, workflow examples, and troubleshooting, highlighting the dominance of \emph{How} questions, similar to other domains like \emph{ChatBot} \citep{abdellatif2020challenges} and \emph{Mobile Development} \citep{rosen2016mobile}. Our findings underscored the need for step-by-step instructions and real-world scenarios to help developers achieve their goals in SWSs implementation.
    
    \item \textbf{RQ3: To what extent do developers perceive the revealed challenges in terms of difficulties?} To assess the difficulty of each topic, we used two metrics: the percentage of unresolved posts/issues/pull requests and the median time to resolve them. This approach followed the methodologies of prior studies \citep{rosen2016mobile, bagherzadeh2019going, yang2016security, li2021understanding, scoccia2021challenges, abdellatif2020challenges}. In \emph{RQ3}, our goal is to identify the most challenging topics to resolve, as these are likely to pose significant difficulties for all but the most experienced developers. We identified workflow execution as the most challenging topic on SO due to its complexity in managing distributed resources, handling large-scale data processing, ensuring fault tolerance, and optimizing performance across diverse environments. Similarly, on GitHub, Kubernetes deployments for SWSs frameworks pose significant challenges, primarily due to difficulties in ensuring scalability, fault tolerance, and efficient resource allocation required for managing complex and distributed workflows. This analysis highlighted areas within SWSs development where additional support and resources are essential.
\end{itemize}

\textbf{Paper Organization: }The remainder of this paper is structured as outlined as follows: In Section \ref{background-and-related-work}, we briefly discussed Scientific Workflow Systems(SWSs), Scientific Workflows, Topic Modeling, and related work. Moving on to Section \ref{study-design}, we detailed our study's methodology. Our findings are presented in Section \ref{case-study-results}, and Section \ref{discussion-implications} illustrates the implications. Section \ref{threats-to-validity} addresses potential threats to the validity of our results. Finally, Section \ref{conclusion} concludes the paper, highlighting directions for future research.

\section{Background and Related Work}\label{background-and-related-work}
SWSs emerged as a response to the growing complexity and scale of scientific research \citep{atkinson2017scientific}. As computational power increased and data generation exploded across various scientific domains, researchers needed more efficient ways to manage, automate, and reproduce their experiments. Traditional manual methods were inadequate for handling the intricate sequences of data processing tasks required. SWSs provide a structured framework for designing, executing, and sharing multi-step computational processes. They integrate diverse tools and data sources, facilitating collaboration and enhancing the reproducibility of scientific research. We aimed to understand developers' challenges in developing SWSs by analyzing SO posts, GitHub issues, and pull requests. In this section, we provided an overview of the background and previous research associated with SWSs development. Additionally, we explored prior work that utilized topic modeling techniques to gain insights into developer perspectives.

\subsection{Scientific Workflow Systems}
An SWS is a specialized software platform designed to automate, manage, and execute complex sequences of computational or data processing tasks, often referred to as workflows, in scientific research whose execution order is driven by a computerized representation of the workflow logic \citep{lin2009reference}. Researchers categorized existing SWSs into five distinct niches, each representing a specific approach to handling workflows in scientific computing \cite{molder2021sustainable}. Below, we describe these niches and provide highlighted examples of systems that exemplify each category.

First, SWSs like KNIME \citep{berthold2009knime}, Galaxy \citep{giardine2005galaxy}, Apache Taverna \citep{oinn2004taverna}, and Watchdog \cite{kluge2020watchdog} offer graphical user interfaces for the composition and execution of workflows. The shallow learning curve makes these systems accessible to everyone, even those without programming skills. Second, with systems like  Anduril \cite{cervera2019anduril}, Balsam \cite{salim2019balsam}, SciPipe \cite{lampa2019scipipe}, and JUDI \cite{pal2020bioinformatics}, workflows are defined using a set of classes and functions in generic programming languages such as Python, Scala, and others. These systems have the advantage of being usable without a graphical interface (e.g., in a server environment) and allow workflows to be easily managed with version control systems like Git (\href{https://git-scm.com/}{https://git-scm.com}). Third, SWSs like Snakemake \citep{koster2012snakemake}, Nextflow \cite{di2017nextflow}, Bpipe \cite{sadedin2012bpipe}, ClusterFlow \cite{ewels2016cluster}, and BigDataScript \cite{cingolani2015bigdatascript} use domain-specific language (DSL) to specify workflows. These DSLs offer the advantages of the prior category while enhancing readability by modeling key workflow management components directly, reducing the need for extraneous operators or boilerplate code. For Nextflow and Snakemake, which extend generic programming languages (Groovy and Python), the full power of the underlying language is retained, allowing for features like conditional execution and configuration handling. Fourth, systems like Popper \cite{jimenez2017popper} use purely declarative approaches for workflow specification through configuration file formats like YAML \cite{ben2009yaml}. While offering the concision and clarity of the third niche, these systems are exceptionally readable for non-developers. Fifth, system-independent workflow specification languages like CWL \cite{amstutz2016common} and WDL \cite{vossfull} use declarative syntax for defining workflows, which can be executed by various executors (e.g., Cromwell (\href{https://cromwell.readthedocs.io/en/latest/}{https://cromwell.readthedocs.io/en/latest/}), and Toil \cite{vivian2017toil}). While these languages, like those in the fourth niche, often lack integration with imperative or functional programming, they offer the advantage of executing the same workflow across different execution backends, enhancing scalability. Additionally, they facilitate interoperability among various workflow languages.

 SWSs encompass several key aspects: workflow design, specification, execution, monitoring and control, resource management, data management, reproducibility and sharing, and collaboration. These systems significantly enhance productivity, collaboration, and reliability in scientific research while optimizing the efficient use of computational resources.

\subsection{Prior Studies on SWSs}
SWSs have been widely studied regarding reproducibility, reusability, usability, performance, and collaborative development. These studies provide valuable insights into the capabilities and limitations of current systems, and several findings directly motivate the goals of our work. Several studies highlight that ensuring reproducibility remains one of the most pressing challenges in computational science. For example, reproducibility is often cited as a core benefit of SWSs, yet studies reveal that ensuring reproducibility in practice remains challenging. Molder et al. \citep{molder2021sustainable} emphasized the importance of sustainable practices in workflow development, showing that inconsistent environment setups often hamper reproducibility. Studies such as Perez et al. \citep{santana2015towards} identified that reproducibility issues arise due to poorly documented workflows and unclear execution contexts. Chirigati et al. \citep{chirigati2016reprozip} proposed \emph{ReproZip} as a solution to improve computational reproducibility, though they acknowledge the lack of standardization across platforms as a limitation.

Reusability is another major concern in SWSs. Garijo et al. \cite{garijo2017abstract} proposed the Open Provenance Model for Workflows (OPMW), an end-to-end framework to publish and share workflows using Linked Data principles. Their work showed that many workflows lack rich metadata, making discovery, understanding, and reuse difficult. This finding aligns with our observation that many developer discussions on GitHub and Stack Overflow center on difficulties in reusing or adapting existing workflows due to unclear structure or insufficient context. Similarly, Alam et al. \citep{10479414} examined the barriers to reusability in scientific workflows, noting that workflow portability across platforms remains a significant challenge. 

Collaborative enhancements in SWSs have been explored by the Galaxy project team \citep{galaxy2022galaxy}, who emphasized the importance of community-driven infrastructure and tool interoperability to support large-scale, distributed analyses. Jagla et al. \citep{jagla2011extending} demonstrated how SWSs could be extended to handle next-generation sequencing (NGS) tasks, highlighting the need for domain-specific customization and robust data-handling capabilities. Evaluation frameworks for SWSs are reviewed in studies such as Kiran et al. \citep{kiran2023criteria} and Sethi et al. \citep{sethi2017scientific}, both of which identified key criteria, including usability, reproducibility, scalability, and fault tolerance as essential for assessing workflow system effectiveness.

Di Tommaso et al. \citep{di2017nextflow} specifically addressed reliability in the context of \emph{Nextflow}, showcasing how the use of containers and reproducible environments can support dependable execution, although debugging and performance tuning remain challenges. The design and adoption of the Common Workflow Language (CWL) are examined by Amstutz et al. \citep{amstutz2016common} and Kotliar et al. \citep{kotliar2019cwl}, who emphasized CWL's role in standardizing workflow portability across diverse computing environments. Jain et al. \citep{jain2015fireworks} introduced \emph{FireWorks} \citep{jain2015fireworks} as a workflow platform tailored for high-throughput computational tasks, addressing the need for dynamic scheduling and job prioritization. Ahmad et al. \citep{ahmad2021scientific} investigated workflow scheduling challenges, noting that optimizing resource allocation and task dependencies remains a complex problem in large-scale SWS deployments.

Hundreds of SWSs have been developed to support the design, execution, and management of computational workflows across scientific domains \cite{replicationpackage}. A substantial body of research has examined the core challenges in building and using these systems effectively, including issues of provenance tracking, usability, scalability, and reproducibility \cite{zhao2008scientific, zhao2011opportunities, davidson2008provenance, gil2007examining, mork2015contemporary}. Recent studies have also begun exploring the integration of Large Language Models (LLMs) with SWSs \citep{gu2023plan, sanger2024qualitative, procko2023towards}, identifying the potential of LLMs, such as GPT models to automate workflow construction, streamline data manipulation, and foster more intelligent, collaborative environments for scientific research. These efforts highlighted the growing convergence between AI and workflow technologies, suggesting promising directions for scalable and user-friendly scientific workflow development. 

While prior research has extensively investigated various aspects of SWSs such as reproducibility challenges \citep{chirigati2016reprozip}, usability and collaborative development \citep{galaxy2022galaxy, mork2015contemporary} related issues, and workflow portability and reusability limitations \citep{garijo2017abstract, jain2015fireworks}, these studies have primarily relied on expert reviews, case studies, or controlled experiments. Though such approaches offer valuable system-level insights, they may not fully reflect the day-to-day challenges developers face in practice when designing, maintaining, and adapting SWSs.

To address this gap, our study systematically mined large-scale, real-world developer discussions from GitHub and Stack Overflow to capture practical, community-driven perspectives on SWS development. Notably, we identified topics such as \emph{Workflow Execution} and \emph{Data Structures and Operations} as particularly difficult, reflecting prior concerns around reproducibility and data management\citep{molder2021sustainable, santana2015towards}, but we showed new depth by revealing how these difficulties are operationalized and perceived by developers. In addition, we surfaced procedural pain points, such as \emph{Distributed Task Management}, \emph{System Redesign} and \emph{Managing Rules and Inputs}, which remain underrepresented in existing system-level literature but are essential for the operational success of SWSs deployment and maintenance. Furthermore, unlike previous studies that focus on individual systems, our analysis spans a diverse range of SWSs, enabling cross-cutting insights into common bottlenecks and recurring developer questions. This broader scope helps explain observed discrepancies with earlier findings and allows us to offer a more comprehensive understanding of developer needs. By identifying and analyzing recurring pain points through topic modeling, we contributed a complementary, developer-centric perspective for designing more robust, usable, and adaptable workflow systems, currently underrepresented in the literature.

\begin{figure}[htbp]
  \includegraphics[width=\linewidth]{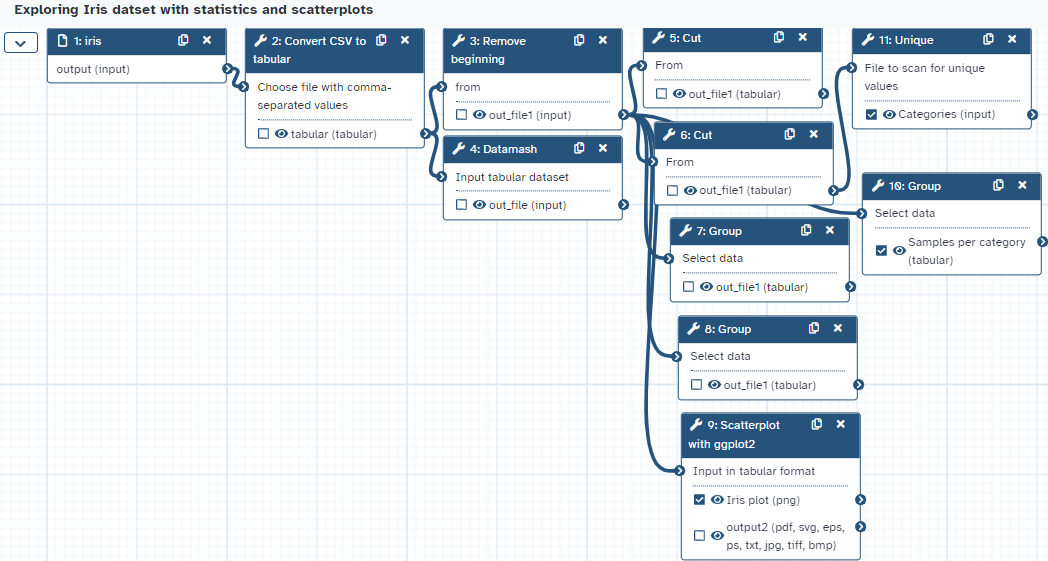}
    \vspace{0.01em}
  \caption{An example of a scientific workflow using Galaxy SWS}
  \label{fig:sw-example}
  \vspace{-1.5em}
\end{figure}
\subsection{Scientific Workflows}
Modern scientific collaborations have opened up the opportunity to solve complex problems that demand multidisciplinary expertise and large-scale computational experimentation. These experiments typically involve a sequence of processing steps that need to be executed on selected computing platforms. However, executing these experiments can be difficult due to the complexity and variety of applications, the diversity of analysis goals, the heterogeneity of computing platforms, and the volume and distribution of data. A common strategy to make these experiments more manageable is to model them as workflows and to use an SWS to organize their execution \citep{liew2016scientific}. A workflow defines the sequence of activities, their dependencies, and the data flow between them, enabling researchers to automate, manage, and reproduce complex scientific computations \cite{barker2007scientific}. It is akin to a cooking recipe, where ingredients represent datasets, and instructions describe the sequence of operations to produce a result (e.g., a plot or new dataset). The order of operations is important, as each step often uses the output of the previous one, similar to sifting flour before mixing it with eggs when baking a cake. This sequence of steps is referred to as a pipeline. Multiple pipelines can combine to form a workflow, much like a menu composed of various recipes. Workflows allow the consistent application of procedures to different datasets by changing the inputs, ensuring adaptability and reproducibility.
\begin{figure}[htbp]
  \includegraphics[width=\linewidth]{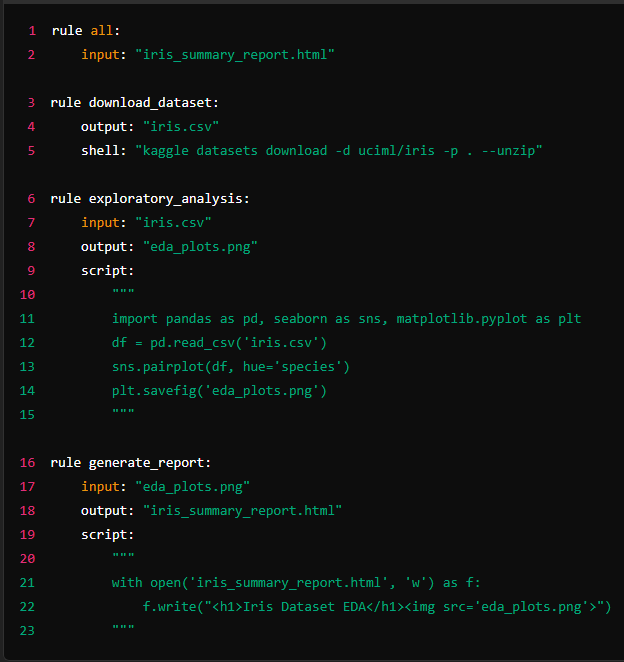}
  \vspace{0.01em}
  \caption{An example of a scientific workflow using Snakemake SWS}
  \label{fig:sw-example-snakemake}
  \vspace{-1.5em}
\end{figure}
Practitioners can share their scientific workflows (from next, we consider workflow as scientific workflows) in several workflow repositories (e.g., WorkflowHub \citep{workflowhub}, myExperiment \citep{myexperimentrepo}, Snakemake workflows repository \citep{snakemakerepo}). These repositories contain thousands of workflows using many SWSs. Figure \ref{fig:sw-example} presents a simple graphic-based workflow that explores the \href{https://www.kaggle.com/datasets/uciml/iris}{Iris dataset} using Galaxy SWS. This workflow performs statistical analysis and creates scatterplots to visualize relationships between its features. By following these steps, we can gain a comprehensive understanding of the Iris dataset's structure and the relationships between its features.

We showed a script-based (DSL) workflow using Snakemake SWS that explores the Iris dataset from Kaggle in Figure \ref{fig:sw-example-snakemake}. This workflow downloads the dataset, performs basic exploratory data analysis (EDA), and generates a summary report.

\subsection{Topic Modeling}
Topic modeling is a statistical modeling technique used to uncover the underlying structure and main themes within a large collection of documents, making it easier to organize, search, and understand the content \citep{jelodar2019latent}. Its applications span various text analysis tasks \citep{wang2023identifying}, including document clustering \citep{yau2014clustering}, information retrieval \citep{yi2009comparative}, recommendation systems \citep{luostarinen2013using}, language translation \citep{eidelman2012topic}, and content summarization \citep{belwal2023extractive}. Researchers have extensively explored topic modeling, publishing numerous articles in software engineering, political science, medicine, and linguistics. Extensive utilization of topic modeling is observed in the software engineering research community \citep{asuncion2010software}, particularly in domains like mining software repositories \citep{bagheri2014adm, jo2011aspect, zhai2011constrained, chen2012explaining, chen2016survey, thomas2011mining, thomas2011modeling, tian2009using}, source code analysis \citep{gethers2010using, linstead2007mining, linstead2008application, lukins2010bug, savage2010topic, tian2009using}, spam detection \cite{10527342}, and recommendation systems \citep{cheng2016effective, kim2014twilite, lu2015twitter, zhao2016personalized, zoghbi2016latent}, among others. While numerous papers contribute to this field, noting all of them is beyond our scope. Hence, we highlighted a selection of significant papers. Several methods, such as BERTopic \citep{grootendorst2022bertopic}, Latent Dirichlet Allocation (LDA) \citep{blei2003latent}, Non-Negative Matrix Factorization (NMF) \citep{lee2000algorithms}, Latent Semantic Analysis (LSA) \citep{landauer1998introduction}, Correlated Topic Model (CTM) \cite{blei2006correlated}, Dynamic Topic Models \citep{blei2006dynamic}, Word Embedding-Based Models \citep{blei2006dynamic}, and Biterm Topic Model (BTM) \citep{yan2013biterm}, contribute to the advancements of topic modeling.

Many of the previous studies have employed the LDA algorithm \citep{blei2003latent} or its variants for topic extraction. However, recent research has increasingly turned to BERTopic \citep{grootendorst2022bertopic} for similar analyses, such as the analysis of scientific papers \citep{10285737}, spam detection \citep{10527342}, research trends on language models \citep{10487248, doi2024topic}, natural language processing \citep{9854488}, and the integration of machine learning \citep{atzeni2022systematic}. While LDA has been a foundational and widely used method in topic modeling, it comes with significant limitations. For instance, LDA requires the number of topics to be fixed in advance, which can be difficult to determine and often necessitates manual tuning. Its reliance on a bag-of-words representation overlooks word order and context, leading to the potential loss of crucial semantic information. Moreover, as the size of the dataset or the number of topics increases, LDA's computational demands rise, making it less scalable. The algorithm is also prone to generating incoherent topics, especially in the presence of noisy or extensive datasets, and its performance is highly sensitive to hyperparameter settings, requiring extensive experimentation.

In contrast, BERTopic \citep{grootendorst2022bertopic} addresses these limitations with several advanced features. It automatically determines the number of topics, captures contextual information through transformer-based models like BERT, and is capable of handling larger datasets. BERTopic also employs hierarchical clustering and dynamic topic modeling, allowing it to adapt to changes in the data and generate more granular and interpretable topics. These strengths make BERTopic particularly well-suited for applications where a deep semantic understanding is essential. Given the short and highly technical nature of SO posts, GitHub issues, and pull requests, BERTopic is the best choice for these data. Its ability to capture nuanced semantics makes it superior for identifying contextually relevant topics. Therefore, this study plans to utilize BERTopic \citep{grootendorst2022bertopic} to leverage these strengths.
\subsection{Topic Analysis of Technical Q\&As}
Topic modeling has significant applications in technical Q\&A data, facilitating the analysis and categorization of large volumes of questions and answers. By identifying underlying themes and topics within the content, topic modeling helps to organize and index data, making it easier for users to find relevant information. For instance, in forums such as Stack Overflow, topic modeling can cluster similar questions, aiding in quicker retrieval of solutions and reducing redundancy \citep{nie2017data}. Additionally, it can track trending topics, monitor user interests, and identify common issues or knowledge gaps, proving valuable for both users seeking information and developers aiming to improve their products or services \citep{fiscus2002topic}. Topic models are extensively used in many studies to understand the topics of general SO posts and track topic trends \citep{barbosa2020software, wang2013empirical, chen2019modeling, allamanis2013and}. Prior work also leverages topic modeling in specific application development domains, such as mobile application development \citep{rosen2016mobile}, machine learning application development \citep{alshangiti2019developing}, concurrency development \citep{ahmed2018concurrency}, security-related development \citep{yang2016security}, DevOps development \citep{10371495}, chatbot development \citep{abdellatif2020challenges}, classification of database technology \citep{9509047}, and understanding nonfunctional requirements \citep{zou2017towards}. To the best of our knowledge, topic analysis of SWSs-related posts on SO is still pending, but SWS is a promising field of software engineering. Therefore, we recognized the need to identify the topics of SWSs, and we utilized BERTopic \citep{grootendorst2022bertopic} because of its contextual understanding capacity and high accuracy.
\subsection{Topic Analysis of GitHub Data}
Topic modeling with GitHub data involves analyzing text from repositories, issues, pull requests, and comments to identify prevalent themes or topics. This technique can uncover trends, common issues, or popular features within software development communities. Researchers apply algorithms like LDA, BERTopic, and others to categorize the vast amount of unstructured text data on GitHub, helping developers and organizations better understand user needs, project focus, and collaboration patterns. Previous studies have utilized topic modeling to analyze GitHub data, revealing prominent challenges developers face, user requirements, and emerging trends in software projects. This insight assists project managers in task prioritization and resource allocation while promoting streamlined collaboration and well-informed decision-making. By delving into the dynamics of software development communities, researchers contribute to improving software engineering methodologies. For example, Scoccia et al. \citep{scoccia2021challenges} used LDA topic modeling to identify challenges faced by web developers, correlating real-time GitHub issues with queries on SO to explore common areas in web app development. Similarly, Li et al. \citep{li2021understanding} applied LDA topic modeling to understand challenges in quantum software engineering by analyzing SO posts and GitHub issue reports. Several other studies \citep{dhasade2020towards, jokhio2021mining, campbell2015latent, chen2016survey, wang2019does, treude2019predicting} have utilized topic modeling with GitHub datasets. Given that GitHub issues and pull requests are often brief and context-specific, BERTopic's embedding-based approach is particularly well-suited for analyzing this data. Therefore, we chose to utilize BERTopic \citep{grootendorst2022bertopic} for our analysis.

\section{Study Design}\label{study-design}
\begin{figure}[htbp]
  \includegraphics[width=\textwidth]{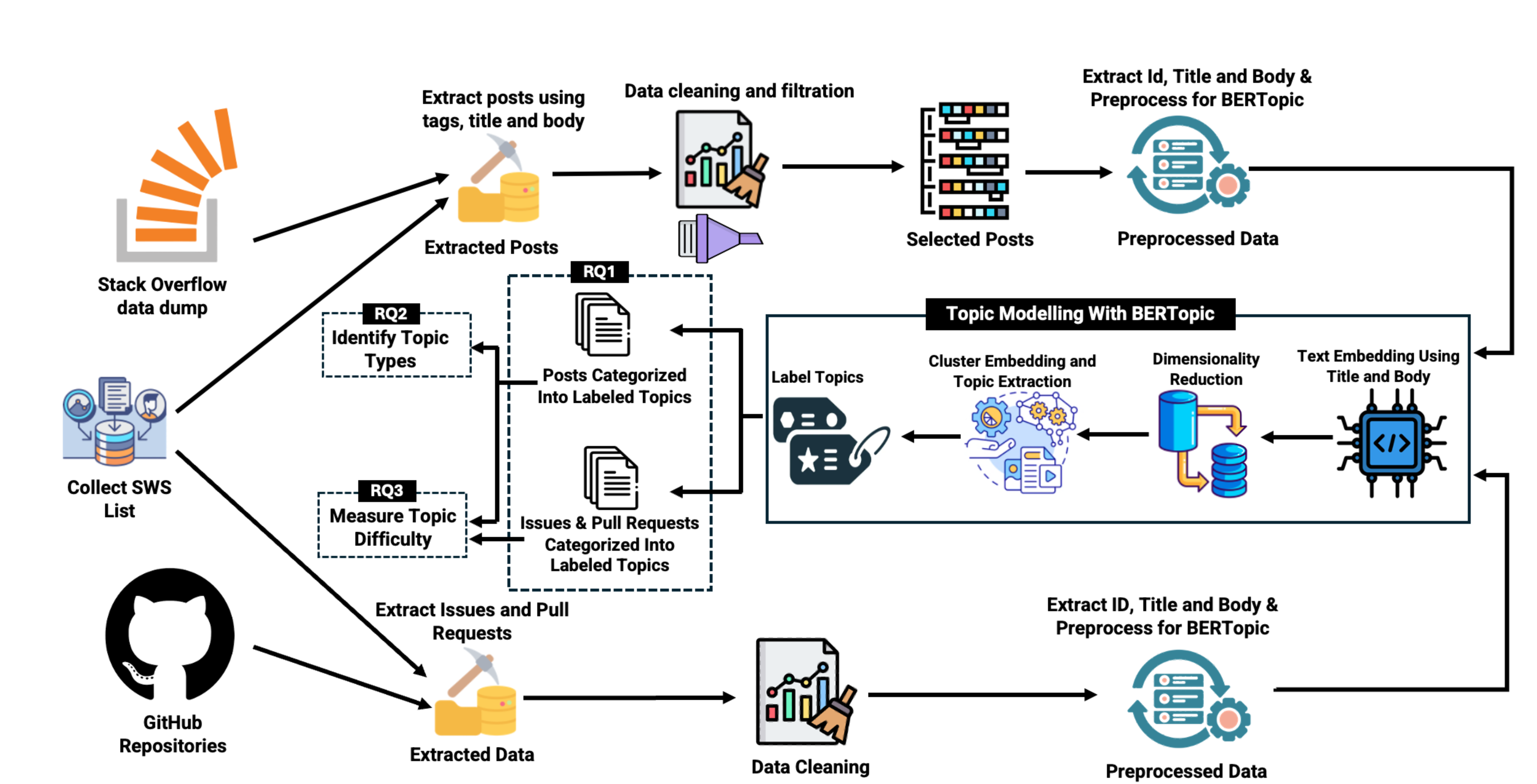}
  \caption{Overview of the methodology of our study}
  \label{fig:studydesign}
  \vspace{-1.0em}
\end{figure}
The main goal of our study is to examine what SWSs developers are asking about. To achieve this goal, we analyzed discussions on SO posts, and issues \& pull requests on GitHub. These platforms provide rich datasets, frequently used in research across various domains, including quantum software engineering \cite{li2021understanding}, cryptography APIs \cite{nadi2016jumping}, concurrency \cite{ahmed2018concurrency}, and deep learning \cite{han2020programmers}. SO offers structured data with questions, answers, and metadata (e.g., accepted answers, scores) but lacks fine-grained topic information related to SWSs. Hence, we first needed to identify posts from SO related to SWSs, group them according to their dominant topic, and then conduct our analysis. For GitHub, we focused on projects related to SWSs identified from our SO data, ensuring consistency across both platforms in the projects we analyzed. Figure \ref{fig:studydesign} shows the overall study design of our research. Below, we described the study design approaches of our work.

\subsection{Stack Overflow Data Collection and Prepossessing} \label{sopreprocessing}
Stack Overflow is a widely used technical Q\&A forum among developers. Therefore, we chose it as one of our data sources. Below, we outlined the steps involved in collecting and preprocessing SO data.
\newline
\textbf{Step 1: Gather Scientific Workflow Systems List. }
Researchers typically use tag-based search functionality on SO to retrieve data, as authors tag their posts with tags  (e.g., javascript, python, arrays) to enhance visibility and increase the possibility of receiving answers \citep{barua2014developers}. However, the field of SWSs spans multiple disciplines (e.g., bioinformatics \citep{spjuth2015experiences}, geospatial analysis \citep{jaeger2005scientific}, software engineering \citep{carver2007software}, machine learning \citep{nouri2021exploring}, grid computing \citep{yu2005taxonomy}), leading to diverse terminologies used by different users. This terminological inconsistency presents a challenge when attempting to conduct searches related to SWSs on SO. Additionally, SWSs have different names (e.g., Galaxy \citep{giardine2005galaxy}, NextFlow \citep{di2017nextflow}, SnakeMake \citep{koster2012snakemake}, Dask \citep{rocklin2015dask}) depending on the specific domain or community, further complicating the process of locating these systems through specific tags on SO. Thus, we plan to gather the SWSs list and use it to identify the posts related to them. We gathered a comprehensive list of SWSs by exploring multiple sources, including curated repositories and official documentation \citep{swfmswiki, swfmspipeline, swfmscwl, wcs}, to ensure broad coverage across various domains and platforms. Upon merging these lists, we identified overlaps, necessitating the removal of duplicates to ensure the integrity and completeness of our compilation. We also added the web link and GitHub link of each SWS wherever possible. Finally, we listed 352 SWS, which can be accessed using the replication package \citep{replicationpackage}.
\newline
\textbf{Step 2: Collecting SWSs-related posts from Stack Overflow. }
We collected posts from SO using the Stack Exchange Data Explorer \citep{stackexchangeforum} for the period between August 2008 and January 2024. As there are no predefined tags available for searching SWSs-related posts, we searched using each system's name, gathered in the \textbf{Step 1}. We extracted data by using SWS names within \emph{Tags}, \emph{Title}, and \emph{Body}. Consider, for instance, an SWS, \emph{celery}. Initially, we filtered posts using the Tags \emph{celery} and subsequently scrutinized whether any titles or bodies contained the term \emph{celery} (this involved utilizing an SQL query to extract the dataset, such as \emph{SELECT * FROM posts WHERE LOWER(Title) LIKE '\%celery\%'}). Throughout this process, we exclusively focused on questions (\emph{identified by posttypeid=1}). We identified posts tagged with celery that also featured celery in their Title or Body. In such instances, we selected only one post to eliminate redundancy in our dataset. In this way, we obtained 9628 posts about celery SWS by January 31, 2024.

We took precautions against potentially noisy data to ensure the accuracy of our dataset and remove any irrelevant information. Despite the possibility of encountering posts with the term \emph{celery} in their title or body unrelated to the celery workflow system, we conducted a meticulous filtering process. By initially narrowing down our dataset using \emph{celery} tags, we identified 9,499 posts specifically discussing the \emph{celery} workflow system. This validation was reinforced by cross-referencing the tags' descriptions provided by SO.
Subsequently, we were left with 139 posts, each potentially containing content unrelated to the \emph{celery} SWS. These posts varied, with some mentioning the term \emph{celery} in contexts unrelated to our focus, such as referring to the \emph{vegetable, a project name, a part of a filename, or a list ID}. To meticulously ensure the precision of our dataset, each of these remaining 139 posts underwent manual scrutiny. Initially, the first author, having over four years of research experience with SWSs and nine years of experience in software development, meticulously reviewed each post, categorizing them as either related or unrelated to the celery workflow system. Following this initial review, the second author, with over eight years of research experience with SWSs, further scrutinized the posts. We only included posts in our final dataset once we all agreed they were relevant. This rigorous process enabled us to identify and exclude six posts unrelated to the celery workflow system (provided in the replication package). Consequently, we finalized our dataset with 9622 posts exclusively pertinent to the \emph{celery} workflow system.

In some cases, we encountered posts where all content was directly pertinent to SWSs. For instance, in the case of \emph{SnakeMake}, our extraction process yielded a total of 1,696 posts, meticulously considering tags, Title, and Body. Remarkably, 1,695 of these posts were tagged with \emph{snakemake} SWS, solidifying their relevance to the \emph{SnakeMake} SWS. The one remaining post not tagged with SnakeMake also discussed the SnakeMake SWS, further affirming its alignment with our focus. Similar outcomes were observed for \emph{KNIME}. A thorough investigation led us to identify a grand total of 308 posts related to KNIME, with 305 of them bearing the \emph{knime} tags. The remaining three, lacking the \emph{knime} tags, were inherently related to KNIME, underscoring their inclusion in our dataset. Posts that were not explicitly tagged with \emph{knime} and \emph{snakemake} underwent the same rigorous manual review process that we applied for \emph{celery}, ensuring their relevance to the respective SWSs.

Utilizing similar approaches, we explored the \emph{Galaxy} SWS related posts. Our search encompassed Tags, Title, and Body, yielding a substantial total of 5652 posts. Next, we examined the co-occurring tags to identify potential noise. For the Galaxy dataset, we observed that many posts were also tagged with tags such as \emph{android} that were unrelated to Galaxy SWS. Consequently, we opted to discard these posts, amounting to the exclusion of 4378 entries. Subsequently, from the remaining 1274 posts, we further refined our dataset by filtering out posts tagged with \emph{samsung, mobile}, or containing these terms in their title and body. This process led to the removal of 510 additional posts. With 764 posts remaining, we then focused specifically on those related to Galaxy SWS by filtering with the term ansible, associated with Galaxy. This step resulted in the identification of 165 posts relevant to Galaxy. Continuing our scrutiny, we observed 26 posts concerning PHP, 35 posts discussing Java, and 77 posts referencing Galaxy S4-S7. These posts were deemed irrelevant to Galaxy SWS and were consequently excluded. From the remaining 461 posts, following a similar approach discussed for \emph{celery}, the first author assessed each post, determining their relevance to Galaxy SWS. Subsequently, the second author verified these assessments. Only posts that were unanimously agreed upon as relevant to Galaxy SWS were retained. Ultimately, this rigorous vetting process yielded ten posts directly related to Galaxy SWS. Thus, our finalized dataset comprises a total of 175 posts aligned with Galaxy SWS.
\begin{table}[htbp]
    \centering
    \caption{Posts and Issues \& Pull Requests Distribution of SWSs}
    \label{tab:swsposts}
    \begin{tabular}{r|c|r|r|r}
        \toprule
        SL. & SWS Name & Total Posts & Filtered Posts &No. of Issues\& Pull Requests\\
        \midrule
         1 & Airflow & 10602 & 10589 & 30168 \\
         2 & SnakeMake & 1696 & 1696 & 5945 \\
         3 & Celery & 9628 & 9622 & 12512 \\
         4 & NextFlow & 419 & 418 & 5075\\
         5 & KNIME & 308 & 308 & 145 \\
         6 & Galaxy & 5648 & 175 & 38808 \\
         7 & Oozie & 2066 & 2052 & 477 \\
         8 & Drake & 846 & 846 & 31186 \\
         9 & Luigi & 362 & 348 & 3191 \\
         10 & Dask & 4681 & 4662 & 27300 \\
         11 & NiFi & 5187 & 5187 & 8311 \\
        \bottomrule
    \end{tabular}
\end{table}

For airflow, we obtained 10602 posts and we found 10448 posts with airflow SWS tags (This validation was supported by cross-referencing the tag descriptions on SO, confirming that the posts were indeed related to the airflow SWS) and 31 posts having apache in title and body. Then, from the remaining 123 posts we manually checked each posts and found 13 posts unrelated with Airflow SWS. They are mainly talking about heat transfer problems, airflow temperature, respiratory airflow, function name etc. Thus, finally we have 10589 posts related with Airflow SWS.

As there are hundreds of SWS, describing each SWS is beyond our scope, though our approach is identical to that described above. In Table \ref{tab:swsposts}, we showed the SWSs, total posts obtained after the query, and total filtered posts. For simplicity, we display the SWSs where we found more than 100 posts for a particular SWS. Our further analysis is also based on the dataset shown in Table \ref{tab:swsposts}. These systems are publicly available state-of-the-art systems designed for a broad spectrum of scientific analysis tasks. We gather a total of 41,443 posts. After filtration, we identified 5,540 posts unrelated to the scientific workflow system, and the remaining 35,903 posts are about SWS.

\begin{figure}[htbp]
  \includegraphics[width=\textwidth]{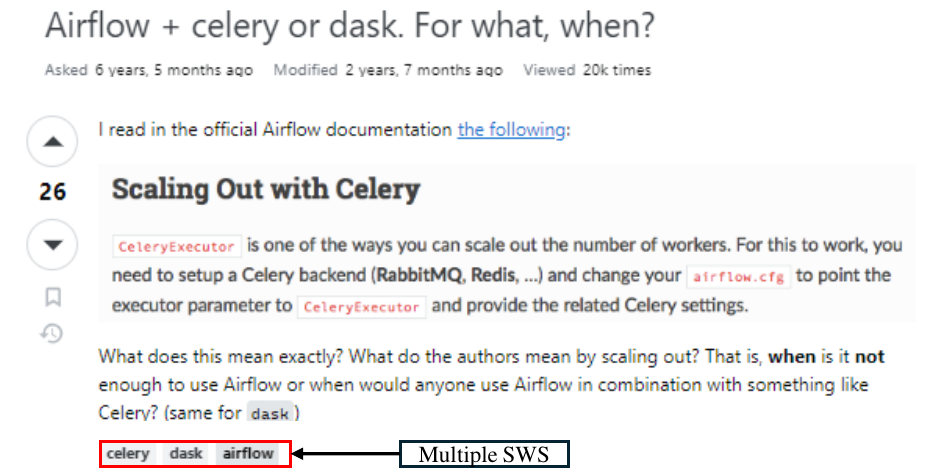}
  \vspace{0.01em}
  \caption{A post discussed about three SWSs}
  \label{fig:multiplesws}
  \vspace{-1.5em}
\end{figure}

We identified several posts discussing multiple SWSs, causing the same post to appear across various workflow platforms. As illustrated in Figure \ref{fig:multiplesws}, one such instance involves a post discussing three SWS: Airflow, Dask, and Celery. 
As our data-gathering process queries each SWS individually, these posts are visible across all three systems. To maintain impartiality in our analysis, we selected only one instance of such posts for our examination, and we finally gathered 35,619 unique posts related to SWS. The collected data are available in the replication package \citep{replicationpackage}.
\newline
\textbf{Step 3: Preprocessing Scientific Workflow Systems' posts. }We first filtered out the irrelevant information before applying topic modeling techniques. While the post Title concisely summarizes the question, the Body provides crucial context and details that help us accurately identify the topic being discussed. Therefore, in this analysis, we focused on both the post Title and Body. However, the Body content might introduce noise to our analysis, so we removed quotes, HTML tags, links, and any code snippets using regular expressions from the Body of the posts. We also removed stopwords, commonly used words in the English language, such as 'how', 'can', and 'at' that do not significantly affect the meaning of a sentence and can introduce bias. We used the NLTK stopwords list \citep{bird2006nltk} for this operation. Then, we performed lemmatization, which involves reducing a word to its canonical form called a lemma, taking into account the word's linguistic context (for example, the word 'scale' is the lemma of the word 'scaling'). All the data pre-processing steps and processed data are provided in the replication package \citep{replicationpackage}. Below, we provide a concise description of the steps we follow to pre-process the SWS-related SO data.
\newline
\textbf{Removing code snippets and HTML tags: }Before applying topic modeling, removing code snippets and HTML tags is essential to ensure cleaner and more meaningful topic extraction. These elements introduce noise by adding structural or syntactic artifacts that do not contribute to the actual discussion. HTML tags, such as \texttt{<div>}, \texttt{<a href>}, and \texttt{<p>}, and programming syntax, like \texttt{\{\}}, \texttt{()}, \texttt{;}, \texttt{:}, and \texttt{if-else} statements, can dominate topics if left unprocessed, leading to misleading results. Retaining code snippets may also cause the model to cluster topics based on programming languages or markup structures rather than the underlying challenges being discussed. Removing these elements allowed the topic model to focus on natural language, improving topic coherence and interpretability. We utilized the \emph{BeautifulSoup} Python library to remove code snippet HTML tags.
\begin{lstlisting}
def clean_html_and_code(text):
    soup = BeautifulSoup(text, "html.parser")
    # Remove code blocks
    for code_block in soup.find_all("code"):
        code_block.decompose()
    # Remove HTML tags
    cleaned_text = soup.get_text(separator=" ")
    return cleaned_text
\end{lstlisting}
\textbf{Stopwords and special character Removal: }To improve the quality of our topic modeling process, we applied stopword removal to eliminate commonly used words in the English language that do not significantly contribute to the meaning of a sentence and may introduce bias. For this, we utilized the NLTK stopwords list \citep{bird2006nltk} and further expanded it by appending names of specific SWSs, such as Nextflow, Galaxy, and Snakemake. This refinement ensures that the topic model focuses on the core issues discussed in the dataset rather than being influenced by frequent mentions of specific workflow system names. To implement this process efficiently, we defined a Python function to systematically remove these stopwords from our data, enhancing the clarity and relevance of the extracted topics.

\begin{lstlisting}
def remove_stopwords(text):
    words = text.split()
    filtered_words = [word for word in words if word.lower() not in stop_words]
    return ' '.join(filtered_words)
\end{lstlisting}
To remove special characters (i.e., \textbackslash n, @, \#), we utilized a regular expression,  \texttt{re.sub(r'\^{}[a-zA-Z0-9\textbackslash s]', '', text)}
\newline
\textbf{Lemmatization: }Lemmatization is the process of converting words to their base form (lemma). One of the most efficient ways to perform lemmatization is by using \emph{spaCy} \citep{vasiliev2020natural}, a fast and optimized NLP library. To achieve this, we loaded the English language model (\emph{en\_core\_web\_sm}) and processed the text with spaCy's pipeline. We transformed each token in the text into its lemma using the \emph{.lemma\_} attribute. We chose this approach because of its high speed and contextual understanding.
\begin{lstlisting}
    def lemmatize_text(text):
    doc = nlp(text)
    lemmatized_tokens = [token.lemma_ for token in doc]
    return ' '. join(lemmatized_tokens)
\end{lstlisting}
\subsection{GitHub Data Collection and Preprocessing}
To complement our analysis of developer challenges on SO, we also examined issue reports and pull requests from GitHub, a widely used platform for collaborative software development. GitHub repositories provide direct insights into real-world problems encountered during the development and maintenance of SWSs. In this subsection, we described our strategy for collecting and preprocessing GitHub data to enable effective topic modeling and challenge identification.
\newline
\textbf{Step 1: Collecting GitHub Issues and Pull Requests. }We utilized the same SWSs listed in Table \ref{tab:swsposts} for our analysis on GitHub. To gather data on issues and pull requests, we employed the GitHub REST API \citep{githubrestapi}. We observed that several SWSs are distributed across multiple repositories during this process. For instance, the \emph{Galaxy} workflow system spans several repositories (e.g., tools-iuc, planemo, galaxy-hub) related to its ecosystem. Similarly, \emph{Nextflow} is associated with multiple repositories (e.g., patterns, rnaseq-nf, nf-validation). Consequently, we aggregated all issues and pull requests from these relevant repositories.

As of January 31, 2024, our initial GitHub dataset comprised 169,692 issues and pull requests. However, we observed that many pull requests  were directly linked to issues, and treating them as independent entries could introduce redundancy into the analysis. To address this, we utilized the GitHub Search API to identify and exclude 6,574 PRs that were associated with existing issues. After this refinement, our final dataset consisted of 163,118 distinct and non-redundant issues and PRs. The distribution of this data is presented in Table~\ref{tab:swsposts}. All scripts used for data collection, PR-issue linkage detection, and filtering are provided in our replication package~\citep{replicationpackage} to ensure transparency and reproducibility.
\newline
\textbf{Step 2: Preprocessing GitHub Data. }To prepare the GitHub issues and pull requests for topic modeling, we applied a series of preprocessing steps to clean and normalize the textual data. First, we concatenated the title and body of each issue and pull request to form a single document, ensuring sufficient context for topic extraction. We then removed code snippets enclosed in backticks using regular expressions, as such blocks often contain syntax elements irrelevant to semantic modeling. Next, we eliminated URLs, which typically reference external documentation or repositories, and cleaned out any remaining HTML tags using regular expressions. To further reduce noise, we removed special characters and punctuation. We also applied stopword removal using the NLTK stopword list and supplemented it by filtering out project-specific terms such as \emph{airflow, celery, and nextflow} to prevent these frequently occurring system names from dominating the topic space. Finally, we performed lemmatization using \emph{spaCy’s en\_core\_web\_sm} model to convert words into their base forms, thus improving topic coherence and reducing lexical variation. 
\newline
\textbf{Removing Code Snippets: }To remove code snippets from GitHub issues and pull requests, we used regular expressions to detect Markdown-style formatting. We identified and removed inline code enclosed within single backticks and multi-line code blocks enclosed in triple backticks. This preprocessing step was essential to prevent programming syntax and formatting elements from distorting the semantic structure of the text during topic modeling.
\begin{lstlisting}
def remove_code_snippets(text):
    if not text:
        return text
    # Remove inline code (e.g., `code`)
    text = re.sub(r'`[^`]*`', '', text)
    # Remove block code (e.g., ```code```)
    text = re.sub(r'```[\s\S]*?```', '', text)
    return text
\end{lstlisting}
\textbf{Remove URLs: }Many issues/PRs contain links to external documentation, repositories, or Stack Overflow discussions. These URLs do not contribute to topic modeling; thus, we removed them using regular expressions.
\begin{lstlisting}
def remove_https_links(text):
    if isinstance(text, str):
        # Regular expression pattern to match HTTPS links
        https_pattern = r'https://\S+'
        # Remove all matches from the text
        return re.sub(https_pattern, '', text).strip()
    else:
        # If not a string, return as is
        return text
\end{lstlisting}
\textbf{Removing HTML Tags: }We removed HTML tags using regular expression.
\begin{lstlisting}
def remove_html_tags(text):
    if text is None:
        return ''
    return re.sub(r'<.*?>', '', text)
\end{lstlisting}
\textbf{Removing Stopwords \& Project-Specific Terms: }We applied the same approach for removing stop words and project-specific terms as described in the SO data preprocessing step.
\newline
\textbf{Removing Special Characters and Punctuation: }We utilized the same approach as we did at SO data preprocessing for this step
\newline
\textbf{Filtering Low-Information Issues and Pull Requests: }After completing all the preprocessing steps, we identified a number of GitHub issues and pull requests containing only one or two words, such as \emph{update, Jpython, typo, or Add Tags}, which lacked meaningful context or insight. To improve data quality, we removed entries where the text length was fewer than eight characters. This filtering step resulted in the exclusion of 636 issues and pull requests, yielding a final curated dataset of 162,169 GitHub issues and pull requests.
\newline
\textbf{Lemmatization: }For this step, we again follow the process we used in SO data lemmatization.
\subsection{Identify Scientific Workflow Systems' topics. } \label{identify-topics}
BERTopic starts with transforming input documents into numerical representations. While there are several methods to accomplish this, \emph{SentenceTransformers}\footnote{\href{https://huggingface.co/sentence-transformers}{https://huggingface.co/sentence-transformers}} stands out as a state-of-the-art technique for generating sentence and text embeddings. Renowned for its ability to capture semantic similarities between documents, it is among the most popular choices for this task. There are many pre-trained models available for \emph{SentenceTransformers}, all hosted on the Huggingface Model Hub \citep{huggingfacemodels}. The \emph{all-*} models were trained on an extensive dataset of over one billion training pairs, making them suitable for general-purpose applications. In addition, there are \emph{Multi-QA} models trained on 215 million question-answer pairs from diverse sources, including StackExchange, Yahoo Answers, and search queries from Google and Bing. Notable models in this category include \emph{multi-qa-MiniLM-L6-dot-v1, multi-qa-distilbert-dot-v1, and multi-qa-mpnet-base-dot-v1}. The \emph{multi-qa-mpnet-base-dot-v1} model excels in semantic search performance, whereas the \emph{multi-qa-MiniLM-L6-dot-v1} model is optimized for speed. We chose to use the \emph{multi-qa-MiniLM-L6-dot-v1} model for our embedding needs because it is trained on question-answer data, which closely aligns with the nature of our dataset.

In BERTopic, the \emph{nr\_topics} parameter allows control over the number of topics by merging similar ones after their initial creation. However, best practices recommend using a clustering model to determine the topic structure more naturally \cite{li2021understanding}. To enable accurate and coherent clustering, we first reduced the dimensionality of the document embeddings, as high-dimensional data can impair clustering performance. Among dimensionality reduction techniques, \emph{UMAP} \citep{mcinnes2018umap} is widely recognized for its ability to preserve local structure when projecting data to lower dimensions. Therefore, we employed UMAP for this step. We then applied \emph{HDBSCAN} \citep{mcinnes2017hdbscan}, a density-based clustering algorithm that complements UMAP well due to its sensitivity to local structure. Unlike traditional clustering methods, HDBSCAN can identify noise and does not force every data point into a cluster, making it especially effective for short, noisy texts such as SO posts and GitHub issues

For BERTopic modeling on both datasets, we carefully tuned UMAP parameters to balance topic granularity and semantic coherence. We experimented with \emph{n\_neighbors} values between 15 and 50 to ensure that semantically similar documents were grouped together while preserving a meaningful global topic structure. The \emph{n\_components} parameter was adjusted between 3 and 20 to retain essential semantic information from the high-dimensional sentence embeddings while enabling efficient clustering. We used \emph{cosine distance} as the metric, as it aligns well with sentence-transformer embeddings and effectively captures angular similarity. For \emph{HDBSCAN}, we tested \emph{min\_cluster\_size} values from 50 to 300 in intervals of 10. We employed \emph{Euclidean distance} for clustering, as it performs reliably in UMAP-reduced space, which is optimized for Euclidean geometry. Additionally, we used the \emph{eom} method for \emph{cluster\_selection\_method} because it identifies stable, well-separated clusters by emphasizing dense core regions. We improved the default representation of topics using the \emph{Countvectorizer} \footnote{\href{https://scikit-learn.org/stable/modules/generated/sklearn.feature\_extraction.text.CountVectorizer.html}{https://scikit-learn.org/stable/modules/generated/sklearn.feature\_extraction.text.CountVectorizer.html}}. It helps to ignore infrequent words and increase the n-gram range. Following previous works \citep{li2021understanding, abdellatif2020challenges}, we used the unigram and bigram models in Countvectorizer. The optimal topic coherence score of 0.64 was achieved for the GitHub dataset with \emph{n\_neighbors} = 20, \emph{n\_components} = 4, and \emph{min\_cluster\_size} = 130, while the SO dataset yielded the optimal score of 0.63 with \emph{n\_neighbors} = 30, \emph{n\_components} = 3, and \emph{min\_cluster\_size} = 210. This configuration enabled BERTopic to generate high-quality, interpretable topics while maintaining scalability across both dataset.

Using the above-mentioned parameters, we obtained 10 topics for SO data and 13 topics from the GitHub data. BERTopic includes a \emph{-1} topic to group documents that do not align well with the identified topics. This situation may arise when a document contains excessive noise, such as stopwords or irrelevant information, or does not strongly connect to the main themes identified by the model. The \emph{-1} topic serves as a 'catch-all' for these outlier documents. However, since we preprocessed the data to eliminate such elements before applying BERTopic, our main topic began with the label \emph{-1}. The scripts for data preprocessing and topic generation for SO and GitHub data are provided in the replication package \citep{replicationpackage}. We described our findings in the following Section.

\section{Results}\label{case-study-results}
In this section, we presented the analysis of the SWS-related data we obtained from SO and GitHub and the topics to answer our research questions.
\newline
\subsection{RQ1: What topics are SWSs developers asking about on Stack Overflow, and which topics are raised in the GitHub projects?} 

\textbf{Motivation: }Scientific Workflow Systems (SWSs) development has unique characteristics that set it apart from traditional software development \cite{liew2016scientific}. For example, SWSs developers need expertise in complex data management and the involvement of specialized software, high-performance resources, and hardware. SWSs should be adaptable to frequent changes, and adherence to community standards, reproducibility, and computational efficiency are crucial. Multidisciplinary team involvement and usage of domain specific languages are often necessary. These particularities make SWS development a complex endeavor, requiring a deep understanding of scientific domains and advanced software engineering practices, which are often unnecessary or overlooked in most conventional software development tasks. Hence, the challenges SWSs developers face may likely differ from those of traditional software development. Developers use Q\&A websites to communicate both problems and solutions, and they rely on GitHub for version control and collaboration, including code review, project management, and collective software development. This research question aims to analyze invaluable data from SO and GitHub to identify the most common and pressing topics related to SWSs and to explore the relationships between these topics on both platforms. By pinpointing the challenges frequently encountered by the SWSs community, we can highlight which topics are gaining traction and which are proving challenging to address. Understanding these trends is crucial for supporting the SWSs community and guiding future development efforts.
\newline
\textbf{Approach: }To identify the key topics discussed by developers, we applied BERTopic to both the Stack Overflow and GitHub datasets, as described in Section~\ref{study-design}. The BERTopic modeling process resulted in 10 distinct topics for the Stack Overflow data, 13 topics for the GitHub data, and the distribution of co-occurring words in each topic. We then manually assigned a meaningful label to each topic. Following established approaches in prior work~\cite{li2021understanding, openja2020analysis, yang2016security, bagherzadeh2019going}, the first author, with over four years of research experience in SWSs and more than nine years of professional software development experience, initially proposed candidate labels. These were derived from the top 15 keywords associated with each topic and a manual review of at least 30 random representative posts, issues, or pull requests per topic. Subsequently, the initial labels were reviewed by the first three authors of the paper in collaborative meetings. The second author contributes over eight years of expertise in SWS research, while the third author brings more than two decades of experience in empirical software engineering. Based on this iterative review and refinement process, we finalized meaningful and contextually accurate labels for all identified topics. Additionally, we examined the most popular SWSs topics among developers. To investigate this, we used two complementary measurements of popularity that have been adopted in prior works \cite{abdellatif2020challenges, ahmed2018concurrency, bagherzadeh2019going, bajaj2014mining, nadi2016jumping}:
\begin{enumerate}
    \item \textbf{The average number of views (avg. views) }of the post received from both registered and unregistered users serves as a valuable metric. A high view count implies significant interest and popularity among SWSs developers. This metric effectively measures community engagement by indicating how frequently a post is viewed.
    \item \textbf{The average score (avg. scores)} of the posts received from the users serves as another important metric for measuring popularity. SO allows its members to upvote posts they find interesting and valuable and downvote posts to maintain high content standards. These votes are aggregated into a score, serving as a metric of the post's perceived community value.
\end{enumerate}

\begin{table*}[htbp]
    \centering
    \caption{The SWSs topics, keywords, and their popularity for SO data.}
    \label{tab:swssotopics}
    \begin{tabularx}{\textwidth}{r|>{\raggedright\arraybackslash}p{3.6cm}|>{\raggedright\arraybackslash}p{5.3cm}|r|r|r}
        \toprule
        SL. & Topic & Keywords & \# Posts & AvgView & AvgSc \\
        \midrule
        1 & Workflow Creation and Scheduling & dag, run, create, task, try, file, error, operator, use, docker. & 10541 & 2656.19 & 1.95 \\
        2 & Distributed Task Management & task, worker, run, queue, start, work, app, celery worker, try, app. & 9447 & 2616.23 & 3.04 \\
        3 & Processor for Data Processing & processor, use, file, flow, datum, json, flowfile, try, value, attribute. & 4995 & 1281.33 & 0.79 \\
        4 & Data Structures and Operations & column, file, dask dataframe, datum, memory, try, code, array, function, read. & 4703& 1203.47 & 1.74 \\
        5 & Workflow Execution & workflow, job, action, hadoop, hive, run, spark, error, xml, file. & 2020 & 1587.62 & 0.97 \\
        6 & Managing Rules, Input, Output, and Files & rule, file, output, input, wildcard, run, error, pipeline, sample, try. & 1691 & 697.58 & 1.28 \\
        7 & Task Dependencies and Management & use, try, error, run, joint, pydrake, task, file, simulation, code. & 1137 & 519.33 & 1.07 \\
        8 & Data Transformation & jolt, spec, output, input, value, transform,  array, expect, use jolt, transformation. & 448& 513.69 & 0.57 \\
        9 & Learning & process, try, channel, file, output, script, pipeline, input, run, use. & 381 & 729.80 & 1.12 \\
        10 & Automation with Ansible & ansible, role, collection, install, yml, module, instal, git, error, run. & 256 & 3159.54 & 3.02 \\
        \bottomrule
    \end{tabularx}
\end{table*}
\textbf{Results Obtained Using SO: }Following the process described in sub-section \ref{identify-topics}, we identified 10 topics for SO data. Table \ref{tab:swssotopics} presents the obtained ten topic titles along with their associated keywords. It also shows the number of posts related to each topic and the popularity of these topics based on our metrics: views and scores received by developers on SO. As shown in Table \ref{tab:swssotopics}, developers inquire about various topics in SWSs development, with the number of posts varying across topics. Below, we discussed these topics in greater detail.

\textbf{1. Workflow Creation and Scheduling: }This is the most dominant topic, representing 29.59\% of posts. It involves defining and organizing a series of computational or data-processing tasks to achieve a particular scientific goal. Workflow creation entails specifying the individual tasks, their dependencies, and the required resources. Scheduling is the process of allocating these tasks to computational resources over time, ensuring efficient execution while considering factors like task priorities, resource availability, and deadlines.

One example of this topic with about 114k views is \emph{'How to create a conditional task in Airflow'}. In this case, the developer was trying to create a conditional task in an SWS based on a schema using \emph{ShortCircuitOperator and/or  XCom} to manage the condition but was unsure how to implement it and asked for a solution. This process is crucial for optimizing performance and resource utilization in complex scientific experiments and data analysis workflows. The prominence of this topic highlights the necessity of more intuitive ways to create complex workflows and manage task execution efficiently.
\newline
The challenges in Workflow Creation and Scheduling arose primarily from complexities in defining, managing, and executing Directed Acyclic Graphs (DAGs) in workflow engines like Apache Airflow. Users often struggle with dynamic job scheduling, ensuring task dependencies are correctly managed, and handling failures or retries efficiently. Issues related to rescheduling tasks, managing large-scale data workflows, and optimizing execution performance further complicate workflow automation. Additionally, debugging and maintaining DAGs, especially when integrating multiple processing steps, require a deep understanding of task execution order and error propagation, making workflow creation and scheduling a demanding task.

\begin{table*}[htbp]
\footnotesize
\caption{Granular Subtopics under Workflow Creation and Scheduling}
\label{tab:workflow_creation_subtopics}
\centering
\begin{tabular}{p{2.7cm}p{2.5cm}p{0.8cm}p{4.5cm}p{3.5cm}}
\toprule
\textbf{Subtopic} & \textbf{Keywords} & \textbf{Post Count} & \textbf{Description} & \textbf{Representative Example} \\
\midrule

DAG Syntax and Structural Issues & dag, task, use, run, file, try, error, work, code & 6116 & Syntactic or structural issues in DAG definitions, including indentation, file format, and import errors. & 'Run task using Python DAG; task not recognized due to indentation error in load definition.' \\
Cloud Composer Configuration and Integration & composer, cloud, google, bigquery, use, cloud composer & 1113 & Cloud Composer-specific problems involving configuration, authentication, and deployment. & 'Google Cloud Composer job fails, BigQuery hook not installed despite dependencies listed.' \\

DAG Scheduling and Triggering Problems & dag, run, task, schedule, trigger, time& 1013 & DAG scheduling inconsistencies, task triggering issues, and incorrect start or end times. & 'DAGs triggered but task doesn't run, possibly due to overlapping schedule and DAG concurrency limit.' \\
Docker-Based Workflow Orchestration & docker, container, compose, run, docker compose, service & 613 & Problems in Docker Compose environments with service orchestration and path bindings. & 'Inside Docker, unable to mount volume correctly, workflow steps fail at execution time.' \\

Installation and Environment Setup Errors & error, install, try, run, apache & 611 & Environment setup and dependency resolution errors, particularly during installation. & 'Error installing Apache Airflow on Windows—command fails at dependency resolution.' \\
Kubernetes Pod Execution and Management & pod, kubernetes, use, run, execute & 509 & Kubernetes-based execution problems, such as pod lifecycle errors and volume mounting issues. & 'Kubernetes pod completes but logs missing, workflow stalled on resource attachment.' \\
XCom and Cross-Task Data Handling in Airflow & xcom, value, task, use, pass, return, operator & 335 & Cross-task communication failures, especially with return values or Airflow's XCom. & 'XCom returns string list instead of single string, task fails in downstream operator.' \\

Spark Workflow Submission and Runtime Errors & spark, job, submit, use, pyspark & 231 & Difficulties in running Spark workflows, job submission, and Spark environment setup. & 'PySpark task stuck in Spark Standalone Cluster, job never exits execution state.' \\
\bottomrule
\end{tabular}
\end{table*}

\textbf{Granular Analysis of Workflow Creation and Scheduling: }To deepen our understanding of how developers construct workflows, we performed a fine-grained BERTopic analysis on the broader topic of Workflow Creation and Scheduling. This decomposition revealed eight distinct subtopics, detailed in Table~\ref{tab:workflow_creation_subtopics}, covering challenges in DAG syntax, cross-task communication, cloud-based orchestration, and environment configuration. Issues range from syntactic errors in DAG files and misconfigured schedules to complex platform-specific concerns in Spark, Kubernetes, and Docker. This analysis highlights that creating reliable and maintainable workflows requires both domain-specific syntax fluency and robust environment orchestration practices.
\begin{table*}[htbp]
\footnotesize
\caption{Granular Subtopics under Distributed Task Management}
\label{tab:dtm_subtopics}
\centering
\begin{tabular}{p{2.7cm}p{2.4cm}p{0.7cm}p{4.5cm}p{3.5cm}}
\toprule
\textbf{Subtopic} & \textbf{Keywords} & \textbf{Post Count} & \textbf{Description} & \textbf{Representative Example} \\
\midrule

Task Initialization and Worker Setup & task, use, run, try, worker, django, work & 3146 & Configuring task workers, setting up runtime environments, and customizing task initialization logic. & 'Detect and register tasks in Django Python, facing problems with dynamic imports during task loading.' \\

Framework-Level Integration & django, task, run, use, file, error & 2463 & Challenges in Celery-Django integration, especially related to app setup, module loading, and execution. & 'Best way to run an indefinite Celery task within Django app while preserving state?' \\

Scheduling and Dependency Management & task, worker, run, use, chain, queue, execute & 1679 & Managing dependencies and chaining across distributed task queues and async calls. & 'Group result chain stuck in Celery, chained tasks execute out of order.' \\

Message Broker Access and Authentication & rabbitmq, task, queue, worker, message, broker & 1077 & Connection/authentication issues with message brokers like RabbitMQ and Redis. & 'RabbitMQ creates idle queues, no consumer connection, task never received.' \\

State Management via Redis Queues & redis, task, use, run, broker, worker & 488 & Configuring Redis-based brokers for task routing, storage, and communication reliability. & 'Main cause for Redis connection error in Celery when using Django + Redis broker URL?' \\

Fault Tolerance and Retry Mechanisms & retry, fail, exception, handle, task, error, recover & 307 & Handling retry policies, exceptions, and implementing fault-tolerant execution patterns. & 'Handle retry logic after task failure with custom exception handler.' \\

Worker Deployment and Execution Management & celery, task, worker, concurrency, run, queue & 287 & Deploying Celery workers at scale, managing concurrency, and orchestrating worker pools. & 'Run multiple Celery workers in parallel on different queues using Docker.' \\
\bottomrule
\end{tabular}
\end{table*}

\textbf{2. Distributed Task Management: }This is the second most prominent topic, accounting for 26.52\% of the posts. This topic discusses issues such as orchestrating complex task sequences, identifying pending tasks while ensuring fault tolerance and scalability, managing resources efficiently, optimizing performance, and handling memory allocation across distributed environments with large datasets. One notable example with 272K views is \emph{'How to retrieve a list of tasks in a queue in Celery?'}, where a developer executing a series of tasks sought help in retrieving a list of unprocessed tasks. Another example with 61K views is \emph{'How do I use the @shared\_task decorator for class-based tasks?'}, where the developer needed guidance on decorating a class-based task. The high volume of posts on this topic indicates a need for improved task management to enhance the utilization of SWSs.
\newline
The core challenge of this topic arose from the complexity of executing, scheduling, and managing tasks across distributed environments. Developers often struggle with task execution failures, misconfigured workers and queues, and difficulties in periodic task scheduling using SWSs such as Celery. For example, in one post, we found that one developer faced challenges in integrating Celery with Django, especially in configurations and dependency management. Furthermore, scalability and performance bottlenecks emerge as systems grow, requiring careful tuning of workers, queues, and memory allocation. These challenges highlight the need for better tooling, documentation, and automation to streamline distributed task execution in SWSs.

\textbf{Granular Analysis of Distributed Task Management: }To investigate the operational complexities of distributed task systems, we conducted a fine-grained BERTopic analysis on the broader topic of Distributed Task Management. As shown in Table~\ref{tab:dtm_subtopics}, this yielded seven subtopics covering key areas such as worker setup, Celery integration, broker connectivity, and task dependency coordination. Posts reveal persistent difficulties in configuring task runtimes, linking asynchronous jobs, and managing reliability across tools like Celery, RabbitMQ, and Redis. These results highlight the technical overhead associated with scaling and maintaining distributed task workflows, particularly in fault-prone and loosely coupled execution environments.
\begin{table*}[htbp]
\footnotesize
\caption{Granular Subtopics under Processor for Data Processing}
\label{tab:processor_subtopics}
\centering
\begin{tabular}{p{2.7cm}p{2.5cm}p{0.7cm}p{4.6cm}p{3.5cm}}
\toprule
\textbf{Subtopic} & \textbf{Keywords} & \textbf{Post Count} & \textbf{Description} & \textbf{Representative Example} \\
\midrule
General Processor Usage in Java Workflows & processor, use, file, java, apache, datum, flow & 2900 & General usage of processors in Java-based workflows, often involving file operations and flow file manipulation. & 'Record maximum timestamp value from flow file content and save it.' \\
NiFi Execution and Error Troubleshooting & use, error, try, apache, run, work, docker & 828 & Challenges in executing and debugging NiFi processors, including setup errors and environment compatibility. & 'Make run-nifi.bat work on Windows; getting runtime errors with processor flow.' \\
Tabular Data Column Configuration & column, value, use, table, datum, csv, processor & 426 & Issues related to modifying or appending tabular data (e.g., columns) in CSV or table-based processors. & 'Add column to CSV using predefined values via Apache processors.' \\
JSON Conversion and Attribute Processing & json, use, value, convert, attribute, array, parse & 396 & Difficulties in converting JSON data, extracting attributes, and handling multiple nested objects. & 'Split multiple JSON objects from a single string into flow files using EvaluateJsonPath.' \\
Kafka Topic and Message Handling & kafka, topic, message, use, datum, consumer, offset & 187 & Integration issues with Kafka, focusing on consuming messages and managing topics and offsets. & 'Read Kafka message starting from last committed offset and process it in NiFi.' \\
Script-Based Data Routing with Conditions & route, attribute, match, regex, path, type & 166 & Conditional routing using processors like RouteOnAttribute, typically based on regex or flowfile metadata. & 'Use RouteOnAttribute to separate XML from JSON files before PutFile processor.' \\
FlowFile Filename and Output Path Issues & putfile, filename, directory, output, path, original & 92 & File output configuration challenges such as naming conventions and output directory setup in PutFile. & 'Configure PutFile to save using original filename rather than timestamp.' \\
\bottomrule
\end{tabular}
\end{table*}

\textbf{3. Processor for Data Processing: }SWSs provide a rich set of processors covering a wide range of data integration, transformation, routing, and interaction tasks. Practitioners can design sophisticated data pipelines to meet their specific data processing needs by selecting and configuring the appropriate processors. This topic covers 14.02\% of the posts. One example of this topic with about 27K views is \emph{'How to use NiFi ExecuteScript processor with Python?'}. \emph{ExecuteScript processor} allows users to write and execute custom scripts within a data flow. This processor provides flexibility for handling complex processing logic that the built-in processors do not cover. Here, the developer tried to use the \emph{ExecuteScript processor} with a simple Python script but failed to see the output and asked for help to explain how to use it with an example. Another example of this topic with 12K views is \emph{'How to specify output filename with PutFile processor?'}. Here, the problem was that the \emph{PutFile processor} was saving files using a timestamp, but the expectation was to save them with the original filename. Thus, the developer was seeking a solution. This topic indicates developers' challenges in understanding data processing processors.
\newline
This challenge primarily stemmed from the complexity of configuring and optimizing SWS processors for handling large-scale data workflows. Developers struggle with file management, data flow orchestration, and custom processor debugging, as seen in SO posts. Integration with external systems, data consistency, and scalability under high loads further complicate processing pipelines. Misconfigurations in SWS processors often lead to inefficiencies, requiring better tooling and documentation for smoother data processing workflows.

\textbf{Granular Analysis of Processor for Data Processing: }Our analysis identified seven (Table \ref{tab:processor_subtopics}) semantically distinct subtopics, highlighting the diverse processing challenges developers encounter. These range from configuring built-in processors to writing custom scripts. Many posts reflect confusion around general usage, particularly when invoking processors alongside file management tasks in Java-based workflows. Other subtopics point to issues specific to NiFi execution and integration with streaming platforms like Kafka. Notably, concerns such as JSON conversion, attribute handling, and managing FlowFile filenames and output paths reveal the fine-grained complexity of data transformation tasks. Challenges in route configuration and tabular data enrichment further underscore the intricacies of building robust data pipelines in SWSs.

\textbf{4. Data Structures and Operations: }This topic encompasses 13.20\% of posts. In SWSs, data structures (i.e., arrays, dataframes, functions, and matrices) and operations (i.e., data ingestion, cleaning, transformation, analysis, visualization, storage, and parallel processing) are crucial for organizing and managing data. These components collectively facilitate the efficient processing and analysis of scientific data to derive meaningful insights. An example of this topic with 175K views is \emph{'Make Pandas DataFrame apply() use all cores?'}. The concern was that \emph{DataFrame.apply()} is limited to a single core, causing a multi-core machine to underutilize its computing power. Thus, the developer was inquiring how s/he can use all the cores of a machine to run \emph{DataFrame.apply()} on a dataframe to run in parallel. Another example with 47.5K views is \emph{'Convert string to dict, then access key: values?. How to access data in a $<$class 'dict'$>$ for Python?'}. Here, the developer faced issues accessing data from a dictionary and seeking a solution.
\begin{table*}[htbp]
\footnotesize
\caption{Granular Subtopics under Data Structures and Operations}
\label{tab:data_structures_subtopics}
\centering
\begin{tabular}{p{2.7cm}p{2.5cm}p{0.8cm}p{4.5cm}p{3.5cm}}
\toprule
\textbf{Subtopic} & \textbf{Keywords} & \textbf{Post Count} & \textbf{Description} & \textbf{Representative Example} \\
\midrule
DataFrame Usage with Files and Functions & use, dataframe, file, datum, try, error, column & 1294 & Challenges in applying functions or accessing values in dataframes while handling files or large datasets. & 'Apply function to large dataframe; RAM gets overloaded and performance drops.' \\

Basic DataFrame Manipulations and Errors & use, dataframe, try, file, worker, run, code & 3182 & General usage of dataframes involving basic manipulations, error-prone operations, and try-except patterns. & 'Worker throws error shortly after starting; tried using dataframe with large CSV.' \\

Parquet File Handling in Pipelines & parquet, file, read, use, dataframe, write & 227 & Difficulties in reading and writing Parquet files, especially with empty returns or schema mismatches. & 'Read Parquet file returns empty dataframe; trying to debug file structure.' \\
\bottomrule
\end{tabular}
\end{table*}

This challenge occurred primarily due to the complexity of efficiently managing and manipulating large-scale data in distributed computing environments. Developers faced difficulties in optimizing data structures like dataframes and collections across frameworks such as Dask, MongoDB, and Hadoop due to limitations in indexing, partitioning, and parallel processing. Issues include handling large datasets without performance bottlenecks, ensuring seamless operations across distributed systems, and integrating diverse data storage formats. Furthermore, the need for efficient data retrieval, transformation, and storage operations adds another layer of complexity, making it challenging to maintain both performance and scalability.

\textbf{Granular Analysis of Data Structures and Operations: }We decomposed the Data Structures and Operations topic into three coherent subtopics using BERTopic (Table~\ref{tab:data_structures_subtopics}) to gain finer insights into developer challenges. The largest subtopic, Basic DataFrame Manipulations, and Errors, includes common operations such as indexing and transformation, often wrapped in error-prone try blocks. DataFrame Usage with Files and Functions captures issues with applying functions to large datasets, frequently leading to performance or memory constraints. Finally, Parquet File Handling in Pipelines reflects difficulties in reading and writing Parquet files, particularly when faced with schema mismatches or empty outputs. These subtopics highlight recurring pain points in structured data management within scientific workflows.
\begin{table*}[htbp]
\footnotesize
\caption{Granular Subtopics under Workflow Execution}
\label{tab:workflow_execution_subtopics}
\centering
\begin{tabular}{p{2.7cm}p{2.4cm}p{0.7cm}p{3.6cm}p{3.5cm}}
\toprule
\textbf{Subtopic} & \textbf{Keywords} & \textbf{Post Count} & \textbf{Description} & \textbf{Representative Example} \\
\midrule

Execution Job Coordination and Control & job, run, workflow, java, use, action, hadoop & 777 & Challenges in managing execution jobs and controlling task coordination in distributed systems. & 'Access local file system not allowed even after setting instructions in Oozie job.' \\

Workflow Definition and XML Parsing Errors & coordinator, workflow, action, job, run, time & 406 & Errors during XML-based workflow definition or parsing in Hadoop/Oozie systems. & 'Difference between workflow, coordinator, and bundle configurations?' \\

Hive Script and Task Scheduling Issues & hive, workflow, file, script, run, error & 277 & Difficulties with job scheduling, cluster communication, and runtime failures. & 'Hive script with input path fails mid-task due to missing temp directories.' \\

Spark Job Execution and Submission Problems & spark, run, job, jar, submit & 213 & Execution management of actions and triggers in workflow engines like Oozie and Spark. & 'Pyspark actions run using Spark 1.6 instead of expected 2.2 in cluster job.' \\

MapReduce and Environment-Specific Workflow Failures & job, run, reduce, map reduce & 129 & Problems with running workflows inside containerized or secured environments. & 'Set up cron job to run three map-reduce pipelines; results inconsistent.' \\

Failure Recovery and Retry Handling & fail, retry, exception, handle, error, step & 114 & Handling failures, retries, or missed steps in scheduled workflows. & 'How to rerun failed step in Oozie without re-triggering the full workflow?' \\

Shell Script and Job Launch Errors & shell, command, job, error, script, exit code & 104 & Configuration and debugging issues related to launching jobs from shell or command line. & 'Shell script exits with code 1; not able to trigger Spark job via command.' \\
\bottomrule
\end{tabular}
\end{table*}

\textbf{5. Workflow Execution: }Workflow execution in SWSs entails orchestrating a predefined sequence of computational tasks to analyze scientific data, manage dependencies, allocate computational resources, and monitor progress. While workflow execution stands as a pivotal component within every SWS, developers encounter challenges in its implementation. 5.67\% of posts within our dataset highlight this specific issue. An illustrative example in this topic, garnering about 42K views, is titled \emph{'Stop Oozie workflow execution'}. In this scenario, a developer initiated two jobs within a workflow but subsequently decided to terminate them. Despite halting the jobs, the workflow persisted. Consequently, the developer sought a command-line solution to cease the workflow execution effectively.

The Workflow Execution challenge primarily stemmed from the complexity of managing and orchestrating distributed workflows in systems like Oozie and Hadoop. Developers struggle with job scheduling, dependency resolution, error handling, and resource allocation issues. Ensuring workflows execute correctly across multiple nodes, handling failures gracefully, and debugging execution errors pose significant challenges. Additionally, misconfigurations, incompatibilities between job components, and inefficiencies in workflow automation further complicate execution, making it difficult to achieve reliable and optimized workflow performance at scale.

\textbf{Granular Analysis of Workflow Execution: }To better understand the challenges developers face during the execution phase of workflows, we applied BERTopic to further decompose the topic Workflow Execution. This fine-grained analysis yielded seven subtopics that reflect distinct pain points ranging from job scheduling and runtime failures to XML parsing errors and environment-specific execution issues. As summarized in Table~\ref{tab:workflow_execution_subtopics}, developers frequently reported problems managing execution jobs, coordinating triggers, and handling resource and configuration mismatches in distributed environments like Hadoop, Oozie, and Spark. These insights suggest that workflow execution remains a technically demanding phase that often requires low-level debugging and system-specific tuning.
\begin{table*}[htbp]
\footnotesize
\caption{Granular Subtopics under Managing Rules, Input, Output, and Files}
\label{tab:rules_io_subtopics}
\centering
\begin{tabular}{p{2.7cm}p{2.5cm}p{0.7cm}p{4.5cm}p{3.5cm}}
\toprule
\textbf{Subtopic} & \textbf{Keywords} & \textbf{Post Count} & \textbf{Description} & \textbf{Representative Example} \\
\midrule
Rule Execution and File Usage Errors & file, use, run, rule, error, output, try, snakemake & 766 & Difficulties in rule definition and execution in Snakemake or similar DSLs. & 'Access multiple key config file, question on how to structure rules for proper path resolution.' \\

Wildcard Pattern and Rule Matching & file, rule, input, output, wildcard, use, run, snakemake & 599 & Challenges in managing wildcard patterns and rule matching logic. & 'Output file with mixed wildcard stuck, rule not producing expected target.' \\

Cluster Job Submission and Path Configurations & job, cluster, slurm, submit, run, use, node, resource & 141 & Issues with configuring input and output paths across tasks and pipelines. & 'Trying SGE cluster submission, need help managing resource paths for input and output.' \\

Conda Environment and File Dependency Issues & conda, environment, use, rule, module & 94 & Handling file dependencies, extensions, and dynamic naming in output rules. & 'Problem using Python module from conda env within Snakemake rule.' \\

Snakefile Errors and Missing Inputs & snakefile, error, run, file, try, script, use & 91 & Debugging problems related to missing inputs, incorrect file paths, or unresolved rule targets. & 'Run Snakefile from directory containing it, but output not found, rule fails silently.' \\
\bottomrule
\end{tabular}
\end{table*}

\textbf{6. Managing Rules, Input, Output, and Files: }In SWS, a rule defines a unit of work within a workflow, detailing how to generate output files from input files. Each rule typically includes directives for input, output, parameters, resources, and conditions, as well as the shell commands or scripts to be executed. This topic covered 4.75\% of posts in our dataset and asked various issues related to defining and running rules, recommended practices for executing shell commands and scripts, and file naming using wildcards. For instance, a question with approximately 17K views is titled \emph{'How to run only one rule in Snakemake?'}. In this scenario, the developer created a workflow consisting of three rules: A, B, and C, where the output of rule A serves as the input for rule B, and the output of rule B serves as the input for rule C. Initially, the workflow ran smoothly, but the developer later wanted to run only rule C and encountered difficulties, thus sought help on SO.

This challenge primarily arose from the complexity of defining, tracking, and executing workflow rules in SWSs like Snakemake and Luigi. Developers struggle with specifying dependencies, handling dynamic file paths, and ensuring that input and output files are correctly mapped across different workflow steps. Issues such as incorrect rule execution, persistent reruns of completed tasks, and difficulties in debugging file-related errors further complicate workflow management. Additionally, ensuring reproducibility while dealing with different file formats and execution environments adds another layer of difficulty, making it challenging to maintain efficient and scalable workflows.

\textbf{Granular Analysis of Managing Rules, Input, Output, and Files: }To further analyze developer difficulties in defining and managing rules in SWSs, we decomposed this topic into five distinct subtopics using BERTopic. As summarized in Table~\ref{tab:rules_io_subtopics}, these subtopics reflect common challenges in writing, debugging, and executing rule-based logic, particularly in systems like Snakemake. Developers reported issues ranging from wildcard mismatches and file path errors to misconfigured environments and unresolved rule dependencies. These results highlight the complexity of rule management in file-driven workflow systems and the need for clearer error reporting and automation support.
\begin{table*}[htbp]
\footnotesize
\caption{Granular Subtopics under Task Dependencies and Management}
\label{tab:task_dependency_subtopics}
\centering
\begin{tabular}{p{2.7cm}p{2.5cm}p{0.7cm}p{4.5cm}p{3.5cm}}
\toprule
\textbf{Subtopic} & \textbf{Keywords} & \textbf{Post Count} & \textbf{Description} & \textbf{Representative Example} \\
\midrule

Target Files and Dependency Errors & use, error, target, file, try, code, run & 358 & Managing task dependencies, ensuring correct order of execution, and dealing with prerequisites in DAGs. & 'Visualize simulation, generic example installed but can't build target due to missing file dependency.' \\

Input Definitions and Execution Control & use, try, joint, plant, constraint, add & 457 & Using decorators, directives, or annotations to define task inputs, outputs, and execution contexts. & 'Construct simulation in pydrake with manipulation framework, inputs not triggering expected behaviors.' \\

Luigi Task Linking and Execution Failures & task, luigi, run, file, pipeline & 234 & Debugging task failure or non-execution due to missing dependencies or incorrect conditions. & 'Run Luigi task depending on another task, only one executes and the other gets skipped.' \\

Conditional Task Execution with Pydrake & error, python, run, pydrake, import, version & 88 & Handling conditional logic and parameter passing between tasks in dynamic workflows. & 'SystemError on pydrake task run, likely due to incorrect dependency configuration.' \\
\bottomrule
\end{tabular}
\end{table*}

\textbf{7. Task Dependencies and Management: }This topic explores issues such as managing complex inter-task dependencies, scalability, efficient resource allocation, intermediate data reuse, and concurrency. It also discusses challenges like implementing robust error handling and recovery mechanisms to maintain data integrity and ensure reproducibility through consistent environment management and version control. An example of this topic with around 12K views is \emph{'How to reset Luigi task status?'}, where a developer faced issues with a dependency chain ($a->b->c->d$), where task d was executed first, and task a was executed last. Then, the developer needed to rerun the same queue, changing the data. When rerunning the same queue, the Luigi SWS used historical data, so the developer sought a way to force the tasks to run without relying on this previous data.

This challenge arose from the complexity of orchestrating and managing task execution in SWSs like Luigi and Drake. Developers face difficulties in defining task dependencies, handling failures, and ensuring that tasks execute in the correct sequence without unintended bottlenecks. Issues such as unregistered task types, state synchronization failures, and debugging interdependent tasks further complicate the process. Additionally, efficiently managing parallel execution while ensuring resource constraints and data consistency presents another layer of difficulty, making it challenging to maintain reliable and scalable task workflows.

\textbf{Granular Analysis of Task Dependencies and Management: }To better capture the nuanced challenges of managing task coordination in workflows, we applied BERTopic to the broader topic Task Dependencies and Management, resulting in four focused subtopics. As shown in Table~\ref{tab:task_dependency_subtopics}, these subtopics highlight the recurring difficulties developers face when defining task hierarchies, passing parameters, and managing execution dependencies across tools like Luigi and pydrake. Common concerns include task non-execution due to unmet conditions, debugging multi-stage workflows, and integrating conditional logic. These findings underscore the need for clearer dependency resolution mechanisms and improved diagnostic support in task-based workflow engines.

\textbf{8. Data Transformation: }In SWS, this topic discusses converting data from one format or structure to another, enabling easier analysis, visualization, or integration with other datasets. This involves altering the shape or appearance of data objects while preserving their content and, for instance, using libraries like Jolt to transform JSON data from one format to another. This topic covers around 1.26\% of posts of our SO dataset. One example of this topic, with around 10K views, is titled \emph{'Is it possible to concatenate the values of JSON attributes using JOLT?'}. In this instance, the developer inquired whether JOLT could be used to concatenate the values of three JSON attributes into a new attribute and also asked about potential alternative approaches.

The Data Transformation challenge originated from the complexity of structuring, mapping, and converting data formats across different processing frameworks. Developers struggle with transforming nested JSON structures, applying transformation rules using tools like Jolt and NiFi, and ensuring data integrity during the conversion process. Issues such as handling dynamic key-value mappings, concatenating attributes, and efficiently parsing large data sets add to the difficulty. Furthermore, inconsistencies in schema definitions, missing values, and compatibility issues between input and output formats make it challenging to implement seamless data transformation workflows.

\textbf{9. Learning: }This topic captures posts primarily focused on introductory usage, conceptual understanding, and early-stage onboarding with Scientific Workflow Systems. Unlike other topics that address specific technical errors or feature-level questions, posts in this category often reflect users who are new to a particular system and are seeking foundational guidance, such as how to choose between SWSs (e.g., Airflow vs. Luigi), understand basic workflow concepts, or set up a minimal working example. One example of this topic with around 6.4K views is \emph{'How to run Python Script in Nextflow?'}. The developer attempted to run an executable Python script in an SWS (NextFlow) but encountered an error stating \emph{No such file or directory}, though it has the file. Subsequently, the developer asked for a solution for this issue on SO. Another example of this topic, with around 5K views, is titled \emph{'How to generate the output directory for a Nextflow process within the Docker container?'}. In this case, the developer ran two processes using Nextflow inside a Docker container. The first process successfully generated some plots, which were intended as input for the second process. However, the second process could not access the plots, prompting the developer to seek a solution in SO to make these plots available for the subsequent process. This topic highlights a significant gap in available documentation, tutorials, and other guidance resources for using SWSs.

The challenges in Learning stemmed from difficulties in understanding complex workflows, debugging process executions, and handling errors in automation tools like Nextflow. Users often struggle with managing process dependencies, correctly structuring workflow channels, and troubleshooting missing or unexpected outputs. Syntax errors and insufficient documentation further hinder the learning process, making it difficult for users to effectively implement and modify workflow scripts. These obstacles highlight the steep learning curve associated with mastering workflow automation tools and understanding their intricate execution behaviors.

\textbf{10. Automation With Ansible: }This is the least dominant topic, accounting for just 0.72\% of posts. It focuses on the use of Ansible in SWSs. Ansible enhances SWSs by providing robust automation for environment setup, configuration management, workflow execution, and orchestration, ensuring scientific research consistency, reproducibility, and efficiency. One of the most viewed questions in this topic, with 111k views, is \emph{'How to automatically install Ansible Galaxy roles?'} where Ansible is an open-source automation tool for deploying and configuring large-scale infrastructures \cite{hochstein2017ansible}. In this question, the developer faced the challenge of manually downloading Galaxy roles on each machine and sought a method to automatically identify and download these roles when running Ansible on a new machine. We observed that many questions within this topic pertain to Ansible installation and role management. This trend likely stems from a lack of proper documentation or tutorials on using the tool effectively.

Challenges in this topic primarily originated from issues related to role management, module dependencies, and execution consistency. Users frequently encounter errors when integrating Ansible roles, especially when using Ansible Galaxy for role installation and dependency resolution. Import errors, missing modules, and version mismatches are common roadblocks that hinder automation efficiency. Additionally, structuring roles and invoking them within playbooks often lead to confusion due to inconsistencies in syntax and execution behavior. These challenges highlight the complexities of maintaining modular, reusable workflows and ensuring compatibility across different environments.

\begin{boxK}
SWSs developers inquire about various aspects of SWSs development, including learning resources, automation, workflow creation, scheduling, data transformation, and task management. The most popular topic is distributed task management. This is because it deals with the complexities of coordinating and executing tasks across multiple systems, which is crucial for handling the large-scale data and computational workloads typical in scientific workflows.
\end{boxK}

\textbf{Results Obtained Using GitHub: }We identified 13 key topics developers encountered when working with GitHub data associated with SWSs following the process described in sub-section \ref{identify-topics}. Our analysis of GitHub issues and pull requests revealed that these challenges often stem from the inherent complexity and interdisciplinary nature of these systems. Table \ref{tab:sws_topics_github} presents a summary of the identified topics, their associated keywords, and a short description of each topic. Below, we provided a detailed description of each topic.

\begin{table*}[htbp]
\footnotesize
    \centering
    \caption{Topics derived from SWSs on GitHub}
    \label{tab:sws_topics_github}
      \begin{tabularx}{\linewidth}{r|>{\raggedright\arraybackslash}p{3.5cm}|>{\raggedright\arraybackslash}p{5.3cm}|>{\raggedright\arraybackslash}p{5.0cm}}
        \toprule
        SL. & Topic & Keywords & Description \\
        \midrule
        1 & Errors and Bugs Fixing & use, file, add, change, run, issue, task, bug, case, code, error, fix, update, follow, work.& Addressing errors and bugs in code, including fixes, version updates, and test cases. \\
        2 & Documentation & documentation, fix, case, change, add, issue, test, code, license, version, follow, commit, make, tool, check & Modifying and improving project documentation, including version updates and test cases. \\
        3 & Dependencies & dependency, file, change, license, update, build, applicable, branch, ensure, tracking, jira number, jdk, contribution, submit, apache jira. & Managing and updating project dependencies and related files. \\
        4 & Visual and Branding Enhancements & vignette, add, slide, readme, image, badge, update, logo, pkgdown, blog post, fix, use, package, make, link. & Focuses on improving the appearance, structure, and consistency of user-facing content in SWSs, such as logos, badges, images, and vignette-style tutorials. \\
        5 & Automated Performance and Tool Update Integration & perf, automatic update, lastdb, bismark, sratools, fasterqdump, picard, samtools, snpsift, annotate, bwa, samse, bcftools filter, bwa mem. & It involves routine updates to tools like BWA, Bismark, and Picard using automated pipelines, reducing manual effort, and ensuring workflows remain efficient and up to date. \\
        6 & CRAN Release Management and Package Validation & cran, release, check, package, submit cran, wait cran, accepted, news, approve, prepare release, release check, blog & This topic focuses on preparing, validating, and submitting R packages to CRAN, ensuring they meet strict compliance, compatibility, and documentation standards. \\
        7 & Managing Releases & release, create, generate release, beep, documentation, bug fixes, chore-release, boop. & Handling the creation and management of software releases. \\
        8 & Browser Compatibility and HDFS Integration Issues & hdfs, chrome, selenium, firefox, version, error, behaviour, chromedriver, selenium server, use, geckodriver, begin, work, run, server version. & Discusses challenges arising from integrating browser automation tools with backend systems like HDFS, often leading to version mismatches and environment-specific errors. \\
        9 & Chord Execution and Task Coordination Issues & chord, ropensci, task, chain, result, group, callback, behavior, reproduce, redis, header, execute, backend, expect,  chord header, chord body, report, version. & Improving code quality through refactoring, and error fixes. \\
        10 & Data Shuffling and Partitioning Strategies &shuffle, p2p, task, add pass, partition, tests, worker, disk, dataframe, datum, use, scheduler, algorithm, operation. & Discusses the optimization of data shuffling and partitioning in distributed workflows, focusing on performance, correctness, and resource use. \\
        11 & Image Compression \& Optimization & optimisation, calibre, image actions, reduced, compression, filename, improvement, auto. & Optimizing images for performance and efficiency. \\
        12 & Automating Code Quality Checks & precommitci, automatic update, precommitci start, end, rebase freeze, biotools. & Implementing automated code quality checks and continuous integration processes. \\
        13 & System Redesign and API Migration & new design, migrate, example dags, introduce aip47, migration list, progress migration, automatically release, provider create, create track, ci process & Focuses on refactoring and migrating legacy components to align with modern design standards and improvement proposals. It involves systematic changes to APIs, DAGs, and CI processes to enhance modularity, scalability, and maintainability.\\
        \bottomrule
    \end{tabularx}
\end{table*}

\textbf{1. Errors and Bugs Fixing: }This topic encompasses a significant portion of all issues and pull requests in the dataset, highlighting its significance in the development and maintenance of SWSs. Addressing bugs and errors is essential for ensuring the reliability, accuracy, and performance of these systems. Bugs can disrupt functionality, lead to unexpected behaviors, or degrade the user experience. Common keywords such as \emph{error, fix, issue, run, code, and update} reflect the technical nature of this category, which typically includes runtime troubleshooting, patching faulty logic, and refining error-handling mechanisms. For instance, one issue involves \emph{'fixing invocation serialization when no state is set'}, while another addresses a computational edge case: \emph{'fix bug for reduction of zero dimensional array'}. These examples underscore the complexity of maintaining correct behavior in modular and distributed workflows. Developers strive to create robust and dependable tools by resolving such problems promptly, minimizing workflow disruptions, and enhancing the long-term stability of SWSs.
\begin{table*}[htbp]
\footnotesize
\caption{Granular Subtopics under Errors and Bug Fixing}
\label{tab:error_fixing_subtopics}
\centering
\begin{tabular}{p{3cm}p{2.5cm}p{5cm}p{3.0cm}}
\toprule
\textbf{Topic Name} & \textbf{Keywords} & \textbf{Description} & \textbf{Representative Example} \\
\midrule
Automated Release and Update Failures & release, update, perf, automatic, version & Challenges related to automation in releases or performance updates, often resulting in regressions, incomplete updates, or unintended changes. & 'set call custom initialization function running...' \\

File Modification and Error-Prone Fixes & use, add, file, change, test, case, work, error & Posts dealing with file-level modifications (e.g., adds, changes) that introduce or attempt to fix bugs, highlighting common pitfalls during patching. & 'python imaplib ssl error use celeryd queue' \\

Commit Management and Change Validation & ensure, change, update, commit, file, version & Issues around ensuring that code changes are properly committed and updated, especially when handling file versions or rollbacks. & \textit{schedule Task execute specific time use celery?} \\

Python Dependency Upgrade Errors & python dependency, update python, dependency, pip & Errors introduced during updates to Python dependencies, often due to incompatibilities, missing versions, or unmet requirements. & \textit{RabbitMQ give access refuse, login refuse use...} \\

Package Backport and Compatibility Issues & heavycheckmark, package, backport, package example, provider & Developer struggles related to backporting packages or maintaining compatibility across library versions and platforms. & \textit{connect Saving Data Redis inside Task} \\
\bottomrule
\end{tabular}
\end{table*}

The underlying cause of this topic often stemmed from the complexity and modularity of SWSs. Errors frequently arise due to misconfigured task dependencies, unhandled edge cases (e.g., zero-dimensional arrays), or missing state during execution. Additionally, evolving software environments, insufficient testing, and interoperability between heterogeneous tools contribute to runtime failures that developers must resolve to maintain system stability.

\textbf{Granular Analysis of Errors and Bugs Fixing: }To gain deeper insight into the specific challenges developers face during error resolution and bug fixing, we conducted a fine-grained analysis of this topic using BERTopic. This second round of topic modeling allowed us to extract semantically coherent subtopics, which were further refined through manual validation to ensure clarity and alignment with real-world developer concerns. The analysis resulted in five distinct subtopics, each representing a critical aspect of the bug-fixing process, as summarized in Table~\ref{tab:error_fixing_subtopics}. These include failures in automated release workflows, error-prone file modifications, version control and commit inconsistencies, dependency upgrade issues, and compatibility challenges in backporting packages. This granular decomposition offers a more nuanced perspective on how developers diagnose and address faults, providing actionable insights into the operational and systemic complexities of SWSs development.

\textbf{2. Documentation: }This topic discusses various issues and pull requests essential for maintaining and enhancing SWSs. It involves removing outdated content, enriching documentation with practical examples, and offering essential support to users and contributors. Activities under this topic include updating READMEs, configuring tools, refining tutorials to reflect new features, and setting up environments to guide both novice and experienced users. Frequent keywords such as \emph{documentation, fix, case, add, tool, and check} indicate continuous efforts to align documentation with evolving workflows and tooling. For instance, a developer raised the issue emph{'Set up multiple Airflow instances under the same domain with different sessions'} when facing challenges related to running multiple instances of an SWS and sought documentation support. Another example includes a contribution to \emph{'Update documentation for CI jobs'}, highlighting the need for accurate procedural guidance during automation and testing. Proper documentation not only aids in understanding and implementation but also ensures reproducibility, promotes best practices, and reduces onboarding friction. Regular updates and clear communication about system changes play a vital role in fostering a collaborative and productive development environment, ensuring that all stakeholders remain informed and equipped to use and contribute to the system effectively.
\begin{table*}[htbp]
\scriptsize
\caption{Granular Subtopics under Documentation}
\label{tab:documentation_subtopics}
\centering
\begin{tabular}{p{2.9cm}p{2.6cm}p{5cm}p{3.0cm}}
\toprule
\textbf{Subtopic} & \textbf{Keywords} & \textbf{Description} & \textbf{Representative Example} \\
\midrule
Test-Driven Documentation Edits & use, add, change, test, case, fix & Posts that focus on updating or adding documentation while simultaneously modifying or testing code to ensure correctness and traceability. & \emph{'register\_type does not work with subclasses ...'} \\

Clarifying Package and Version Usage & use, version, run, package, task, install & Documentation efforts aimed at clarifying how to use specific versions of packages, manage environments, or address usage inconsistencies. & \emph{'Pin hypothesis to latest version 6.97.4'} \\

License and Contribution Note Updates & change, case, license, add, commit, copyright & Changes targeting licenses, contributor acknowledgements, or standardizing language around usage and legal metadata. & \emph{'Update hypothesis requirement from $\leq$6.97.1 to..'} \\

Typo and Link Corrections & link, typo, fix, translation, doc & Minor but common documentation errors such as typos and broken links, including efforts to improve clarity or localization support. & \emph{'docs: fix simple typo, propery → property'} \\

Framework-Specific Error Documentation & django, redis, task, file, log & Bug reports and fixes focused on errors encountered in specific frameworks (e.g., Django or Redis), which require better documentation. & \emph{'Redis - Losing Unacked Messages - WorkerLostError...'} \\

Restructuring and Refactoring Documentation & airflow, doc, refactor, clean, example & Posts that aim to clean up, reorganize, or refactor documentation for better structure, navigation, and usability. & \emph{'Lack of ZeroMQ in transport comparison chart'} \\
\bottomrule
\end{tabular}
\end{table*}
The underlying cause of documentation challenges in SWSs was their inherent complexity, rapid evolution, and multidisciplinary nature. SWSs often integrate diverse tools, domain-specific configurations, and custom environments, making it difficult to maintain up-to-date and comprehensive documentation. Contributors may prioritize code over documentation, and inconsistent contribution practices can lead to fragmented or outdated guides. Additionally, as workflows evolve, tutorials and examples often lag behind, creating confusion for users. This complexity, coupled with the lack of standardized documentation practices, results in usability gaps and hampers effective onboarding and workflow reproducibility.

\textbf{Granular Analysis of Documentation}
To better understand the nature of documentation-related challenges in SWSs, we conducted a fine-grained analysis of the broader \emph{Documentation} topic using BERTopic. By applying a second round of topic modeling and validating representative posts manually, we identified six semantically coherent subtopics that capture distinct aspects of how developers engage with, revise, and troubleshoot documentation. These subtopics span concerns such as documentation updates tied to code changes, clarifying package usage, correcting metadata and typos, framework-specific error documentation, and broader restructuring efforts. A summary of these subtopics, along with their keywords and representative examples, is presented in Table~\ref{tab:documentation_subtopics}. This decomposition offers a more nuanced understanding of the documentation practices and pain points encountered in real-world SWS development.

\textbf{3. Dependencies Management: }Dependencies management is a critical yet challenging aspect of SWS development, as these systems rely on a vast ecosystem of libraries and external tools to function correctly. This topic gathers issues and pull requests related to the integration, updating, and configuration of such software dependencies. For example, one issue titled \emph{'Update elastic-transport requirement from  \(\leq 8.10.0 \) to  \( \leq 8.11.0 \)'},  illustrates the need to maintain compatibility with evolving third-party libraries. elastic-transport, a Python library, plays a key role in managing HTTP communication and request/response workflows. Similar examples include \emph{'Fix python dependencies in Parse Document'}, reflecting efforts to address conflicts or runtime errors stemming from outdated or mismatched dependencies. . SWSs utilize thousands of libraries like \emph{elastic-transport, pymongo, mypy, and conda}, which can present various challenges during their implementation, which can introduce challenges related to version control, transitive dependencies, and platform-specific behavior. Effective dependency management, including clear documentation and automated installation scripts, is essential for maintaining a reliable and reproducible workflow environment.
\begin{table*}[htbp]
\footnotesize
\caption{Granular Subtopics under Dependencies}
\label{tab:dependency_subtopics}
\centering
\begin{tabular}{p{3.1cm}p{2.4cm}p{5cm}p{3.0cm}}
\toprule
\textbf{Subtopic} & \textbf{Keywords} & \textbf{Description} & \textbf{Representative Example} \\
\midrule
Ensuring File Consistency During Dependency Updates & update, ensure, change, file, add, missing & Addresses the need to ensure that all relevant files are consistently updated when modifying or upgrading dependencies. & \textit{Convert readthedocs link for their .org → .io migration} \\
\addlinespace

Dependency Tracking with Apache JIRA & apache, tracking, jira, update & Captures issues related to dependency tracking, often involving coordination through Apache JIRA systems and external update logs. & \textit{Pin hypothesis to latest version 6.97.3} \\
\addlinespace

Change Commit Validation During Dependency Synchronization & change, update, ensure, commit, file, correct & Focuses on validating and ensuring that code changes related to dependency upgrades are correctly committed and versioned. & \textit{Bump kafka deps versions \& fix integration test failures} \\
\addlinespace

License File Updates in Dependency Distributions & license, file, change, distribute, usage & Reflects updates to licensing metadata and distribution files when dependencies are changed, added, or redistributed. & \textit{Update runtime.txt to include Django 5.0} \\
\bottomrule
\end{tabular}
\end{table*}
The challenge of dependencies management in SWSs stems from their reliance on a complex and rapidly evolving ecosystem of software libraries. SWSs integrate a wide array of tools, each with its own versioning, compatibility requirements, and transitive dependencies. Frequent updates, deprecations, and platform-specific behaviors can lead to conflicts or break existing workflows. Moreover, ensuring compliance with licensing policies (e.g., Apache guidelines) and aligning contributions with issue-tracking systems like JIRA adds procedural overhead. This dynamic and interconnected environment makes managing dependencies both technically and organizationally demanding.

\textbf{Granular Analysis of Dependencies: }To uncover detailed insights into how developers manage dependencies in SWSs, we applied a second round of BERTopic modeling to the broader Dependencies topic. This fine-grained analysis allowed us to identify semantically coherent subtopics, each representing a distinct aspect of dependency management workflows. Through manual validation of representative issues and pull requests, we confirmed the relevance and clarity of each subtopic. These findings expose nuanced challenges developers face when synchronizing dependencies, validating updates, maintaining metadata, and coordinating across issue-tracking systems. A summary of the subtopics, along with their descriptions and representative examples, is presented in Table~\ref{tab:dependency_subtopics}. This decomposition contributes to a richer understanding of how dependency-related tasks intersect with maintainability, collaboration, and release engineering in SWS development.

\textbf{4. Visual and Branding Enhancements: }This topic captures issues and pull requests focused on improving the visual presentation, structure, and branding elements within Scientific Workflow Systems (SWSs). Activities include adding or correcting images, logos, and badges, updating blog posts and vignette-style tutorials, and ensuring consistent styling across README files and other public-facing materials. Keywords such as \emph{vignette, slide, image, badge, logo, and pkgdown} reflect an emphasis on improving the aesthetic and communicative clarity of project assets. For example, one issue mentions \emph{'Add some logos and other changes for ARTIC page'}, where another pull request asks \emph{'Add News/Event: helper bots event'}. These visual updates play an important role in user engagement, project identity, and the overall usability of SWSs.

The underlying cause of this challenge lied in the fact that visual design and branding are often deprioritized in SWSs, which tend to emphasize functionality over presentation. Contributors may lack expertise in user experience (UX) or web design, leading to inconsistent visuals, outdated logos, or poorly structured content. Additionally, as SWS projects evolve rapidly and include contributions from diverse teams, maintaining coherent and appealing visual assets across documentation, tutorials, and web pages becomes difficult. This results in fragmented user-facing materials that can hinder engagement, usability, and the professional image of the project.

\textbf{5. Automated Performance and Tool Update Integration: }This topic encompasses issues and pull requests focused on enhancing the performance of SWSs through the automated integration and updating of key bioinformatics tools. The presence of tools such as \emph{BWA, Bismark, Picard, bcftools, samtools, and sratools }, combined with keywords like \emph{perf} and \emph{automatic update} indicates a systematic effort to keep tool versions current and optimize their use within workflows. Many of these updates are carried out using automated pipelines (e.g., GitHub Actions), reducing manual intervention and ensuring compatibility with the latest improvements and bug fixes. For example, several issues consistently note “precommit autoupdate update GitHub actions,” showing the automation of routine update processes. These activities are critical for maintaining workflow efficiency, ensuring up-to-date analyses, and supporting scalability in computational biology environments.

The main reason for this challenge stemmed from the rapid evolution and diversity of bioinformatics tools. These tools are frequently updated to improve accuracy, speed, and compatibility, but keeping workflows aligned with the latest versions is complex and error-prone. Additionally, SWSs often integrate multiple interdependent tools, making even minor version changes potentially disruptive. Without automation, managing these updates manually would be time-consuming and risk inconsistencies, especially in collaborative or large-scale environments.


\textbf{6. CRAN Release Management and Package Validation: }This topic encompasses issues and pull requests related to the preparation, submission, and validation of software packages for CRAN (Comprehensive R Archive Network). Keywords such as \emph{release, check, cran, submit, approve, prepare, and blog post} reflect tasks like fixing CRAN check errors, preparing changelogs, polishing documentation, and handling release approvals. For example, one issue notes: \emph{'Fix CRAN error CRAN package check result Solaris…'}, highlighting the debugging process tied to platform-specific CRAN checks. These contributions ensure that R packages used in scientific workflows meet CRAN’s standards and remain accessible, stable, and reproducible for the broader research community.

The underlying cause of this challenge lied in the strict and evolving standards enforced by CRAN, combined with the diversity of system environments where packages are tested. Developers must ensure that their packages pass rigorous automated checks across multiple operating systems, handle backward compatibility, and meet documentation and metadata requirements. This process is further complicated by frequent updates to CRAN policies and dependencies and the manual review and approval workflow, which can introduce delays and uncertainty. As a result, preparing and maintaining CRAN-compliant packages demands meticulous attention and ongoing effort.

\textbf{7. Managing Releases: }Managing releases is a critical aspect of SWS development, marking the consolidation and deployment of new features, bug fixes, and maintenance updates into stable, production-ready versions. A release signifies that a tested and validated version of the software is ready for broader use, typically supported by automated workflows that handle packaging, changelog generation, and deployment. This topic includes issues and pull requests related to version tagging, integration of user-reported fixes, and routine maintenance tasks, often managed through bots or CI pipelines as seen in patterns like \emph{'bot create release'} or \emph{'trigger new release to pull in latest snakemake'}. Key components of the release process include not only addressing bugs and integrating enhancements but also ensuring updated documentation for user clarity. For example, the issue \emph{'Add note on fastapi to admin release notes'} reflects the importance of communicating recent changes, while \emph{'Release Celery v5.3.5'} underscores the role of timely bug resolution in maintaining release stability. Regular chore releases, such as dependency updates and configuration adjustments, further ensure that workflows remain up-to-date and reproducible. This structured, automated release strategy enhances consistency, reduces manual overhead, and supports sustainable development in collaborative environments.

The underlying cause of challenges in release management within SWSs stemmed from the complexity and coordination required to consolidate diverse updates, including bug fixes, new features, dependency changes, and documentation, into a stable, deployable version. Frequent contributions from multiple developers, fast-evolving tools, and the need to maintain reproducibility and compatibility across versions make the release process prone to errors if not carefully managed. Additionally, the reliance on automation tools (e.g., bots, CI/CD pipelines) introduces its own challenges, such as configuration mismatches or incomplete changelogs, especially when updates span multiple components or repositories.

\textbf{8. Browser Compatibility and HDFS Integration Issues: }This topic encompasses issues and pull requests related to the interaction between web browsers (e.g., Chrome, Firefox) and backend systems, particularly focusing on errors involving Selenium drivers, browser versions, and HDFS (Hadoop Distributed File System) access. Keywords like \emph{chromedriver, geckodriver, selenium server, hdfs, and version error} indicate recurring challenges in setting up compatible environments for automated testing or distributed data access. For instance, one issue reports: \emph{'browser tab notification crash chrome seem not work correctly'}, while another notes: \emph{'importerror find share library libhdfs3so'}, reflecting failures in loading required libraries. These challenges often arise from mismatched versions, missing dependencies, or evolving browser behaviors, which can significantly affect the stability and testability of scientific workflows that rely on browser automation or distributed data handling.

The reason for this challenge lied in the fragility of cross-component dependencies and version mismatches across browser automation tools and distributed file systems. Tools like Selenium, Chromedriver, Geckodriver, and HDFS clients are frequently updated, and small changes in browser versions, library paths, or environment configurations can lead to incompatibilities or runtime errors. Additionally, these components often interact across different layers, browsers, OS, and distributed systems, making it difficult to maintain a stable and reproducible setup, especially in CI/CD or multi-platform environments.

\textbf{9. Chord Execution and Task Coordination Issues: }This topic captures issues related to the orchestration and coordination of asynchronous task execution using chord constructs in distributed systems, such as Celery with Redis backends. Keywords like \emph{chord, header, callback, chain, task, group, execute, and backend} highlight challenges in managing complex task dependencies, particularly where a group of tasks (the \emph{header}) must complete before triggering a follow-up task (the \emph{body}). For example, an issue notes: \emph{'Chord raise ChordError task header even error…'}, reflecting failures in chaining tasks or managing callback execution. These problems often stem from version mismatches, backend misconfigurations, or inconsistent task states, and they can lead to silent failures or incorrect aggregation of results, making them particularly difficult to reproduce and debug.

The underlying cause of Chord Execution and Task Coordination Issues lies in the inherent complexity of managing asynchronous workflows in distributed systems. Chord constructs, which depend on precise coordination between multiple concurrent tasks and a final callback, are highly sensitive to backend reliability (e.g., Redis or RabbitMQ), version compatibility, and task state consistency. Failures often result from race conditions, misconfigured result backends, or improper serialization of task metadata, making these issues difficult to detect, reproduce, and resolve in real-world deployments.

\textbf{10. Data Shuffling and Partitioning Strategies: }This topic focuses on the design, testing, and optimization of data shuffling and partitioning mechanisms within distributed data processing workflows. Keywords such as \emph{shuffle, p2p, partition, task, disk, dataframe, scheduler, and algorithm} indicate work on improving how large datasets are split, moved, and reassembled across workers in systems like Dask. These operations are central to parallel computation, and issues under this topic address performance bottlenecks, the correctness of shuffle algorithms, and infrastructure components like disk-based operations. For example, titles such as \emph{'adjust shuffle test daskexpr...'} and \emph{'drop duplicate produce incorrect result...'} reveal ongoing challenges in ensuring data integrity and efficiency when scaling task-based execution over large distributed datasets.

The root cause of this challenge arose from the inherent complexity of efficiently redistributing large-scale data within distributed systems.. Shuffling involves moving data between workers based on partitioning logic, which can lead to performance bottlenecks, disk I/O overhead, memory pressure, and even task failures, especially when working with large dataframes or unbalanced workloads. Additionally, implementing and maintaining correct shuffle algorithms is non-trivial, as they must account for task dependencies, serialization, fault tolerance, and backend constraints (e.g., disk or network bandwidth). These factors make shuffle optimization both technically challenging and critical for scalable workflow execution.

\textbf{11. Image Compression \& Optimization: }This topic addresses the automated compression and optimization of image assets in SWSs, with the goal of improving performance, reducing storage requirements, and enhancing the efficiency of documentation and web-facing content. Keywords such as \emph{compress, image, optimisation, auto, calibre, and filename} reflect ongoing efforts to streamline media handling using tools like Calibre and automated scripts. In SWSs, image optimization plays a vital role in accelerating data transfer, improving load times, and conserving system resources by reducing file sizes without significantly compromising visual quality. This process often involves identifying essential image components while discarding unnecessary data, as well as applying format conversion and resizing techniques. Furthermore, improving filename structures through descriptive and standardized naming enhances organization and accessibility across workflows. A recurring issue titled \emph{'Auto Compress Images by Calibre's image-actions'}, exemplifies this topic, demonstrating how automated workflows are used to optimize images and maintain lightweight, maintainable project assets. Collectively, these strategies contribute to more efficient data processing and resource management in complex, scalable workflow environments.

This issue stemed from the absence of standardized practices and automated solutions for effectively managing media assets. As workflows grow in complexity and scale, large or unoptimized images can lead to increased storage costs, slower data transfer, and reduced system performance. Additionally, contributors may use inconsistent image formats or naming conventions, making maintaining a cohesive and efficient system difficult. Without automated tools or clearly defined optimization workflows, these issues accumulate and hinder the overall usability and scalability of the system.

\textbf{12. Automating Code Quality Checks: }In SWSs, maintaining high code quality and consistency is essential for reliable data analysis and reproducibility. Tools like pre-commit ci are invaluable in this process, as they automate code quality checks and enforce coding standards. By running a series of predefined checks, such as linting, formatting, and security scans, automatically whenever code changes are made, pre-commit ci ensures that any issues are identified and addressed promptly. This mechanism, often called \emph{precommitci autoupdate}, keeps the codebase aligned with the latest best practices and standards by automatically updating the checks.
An example of this is seen in the pull request titled \emph{'TS config and linting tweaks'}, where pre-commit ci was used to enforce code quality standards. This pull request illustrates how automated checks can detect potential issues early in the development process, ensuring that the code adheres to the project's guidelines before being merged. This pull request demonstrates how automated checks can catch issues early, ensuring the code adheres to the project's guidelines before merging.

This topic surfaces challenges with setting up and maintaining pre-commit hooks, CI pipelines, and style enforcement tools. Developers encounter integration friction between tools like flake8, pre-commit, and GitHub Actions, especially when working across teams with different environments. Misconfigurations and inconsistent enforcement reduce the effectiveness of automation.

\textbf{13. System Redesign and API Migration}This topic includes issues and pull requests related to large-scale system refactoring, API standardization, and migration efforts, often guided by formal proposals like AIP (Airflow Improvement Proposals). Keywords such as \emph{migrate, design, AIP47, example DAGs, and migration list} highlight efforts to transition existing components (e.g., models, operators, and CI processes) to newer architectures or frameworks. For example, one issue notes: \emph{'AIP44 Migrate Task Instance… and refactor local task join logic'}, reflecting a migration that aims to decouple and improve task execution logic. These activities are essential for modernizing SWSs, ensuring scalability, maintainability, and alignment with long-term design goals.

The underlying cause of this challenge stemmed from the need to modernize legacy components while preserving compatibility and functionality. As SWSs evolve, older architectures and APIs often become inefficient, tightly coupled, or misaligned with new design principles (e.g., modularity, standardization). Migrating to updated frameworks, such as those introduced via formal improvement proposals (e.g., AIPs), requires careful refactoring, dependency tracking, and testing across complex components like DAGs, operators, and CI pipelines. These transitions are technically demanding and risky, often disrupted by incomplete migration paths, undocumented behaviors, or inconsistent integration patterns.
\begin{boxK}
     Tables \ref{tab:swssotopics} and \ref{tab:sws_topics_github} present topics related to SWSs but from different sources, Stack Overflow and GitHub. Based on the findings, the challenges developers face in SWS development are diverse and span both SO and GitHub platforms. The topic modeling analysis uncovered ten dominant topics on SO and thirteen on GitHub. On Stack Overflow, developers frequently struggle with issues like \emph{workflow creation and scheduling, distributed task management, and processor configuration for data processing}, with challenges rooted in orchestrating tasks, managing DAGs, and debugging execution environments. GitHub, on the other hand, revealed additional pain points in areas such as \emph{error and bug fixing, documentation, and dependency management}, which are critical during real-world SWS development and maintenance. Several topics, like data structures and operations, task coordination, and workflow scheduling, emerged as common across both platforms, suggesting a shared concern between learning and implementation contexts. These challenges reflect the complexity and multidisciplinary nature of SWS development, where developers must navigate domain-specific tools, distributed environments, and evolving community standards. Understanding these topics provides a foundational step toward designing tools and practices that better support SWS developers.
\end{boxK}

\subsection{RQ2: What types of SWSs development questions are asked on technical forums?}
\textbf{Motivation: }Building upon identifying key topics in \textbf{RQ1}, a critical question emerges: \emph{What type} of questions are SWSs developers asking? Answering this is crucial as it deepens our understanding of the role SO plays for SWSs developers. This investigation shifts the focus from just what developers ask to why they ask it. By analyzing the types of posts created within each SWS topic, we aim to uncover the nature of the challenges developers face during SWSs development. Prior research \cite{rosen2016mobile, abdellatif2020challenges} highlights that developers typically pose different types of questions—such as 'How', 'Why', and 'What'—to address specific challenges. This detailed analysis will not only highlight the common problems encountered but also expose the underlying complexities and pain points within the SWSs development process. Categorizing these questions will provide insight into developers' cognitive processes and problem-solving strategies, helping to identify areas where additional support or resources are needed. Ultimately, this work offers a deeper understanding of the reasons SWSs developers turn to SO instead of relying solely on official documentation.

\textbf{Approach: }To achieve that, we followed a similar approach used by prior works to identify the types of posts on SO \cite{rosen2016mobile, abdellatif2020challenges, treude2011programmers}. In particular, we randomly sampled SO posts from each of the ten topics with a confidence level of 95\% and a confidence interval of 5\%. Our random sample size of each topic yields a total of 2933 posts: 371 Workflow Creation and Scheduling, 370 Distributed Task Management, 357 Processor for Data Processing, 356 Data Structures and Operations, 323 Workflow Execution, 314 Managing Rules, Input, Output, and Files, 288 Task Dependencies and Management, 208 Data Transformation, 192 Learning, and 154 Automation with Ansible. We reviewed each question's title and body to determine its type (at least two authors review each post). We manually analyzed and categorized each sample question. Each question is assigned a single label; if it could fall under multiple labels, we selected the most relevant one. Finally, each annotator labeled the post using one of the following types that were used by prior works \cite{rosen2016mobile, abdellatif2020challenges}. In cases where annotators couldn't reach a consensus, they revisited and discussed the questions together to achieve an agreement. Despite the annotators' excellent comprehensive understanding of SO data, we carefully review all the labeled data from the replication package of \cite{abdellatif2020challenges} where a similar task was done for chatbot development, together to ensure a shared understanding of the labeling process. Overall, the annotators achieved almost perfect agreement \citep{blackman2000interval}, with a kappa value of 0.82, on the 2,933 classified posts. This high kappa value signifies a very strong level of consistency among the annotators, indicating minimal discrepancy in the assessments. Such a result underscores the reliability of the classification process and suggests that the annotated data is robust, with the raters largely aligning in the judgments.

\begin{itemize}
    \item \textbf{How: }This category includes posts where developers seek guidance on specific methods or techniques to accomplish a particular task \cite{rosen2016mobile, abdellatif2020challenges}. Unlike 'Why' posts that inquire about underlying reasons or explanations, these posts are goal-oriented. Developers ask for precise, step-by-step instructions to achieve their specific objectives \emph{(e.g., 'How to remove default example DAGs in Airflow')}.
     \item \textbf{Why: }This type includes posts where developers inquire about certain phenomena' reasons, causes, or purposes \cite{rosen2016mobile, abdellatif2020challenges}. 'Why' posts often pertain to troubleshooting, where developers seek explanations for specific behaviors or issues \emph{(e.g., 'Why does Celery need a message broker?')}.
    \item  \textbf{What: }This category includes posts where developers seek specific information \cite{rosen2016mobile, abdellatif2020challenges}. These inquiries are often aimed at clarifying doubts to make more informed decisions. For example, a developer asked, \emph{'What is the best approach for adding cron jobs (scheduled tasks) for a particular service in Docker Compose?'} in SO.
    \item \textbf{Others: }We assigned this type to posts that do not fit into any of the above-described types \emph{(e.g., 'Understand the Notify and Wait process in NIFI in my flow?')}.
\end{itemize}

To evaluate the quality of our classification of the random sample, we utilized Cohen's Kappa \cite{mchugh2012interrater} to quantify the level of agreement among the annotators. This statistical measure helps us assess the consistency and reliability of our categorization process.

\begin{table}[htbp]
    \centering
    \caption{SWSs posts types on Stack Overflow}
    \label{tab:swssotopicstypes}
    \begin{tabular}{l|c|r|r|r}
        \toprule
        Topic & \% How & \% Why & \% What & \% Other \\
        \midrule
        Workflow Creation and Scheduling & 51.75 & 21.29 & 23.99 & 2.96 \\
        Distributed Task Management & 61.89 & 19.46 & 17.84 & 0.81 \\
        Processor for Data Processing & 63.31 & 13.45 & 18.77 & 4.48 \\
        Data Structures and Operations & 58.71 & 13.76 & 24.16 & 3.37 \\
        Workflow Execution & 44.89 & 21.98 & 28.79 & 4.34 \\
        Managing Rules, Input, Output, and Files & 65.92 & 16.24 & 16.56 & 1.28 \\
        Task Dependencies and Management & 60.42 & 20.49 & 17.71 & 1.38 \\
        Data Transformation & 76.92 & 10.58 & 9.13 & 3.37 \\
        Learning & 71.35 & 14.58 & 14.06 & 0.0 \\
        Automation with Ansible & 54.55 & 20.13 & 24.68 & 0.65 \\
        \midrule
        SWSs (all) & 60.97 & 17.2 & 19.57 & 2.26 \\
        \bottomrule
    \end{tabular}
\end{table}

\textbf{Results: }Table \ref{tab:swssotopicstypes} shows the percentage of the posts types for each SWSs topic. 
The analysis of question distributions across various topics in SWSs reveals distinct patterns that underscore developers' needs and knowledge gaps. The predominant focus on 'How' questions, averaging 60.97\% across all topics, indicates a significant demand for procedural guidance among users. For instance, topics such as 'Workflow Creation and Scheduling' (51.75\%) and 'Distributed Task Management' (61.89\%) demonstrate users' strong interest in understanding the execution of specific tasks. Similarly, 'Processor for Data Processing' (63.31\%) and 'Data Structures and Operations' (58.71\%) highlight the necessity for detailed instructions and methodologies. Conversely, the 'Why' questions, constituting an average of 17.2\%, reflect users' curiosity about the underlying reasons behind certain actions, with 'Workflow Execution' and 'Task Dependencies and Management' topics showing notable percentages (21.98\% and 20.49\%, respectively). 'What' questions, accounting for 19.57\% on average, indicate users' need for clarification on the specifics of tasks or concepts, as seen in 'Workflow Execution' (28.79\%) and 'Automation with Ansible' (24.68\%). The 'Other' category, with a minimal average of 2.26\%, suggests that most user inquiries are well-categorized within the primary question types.

We observed that the 'How' type consistently dominates across various domains. For instance, 'How' questions constitute 61.8\% of chatbot-related posts and 59\% of mobile development posts, closely aligning with our domain's 60.97\%. The second most common question type in chatbot and mobile development topics is 'Why' (25.4\% and 29\%, respectively). Interestingly, the second most common question type in our domain is 'What', whereas 'Why' questions rank third. Conversely, 'What' questions are the third most frequent in the chatbot and mobile development domains.

The higher percentage of 'What' type questions compared to the 'Why' type in SWSs posts on SO can be attributed to the practical and task-oriented nature of scientific computing. Users in this domain often seek clarity on specific functionalities, features, or tools necessary for their workflow, which drives a higher frequency of 'What' inquiries. These questions typically focus on understanding particular components, commands, or processes within their workflows. In contrast, 'Why' questions, which delve into the rationale or underlying principles, are less frequent because users are more immediately concerned with the practical implementation and troubleshooting of their tasks. This trend reflects the urgency to solve concrete problems and efficiently utilize tools and methods in SWSs.

\subsection{RQ3: Which topics are most difficult to answer?}
\textbf{Motivation: }Given our understanding of different topics obtained from SO and GitHub, our next step is to investigate the difficulty of answering posts, resolving issues, and pull requests within each identified topic. By determining whether some topics are more challenging to address than others, we can identify areas that require increased community support. Additionally, this analysis can help us highlight topics where there may be a need for improved tools or frameworks to assist developers in tackling SWSs development challenges.
\newline
\textbf{Approach: }We assessed the difficulty of each topic using two distinct metrics. For SO, we considered the percentage of posts within a topic that lack accepted answers and the median time, in hours, for an answer to be accepted. On GitHub, our metrics include the percentage of unresolved issues and pull requests, along with the median time taken to resolve them. This approach is consistent with methods used in previous studies \citep{rosen2016mobile, bagherzadeh2019going, yang2016security, li2021understanding, scoccia2021challenges, abdellatif2020challenges}. The metrics are described in detail below:
\newline
\textbf{Metrics for SO: }
\begin{enumerate}
    \item \textbf{The percentage of posts of a topic without accepted answers (\% w/o accepted answers).} For each SWSs topic, we measured the percentage of posts without accepted answers. Although a post can receive multiple answers, only its author can mark one as accepted if it satisfactorily resolves the original question. Consequently, topics with fewer accepted answers are considered more challenging  \cite{abdellatif2020challenges, rosen2016mobile, bagherzadeh2019going}.
    
    \item \textbf{The median time for an answer to be accepted (median time to answer (hrs.)).} We measured the median time, in hours, for posts to receive an accepted answer. This metric is based on the time the accepted answer is created, not when it is marked as accepted. A longer time to receive an accepted answer indicates that the post is more challenging \cite{abdellatif2020challenges, rosen2016mobile,bagherzadeh2019going}.
\end{enumerate}
\textbf{Metrics for GitHub: }
\begin{enumerate}
\item \textbf{The percentage of unresolved issues and pull requests of a topic (\% w/o solutions).} The unresolved issues and pull requests percentage within a specific topic indicates the proportion of issues and pull requests that remain unresolved within that particular topic area. For each SWSs topic, we measured the percentage of issues and pull requests that are not closed (when a developer closes an issue or pull request in a software development context, it typically means that they have addressed the problem or implemented the requested feature and believe the issue or pull request is resolved). Therefore, topics with open state are considered more difficult.

\item \textbf{The median time for an issue or pull request to be resolved (Median Time to Resolve(hrs.)).} We calculate the median resolution time in hours for issues and pull requests, considering the duration between the issue's or pull requests' creation and closing times. This metric focuses explicitly on closed issues or pull requests since open issues or pull requests lack a closing time. The greater the duration it takes to be resolved, the more challenging it is considered.
\end{enumerate}
\begin{table}[htbp]
    \centering
    \caption{The difficulty per topic on SO data}
    \label{tab:sotopicdifficulty}
    \begin{tabular}{l|r|r}
        \toprule
        Topic & Posts w/o Accepted (\%) & Median Time (h) \\
        \midrule
        Workflow Creation and Scheduling & \hfill 63.32 & 9.09 \\
        Distributed Task Management & \hfill 61.22 & 9.32 \\
        Processor for Data Processing & \hfill 59.53 & 4.77 \\
        Data Structures and Operations & \hfill 57.71 & 6.88 \\
        Workflow Execution & \hfill 68.50 & 23.58 \\
        Managing Rules, Input, Output, and Files & \hfill 52.43 & 7.73 \\
        Task Dependencies and Management & \hfill 47.13 & 7.06 \\
        Data Transformation & \hfill 34.08 & 1.81 \\
        Learning & \hfill 44.74 & 6.82 \\
        Automation with Ansible & \hfill 58.98 & 4.49 \\
        \bottomrule
    \end{tabular}
\end{table}
\textbf{Results: }Table \ref{tab:sotopicdifficulty} presents the percentage of questions without accepted answers and the median time (in hours) to receive an accepted answer for each identified topic of SO data. The topic of \emph{Workflow Execution} has the highest percentage of unanswered questions and also requires the longest median time to receive an answer. To understand the high rate of posts without accepted answers (68.50\%), we examined the posts within this topic. We discovered that posts without accepted answers often received low scores (666 posts have a score of zero) from developers on Stack Overflow. This may be due to unclear or poorly formulated questions, reducing the likelihood of receiving an accepted answer. Additionally, we explored why this topic has a significantly higher median response time. We found that questions on this topic often involve complex tasks such as chaining jobs, managing processes, running specific jobs on particular nodes, and integrating tools in SWS. These tasks require rigorous testing and workflow implementation as well as execution in SWS, which significantly increases the time needed to provide an accepted answer.

We also found that \emph{Workflow Creation and Scheduling} and \emph{Distributed Task Management} topics have low acceptance rates and high median time requirements to achieve an accepted answer. By exploring the posts, we identified 3,078 and 2,228 posts with zero or negative scores, indicating low-quality questions. The high median time for answers in these areas is due to the highly specific nature of the questions, which are often tied to individual setups, tools, and requirements. Additionally, reproducing these issues without a similar environment is difficult, complicating the process of diagnosing and providing solutions. Many times, offering a meaningful answer demands a thorough understanding of the specific infrastructure and software versions in use.

Conversely, posts related to \emph{Data Transformation} boast the highest percentage of accepted answers and a remarkably low median time of just 1.81 hours to receive an accepted answer. These posts often focus on how to perform data transformations using specific technologies  (e.g., \emph{How to convert string to Datetime with specific format using jolt transformation?}) or transforming JSON, array, or vector data. This topic's close connection to traditional software development likely explains why SO respondents tend to answer these questions more quickly and frequently.

To gain a comprehensive understanding of SWSs posts, we aim to examine whether there is a statistically significant correlation between their difficulty and popularity. Specifically, we used the Spearman Rank Correlation Coefficient \cite{ref1} to assess the relationships between two popularity metrics (average views and average scores) and two difficulty metrics (percentage without accepted answers and median time to answer in hours).
\begin{table}[htbp]
    \centering
    \caption{Correlation of topics popularity and difficulty.}
    \label{tab:correlationofpopularityanddifficulty}
    \begin{tabular}{r|r|r}
        \toprule
        Correlation Coeff./p-value & Avg. Views & Avg. Score \\
        \midrule
        \textbf{\% w/o Accepted Answers} & \hfill 0.673/0.033 & 0.458/0.183 \\
        \textbf{Median Time to Answer(Hrs)} & \hfill 0.144/0.691 & -0.053/0.884 \\
        \bottomrule
    \end{tabular}
\end{table}

We chose Spearman's rank correlation because it makes no assumptions about the normality of the data distribution. As shown in Table \ref{tab:correlationofpopularityanddifficulty}, the analysis of correlations indicates a significant positive relationship between the percentage of unanswered questions and average views on SO, suggesting that topics with more unanswered questions tend to attract greater interest. This phenomenon can be attributed to developers frequently tackling intricate or specialized subjects that fewer people are knowledgeable about, which intrigues users facing similar challenges seeking solutions or insights. Moreover, unanswered questions often foster community engagement as users monitor them in anticipation of potential answers or opportunities to demonstrate expertise. For instance, the question \emph{How to retrieve a list of tasks in a Celery queue?} has garnered over 273K views despite lacking an accepted answer. We found many questions with numerous views remain unanswered or lack accepted solutions. However, there is a very weak positive and statistically insignificant correlation between the median hours to answer and average views, suggesting no meaningful relationship. Similarly, there is a moderate positive but statistically insignificant correlation between the percentage of unanswered questions and average score, and a very weak negative and statistically insignificant correlation between the median hours to answer and average score. These results imply that while the percentage of unanswered questions may influence viewership, the time taken to answer and the average scores are not significantly correlated with either difficulty or popularity metrics.

\begin{table}[htbp]
    \centering
    \caption{Unresolved Issues and Median Time for Different SWSs Topics on GitHub}
    \label{tab:unresolved_issues_median_time}
    \begin{tabularx}{\linewidth}{X|r|r}
        \toprule
        \textbf{Topic Name} & \textbf{Unresolved (\%)} & \textbf{Median Time (h)} \\
        \midrule
        Errors and Bugs Fixing & 9.98 & 52.74 \\
        Documentation & 8.03 & 40.50 \\
        Dependencies & 0.02 & 44.62 \\
        Visual and Branding Enhancements & 5.97 & 18.28 \\
        Automated Performance and Tool Update Integration & 0.29 & 0.1 \\
        CRAN Release Management and Package Validation & 4.37 & 85.53 \\
        Managing Releases & 0.38 & 2.27\\
        Browser Compatibility and HDFS Integration Issues & 25.0 & 95.96 \\
        Chord Execution and Task Coordination Issues  & 9.7 & 91.45 \\
        Data Shuffling and Partitioning Strategies & 10.12 & 75.17\\
        Image Compression \& Optimization  & 0.0 & 0.97 \\
        Automating Code Quality Checks  & 0.01& 16.63 \\
        System Redesign and API Migration & 3.85 & 117.95 \\
        \bottomrule
    \end{tabularx}
\end{table}

Table~\ref{tab:unresolved_issues_median_time} presents the proportion of unresolved issues and the median time to close issues across various identified topics in SWS development based on GitHub issue and pull request data. This analysis helps uncover persistent developer pain points and the relative difficulty associated with resolving them.
Notably, \emph{Browser Compatibility and HDFS Integration Issues} exhibit the highest rate of unresolved issues (25.0\%) and a significantly long median resolution time (95.96 hours), indicating substantial complexity or lack of sufficient support in addressing such problems. Similarly, \emph{Chord Execution and Task Coordination Issues} (9.7\% unresolved, 91.45 hours) and \emph{Data Shuffling and Partitioning Strategies} (10.12\% unresolved, 75.17 hours) also show high-resolution times, suggesting that challenges involving distributed computation and task orchestration remain particularly difficult to resolve.

Conversely, topics such as \emph{Automated Performance and Tool Update Integration, Image Compression \& Optimization}, and \emph{Managing Releases} show low unresolved rates and short median resolution times, indicating they are either well-supported or considered lower priority by the community. Interestingly, even topics like \emph{Documentation} (8.03\% unresolved) and \emph{Errors and Bugs Fixing} (9.98\% unresolved) have non-trivial unresolved rates, reinforcing that even foundational issues can persist over time. \emph{System Redesign and API Migration}, while showing a moderate unresolved rate (3.85\%), has the highest median resolution time (117.95 hours), possibly due to the complex, architectural nature of such tasks.
Overall, the data highlights that unresolved rates and resolution times vary significantly by topic, with issues related to distributed execution, integration, and large-scale redesign emerging as particularly burdensome for developers. These findings point to critical areas where improved tooling, documentation, or community support could help alleviate recurring developer challenges in SWS development.
\begin{boxK}
The analysis of difficulty across various SWSs topics on SO and GitHub reveals distinct patterns. On SO, topics like \emph{Workflow Execution} (68.50\% unresolved, 23.58 hours median time) and {Workflow Creation and Scheduling} (63.32\%, 9.09 hours) have higher unresolved post percentages, indicating their complexity. In contrast, \emph{Data Transformation} (34.08\%, 1.81 hours) seems easier for developers. Similarly, the GitHub analysis revealed that \emph{System Redesign and API Migration} had the highest median time to close issues, at 117.95 hours, followed by \emph{Browser Compatibility and HDFS Integration Issues} (95.96 hours) and \emph{Chord Execution and Task Coordination} (91.45 hours). Notably, \emph{Browser Compatibility and HDFS Integration} also exhibited the highest proportion of unresolved issues (25.0\%), highlighting substantial challenges related to system maintenance and cross-platform compatibility. Other topics, such as \emph{Data Shuffling and Partitioning Strategies and CRAN Release Management}, also demonstrated prolonged resolution times, reflecting persistent developer struggles with performance tuning, dependency handling, and validation workflows within SWS environments.
In contrast, topics like \emph{Image Compression and Optimization and Automated Performance and Tool Update Integration} showed median resolution times below one hour and near-zero unresolved rates, suggesting these areas pose minimal difficulty or are well-supported by existing tooling. Overall, these findings indicate that developers face the greatest difficulty in infrastructure-level orchestration, fault-tolerant execution, and managing breaking system-level changes. Addressing these pain points through enhanced automation, robust deployment support, and improved documentation is essential to reduce the development burden and improve the overall maintainability of SWSs
\end{boxK}

\section{Discussion \& Implications of our Findings}\label{discussion-implications}
In this section, we examined the evolution of SWSs topics and compared our results with those of previous studies. Subsequently, we analyzed the data to identify the dominant topics across various platforms and discuss the significance of our findings.
\begin{figure}[htbp]
  \includegraphics[width=\textwidth]{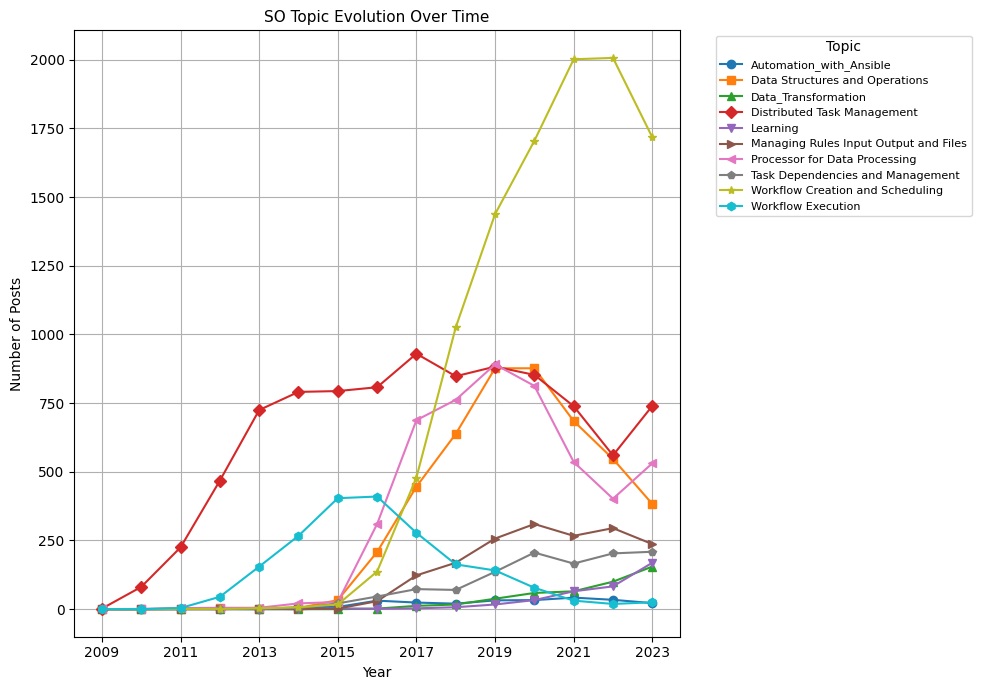}
  \caption{SWSs topics evolution over time on SO}
  \label{fig:absolutegrowhofswstopics}
\end{figure}

\subsection{SWSs Topics Evolution}
SWS is an emerging topic that is attracting increasing attention from developers across various domains, including software engineering \cite{6866038}, machine learning \cite{nadeem2019using}, bioinformatics \cite{wratten2021reproducible}, and data analysis\cite{liew2016scientific}. To examine the evolution of SWSs, we presented Figure \ref{fig:swsdatainqaforums}, which shows the consistent growth of SWSs over time. The upward trend in the number of posts, issues, and pull requests indicates that SWSs topics are gaining significant attention from the community.

\begin{figure}[htbp]
  \includegraphics[width=\textwidth]{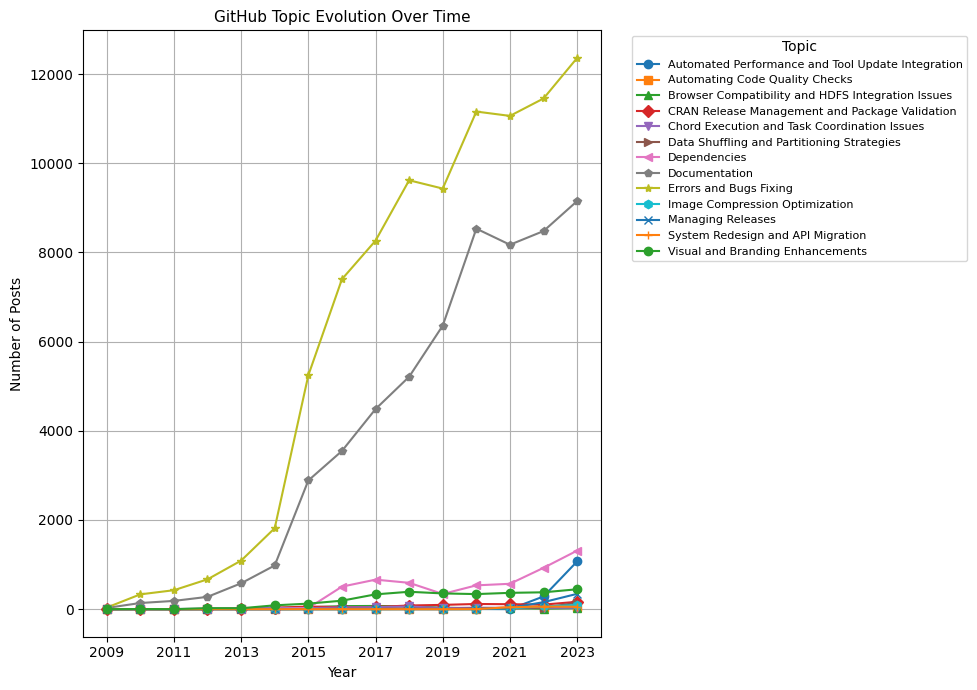}
  \caption{SWSs topics evolution over time on GitHub}
  \label{fig:absolutegrowhofswstopics-github}
  \vspace{-1.5em}
\end{figure}

To analyze how developer challenges in SWSs have evolved over time, we examined the year-wise distribution of all identified topics across both SO (Fig: \ref{fig:absolutegrowhofswstopics}) and GitHub (Fig: \ref{fig:absolutegrowhofswstopics-github}). On SO, early years were dominated by foundational concerns such as \emph{Workflow Creation and Scheduling and Distributed Task Management}, which continued to grow steadily over the past decade. These trends reflect ongoing difficulties in defining pipelines, managing execution orders, and integrating workflow engines like Airflow and Nextflow. In recent years, the rise of topics such as \emph{Processor for Data Processing, Workflow Execution, and Task Dependencies and Management indicates} a shift toward runtime optimization, resource allocation, and fine-grained control over pipeline execution.

In parallel, GitHub exhibited a different yet complementary evolution. \emph{Errors and Bugs Fixing} remained the most persistent topic across all years, underscoring its centrality in practical software maintenance. Documentation also showed a notable increase, reflecting the community’s growing attention to tool usability, onboarding, and reproducibility. Emerging topics like \emph{Workflow Execution and Processor for Data Processing} mirrored the recent SO trends, while GitHub uniquely highlighted platform-specific challenges such as \emph{Browser Compatibility and HDFS Integration Issues and Chord Execution and Task Coordination}. These reflect low-level system dependencies and orchestration concerns that are more likely to surface in code-level issue tracking than in general developer Q\&A forums.
Overall, Stack Overflow topics tend to center on conceptual understanding, best practices, and troubleshooting common errors, whereas GitHub discussions more frequently capture in-depth implementation, system-specific debugging, and long-term maintenance issues. The alignment and divergence of topic trajectories across platforms reveal the layered nature of SWS development, from initial learning and design on SO to deployment and sustainability on GitHub.
\begin{figure}[htbp]
  \includegraphics[width=\textwidth]{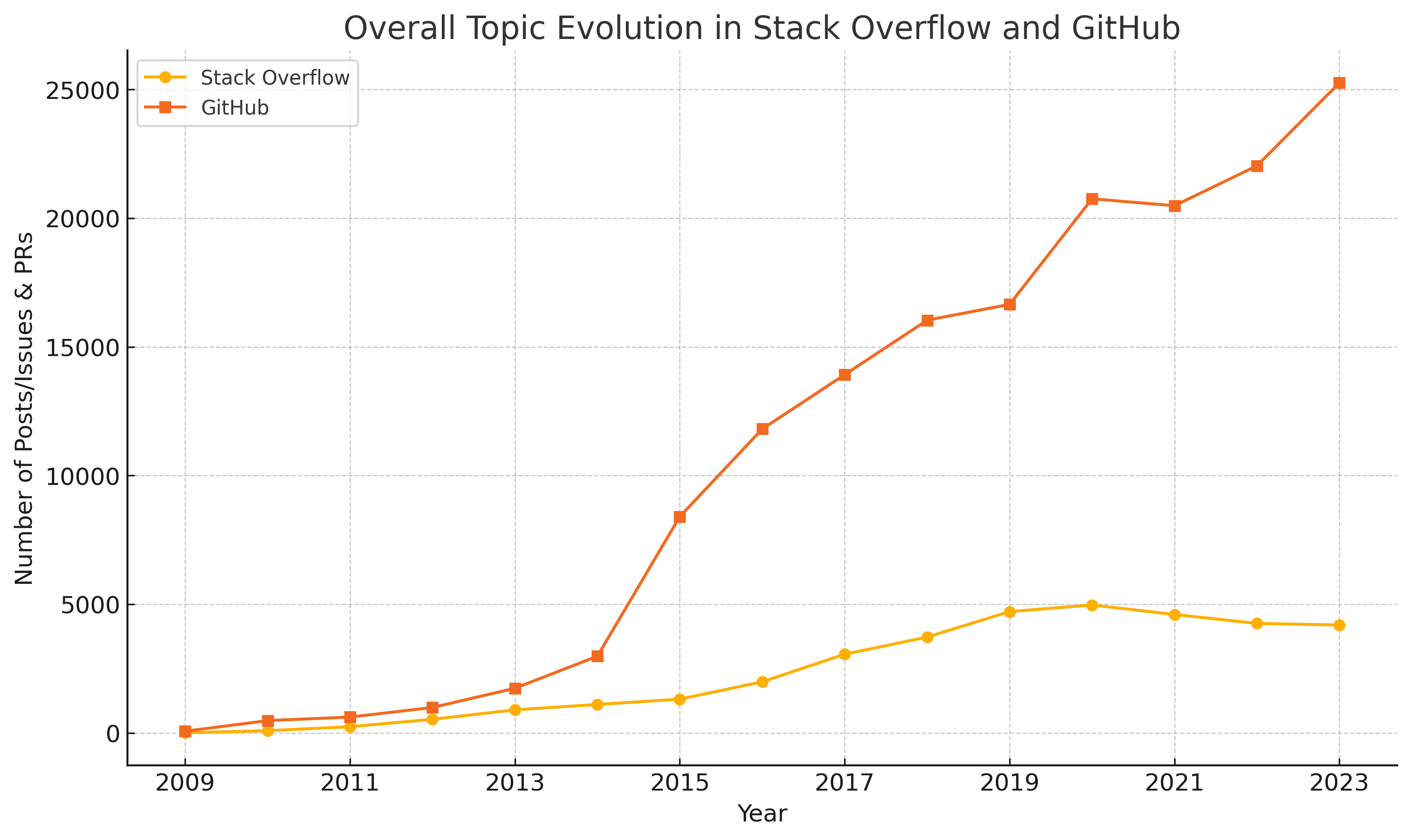}
  \caption{Overall evolution of SWSs Data}
  \label{fig:so_github_topic_evolution_chart}
  \vspace{-1.5em}
\end{figure}
Figure \ref{fig:so_github_topic_evolution_chart} presents the evolution of overall topic-related developer activity on SO and GitHub over time. The number of GitHub issues and pull requests shows a marked increase beginning in 2015, reflecting the growing use of GitHub for collaborative development and issue tracking in SWSs. In contrast, SO activity gradually rose until it peaked around 2020, after which it plateaued or slightly declined. This divergence may indicate that while SO remains a valuable platform for asking general and technical questions, GitHub has increasingly become the central hub for hands-on debugging, issue resolution, and workflow collaboration. The visualization underscores the temporal dynamics of developer engagement and complements our topical analysis by showing how platform usage and topic discussion volume have shifted over time.

\subsection{SWSs compared to Other SE Fields}
SWS is an emerging field, making it essential to investigate how its topics compare to discussions in other software engineering domains such as chatbot \cite{abdellatif2020challenges}, quantum software engineering \cite{li2021understanding}, mobile apps \cite{rosen2016mobile}, big data \cite{bagherzadeh2019going} and security \cite{yang2016security}. To answer this question, we examined the difficulty and popularity of SWSs topics and compared them to other disciplines using data from similar studies on SO focused on chatbots, quantum software engineering, mobile apps, big data, and security. These domains have been previously studied and provide a useful benchmark. This comparison is crucial for identifying trends and gaps, allocating resources, and fostering innovation. Since these studies were conducted at different times, we used their reported keywords to construct an updated dataset and calculate the popularity and difficulty metrics for each field. This approach allows us to get insight into the current landscape and future directions of SWSs about other key areas in software engineering.
\begin{table}[htbp]
    \centering
    \caption{Comparison of popularity and difficulty between different fields}
    \label{tab:differentfiledcompare}
    \begin{tabularx}{\linewidth}{X|r|r|r|r|r|r|r}
        \toprule
        Metrics & SWSs & Chatbot & QSE & WebApp & Mobile& Security & BigData \\
        \midrule
        \# of Posts & 35,619 & 7104 & 634 & 14075 & 2474527 & 169865 & 194356 \\
        Avg ViewCount & 2119.46 & 996.46& 719.06 & 2409.98 & 2792.04 & 3462.07& 2799.27 \\
        Agv Score &1.96  & 0.78 & 1.10& 2.12 & 2.40 & 2.73 & 1.77 \\
       Avg AnswerCount &1.09 & 1.06 & 1.08 & 1.02 & 1.45& 1.41 & 0.84 \\
       \% w/o Answers & 60.03 & 67.69 & 60.57 & 64.31 & 53.21 & 52.33 & 58.83 \\
       \% Duplicate & 0.37 & 0.72 & 0.47 & 0.87 & 1.67 & 1.6 & 1.17 \\
        Med. TimeToAns (Hrs) & 7.67 & 16.16 & 9.17 & 11.38 & 0.91 & 1.75& 3.26 \\
        \bottomrule
    \end{tabularx}
\end{table}
Table \ref{tab:differentfiledcompare} shows the result of the popularity and difficulty metrics among the seven fields. The metrics reveal several compelling reasons for increased focus on the development of SWSs. Despite having 35,619 posts and a high average view count of 2,119.46, indicating substantial interest, the community engagement for SWSs remains moderate. When compared to fields like Chatbot and QSE, SWSs perform better in terms of view count but still has a high percentage of unanswered posts (60.03\%), similar to Chatbot (67.69\%) and QSE (60.57\%). The median time to answer for SWSs is 7.67 hours, which is slower than more mainstream technologies like Mobile (0.91 hours) and Security (1.75 hours) but faster than Chatbot (16.16 hours) and WebApp (11.38 hours). Additionally, SWSs have a lower average score (1.96) compared to fields like WebApp (2.12) and Security (2.73), indicating room for improvement in the quality of responses. Given the critical role that SWSs play in automating and optimizing complex scientific processes, enhancing community support, improving documentation and resources, and implementing measures to ensure quicker and higher quality responses are essential. This focused development can bridge the support gap, promote more effective utilization, and drive scientific research and innovation advancements.

We then showed comparison across different fields—including Chatbots, Quantum Software Engineering (QSE), WebApp, Mobile, Security, and BigData in terms of the percentage of duplicate questions in Table \ref{tab:differentfiledcompare}. The \% Duplicate quantifies the proportion of Stack Overflow questions marked as duplicates within each field. Interestingly, we observe that SWSs have an exceptionally low duplicate rate of 0.37\%, which is significantly lower than more mature domains like Mobile (1.67\%), Security (1.6\%), and WebApp development (0.87\%). This indicates that developers working in SWSs are frequently encountering novel or less-documented problems, suggesting that the community is still in the early stages of building a reusable knowledge base. By contrast, fields like Mobile and Security appear to benefit from more accumulated knowledge, as reflected by their higher duplicate percentages. This aligns with our difficulty analysis, where SWSs exhibit a relatively high percentage of unanswered posts (60.03\%) and a longer median time to answer (7.67 hours) compared to more mature fields. These metrics support the observation that SWS-related questions are more challenging to answer and less likely to have existing solutions, reinforcing the need for more structured community knowledge development.

\subsection{SWSs Challenges compared to SE Challenges: }While some challenges identified in this study, such as dependency management, debugging, and scalability, are common in traditional SE, their manifestations in SWSs present distinct complexities that require further exploration. To ensure clarity and maintain the focus on SWS-specific challenges, we differentiated how these issues uniquely impact SWSs compared to general SE challenges.
\begin{itemize}
    \item \textbf{Dependency Management in SWSs vs. Traditional SE: }In traditional SE, dependency management primarily involves managing software libraries and resolving version conflicts \cite{de2008empirical, cataldo2009software}. However, in SWSs, dependency management extends beyond software to include:
    \newline
    \textbf{Data dependencies: }Ensuring datasets are correctly versioned, linked to workflow executions, and reproducible.
    \newline
    \textbf{Computational dependencies: }Managing task execution across distributed infrastructures such as high-performance computing clusters and cloud environments.
    \newline
    \textbf{Provenance tracking: }Maintaining workflow lineage to facilitate reproducibility, validation, and long-term usability of results.

    \item \textbf{Error Handling and Debugging in SWSs vs. Traditional SE: }Traditional SE debugging focuses on fixing code-level errors \cite{hailpern2002software}. On the other hand, SWSs debugging presents additional challenges, such as:
    \newline
    \textbf{Dynamic execution environments }, where failures can result from missing data, unresolvable dependencies, or scheduling conflicts.
    \newline
    \textbf{Data integrity and checkpointing }to ensure that partial computations can be resumed.
    \newline
    \textbf{Handling failed tasks in multi-node distributed workflows }e.g., failures in HPC or cloud execution.

    \item \textbf{Scalability and Performance Bottlenecks in SWSs: }In traditional SE, scalability challenges mostly arise in terms of handling increased user requests (e.g., scaling web applications) \cite{duboc2007framework}. On the other hand SWSs scalability concerns include: 
    \newline
    \textbf{Efficient parallelization of computational workflows }in data-intensive experiments.
    \newline
    \textbf{Resource allocation optimization }in clusters or cloud platforms.
    \newline
    \textbf{Handling petabyte-scale data processing }in disciplines like bioinformatics and climate modeling.
    \item \textbf{Reproducibility Challenges in SWSs: }Unlike traditional software applications, SWSs require rigorous reproducibility standards for scientific results. Challenges include: 
    \newline
Version control for datasets, computational steps, and execution environments. Ensuring workflows produce the same results across different infrastructures (HPC, cloud, local). Recording metadata and provenance tracking to support research transparency.
\end{itemize}

\subsection{Impact of Emerging Technologies on SWSs Development}
Emerging technologies such as cloud computing, containerization, and distributed computing frameworks have significantly reshaped the development, deployment, and execution of SWSs. While these technologies help alleviate long-standing issues such as scalability, reproducibility, and dynamic resource allocation, they also introduce new challenges related to system integration, orchestration, and data management.

Cloud Computing enables on-demand access to vast computing and storage resources through platforms like Amazon EC2, Google Cloud, and Microsoft Azure. For SWSs, this means workflows can dynamically scale according to workload needs, especially in patterns like scatter-gather that are common in scientific pipelines. SWSs like Pegasus \cite{deelman2015pegasus} have successfully leveraged cloud environments such as Amazon EC2, NERSC’s Magellan, and FutureGrid to enable cross-cloud execution \citep{juve2010scientific}. Clouds address scalability and cost-efficiency challenges by dynamically scaling resources in alignment with workflow stages. Benefits include improved responsiveness, elastic scaling, and support for collaborative workflows. However, integrating SWSs with cloud infrastructure presents architectural and operational challenges, including resource acquisition, monitoring, provenance tracking, and fault tolerance \citep{berriman2013application, ahmad2021scientific, yu2005taxonomy}. Moreover, clouds introduce integration complexities and security concerns, especially in multi-tenant or cross-institutional settings \citep{zhao2011opportunities, foster2008cloud}.

Containerization technologies such as Docker and Singularity have become essential for ensuring reproducibility across diverse computing environments. By encapsulating software, dependencies, and configurations, containers eliminate common dependency conflicts in SWS execution and enable consistent behavior across local, HPC, and cloud environments \cite{di2017nextflow}. Systems like Nextflow and Galaxy have increasingly embraced container-based execution to ensure workflow portability. However, managing large containerized workflows at scale introduces orchestration overhead. Tools like Kubernetes or Apache Mesos are required to manage execution across distributed environments, adding layers of complexity that are not trivial for domain scientists to manage \cite{boettiger2015introduction}.

Distributed data processing frameworks, such as Apache Spark, Dask, and Ray, enhance the performance and parallelism of data-intensive workflows in scientific domains. These systems support large-scale in-memory computations and distributed task scheduling, but they require careful tuning of data partitioning, resource scheduling, and failure recovery strategies \cite{zaharia2010spark}. Misconfiguration or inefficient partitioning can result in suboptimal resource usage and performance bottlenecks.

Serverless computing and microservice-based workflow architectures represent new trends in workflow automation. Platforms such as AWS Lambda and Google Cloud Functions allow execution of stateless, event-driven components, making them appealing for modular and lightweight tasks. When combined with orchestration tools like AWS Step Functions or Argo Workflows \citep{rak2012argo}, they can provide scalable and lightweight alternatives to traditional SWS engines. However,  the statelessness and time constraints of serverless platforms limit their applicability to long-running or stateful scientific workflows \citep{baldini2017serverless}.

Overall, while cloud-native approaches can mitigate some key developer challenges, such as ensuring reproducibility, improving scalability, and reducing infrastructure maintenance, they can also exacerbate integration complexity, orchestration burden and usability challenges. These issues may be addressed through  better abstraction layers, scheduling techniques (e.g., ADAS, IC-PCP), and workflow engines that simplify multi-environment deployment \cite{adhikari2019survey}.

\subsection{Practical Recommendations and Best Practices}
Our findings highlighted critical challenges in the development of SWSs. Here, we provided actionable recommendations to assist developers and practitioners in overcoming these difficulties.
\newline
\begin{itemize}
    \item \textbf{ Workflow Execution and Management: }We found that developers often struggle with optimizing workflow execution due to inefficient scheduling, error handling, and resource allocation. SWSs like  Nextflow, Snakemake, KNIME, and Galaxy face several scheduling challenges, including inefficient resource allocation, lack of fault tolerance, high latency in task execution, and limited adaptability to dynamic workloads. Many systems like Luigi struggle with load balancing, handling failures intelligently, and optimizing multi-cloud or GPU-based resource scheduling. Some systems, like KNIME and Oozie, rely on predefined schedules, making them inefficient in dynamic environments where workloads change over time. Systems like Celery and Dask, which dynamically distribute tasks, may suffer from overloading specific nodes, leading to performance bottlenecks. Additionally, most rely on static or heuristic-based scheduling rather than leveraging AI-driven predictive scheduling for better adaptability. Improvements such as machine learning-based task prediction, failure-aware rescheduling, fine-grained checkpointing, and multi-objective optimization for cost and speed can significantly enhance scheduling efficiency, making these systems more robust and scalable for modern scientific workflows.
    
    \item \textbf{Task Dependencies and Workflow Orchestration: } We found many SWSs like Snakemake,  KNIME, Galaxy, and Luigi face challenges in task dependency management and workflow orchestration, including manual predefined task dependencies, inefficient resolution of large-scale DAGs, and execution order inefficiencies. Many (i.e., Luigi, Galaxy) struggle with handling dependencies, managing long dependency chains, and dynamically adjusting task relationships at runtime. Additionally, high orchestration overhead, poor failure recovery, and limited parallel execution hinder scalability and efficiency. Improvements such as intelligent dynamic dependency resolution, adaptive failure handling, and parallelism-aware scheduling can significantly enhance workflow execution, making these systems more robust and scalable.

    \item \textbf{Data Structures and Operations Optimization: }Many SWSs like Snakemake, Celery, and Nextflow face challenges in data structures and operations optimization, affecting their performance and scalability. Snakemake and Nextflow often encounter data management bottlenecks when handling large datasets due to inefficient data handling architectures. Celery and Dask struggle with concurrency and parallelism, leading to suboptimal resource utilization in distributed task execution. Luigi and Oozie face dependency resolution inefficiencies, increasing execution latency and reducing throughput. Galaxy and KNIME lack dynamic resource allocation mechanisms, making it difficult to optimize operations across diverse computing environments. To overcome these challenges, SWS can implement more efficient data structures (e.g., indexed storage, in-memory databases), enhance scheduling algorithms for better parallel execution, adopt intelligent dependency resolution techniques, and integrate adaptive resource management to dynamically adjust workloads based on real-time requirements. These improvements can significantly enhance the efficiency, scalability, and reliability of SWSs.

    \item \textbf{Debugging and Error Handling: }SWSs face significant challenges in debugging and error handling, impacting workflow reliability and efficiency. Snakemake and Nextflow often produce ambiguous error messages, making it difficult to trace the root cause without detailed logs, especially in HPC environments. Celery and Dask struggle with concurrency-related race conditions, leading to unpredictable execution errors. KNIME and Oozie lack consistent error handling across integrated components, while NiFi provides opaque error messages, complicating troubleshooting. To address these issues, SWSs should adopt comprehensive logging and monitoring, implement standardized error-handling mechanisms, integrate interactive debugging tools, and enhance community-driven documentation and automated testing frameworks. These improvements would significantly boost SWSs reliability, ease of debugging, and operational efficiency.

    \item \textbf{Scalability and Performance Optimization: }We found that many SWSs face scalability and performance optimization challenges impacting their efficiency. KNIME and Galaxy struggle with large dataset processing, leading to slow execution times and I/O bottlenecks. Luigi and Oozie face difficulties in dependency resolution, increasing latency in complex workflows. Celery and Dask, while designed for distributed execution, lack efficient parallel processing and resource management, causing suboptimal performance in large-scale tasks. To resolve these issues, SWS should adopt dynamic resource allocation, improve dependency resolution algorithms, enhance parallel execution capabilities, and optimize data handling mechanisms to reduce performance bottlenecks. These improvements will significantly boost scalability, making SWS more efficient for large-scale scientific workflows.

    \item \textbf{Domain-Specific Considerations: }Scientific Workflow Systems (SWS) are used across various scientific domains, each requiring tailored optimizations to address specific challenges. In bioinformatics and genomics, workflows should prioritize reproducibility and seamless dataset integration to handle large-scale biological data efficiently. For climate science and environmental research, optimization should focus on high-performance computing (HPC) integration and geospatial data processing to manage extensive simulations and satellite data. In machine learning and AI workflows, SWSs must enhance automation for model training, hyperparameter tuning, and distributed computing to support scalable experimentation. Addressing these domain-specific requirements ensures that SWSs are adaptable, efficient, and capable of supporting cutting-edge research in their respective fields.
\end{itemize}

By incorporating these recommendations, we aim to ensure that our study provides both a diagnostic and prescriptive perspective, making the findings more actionable for practitioners working with SWSs across diverse research domains.

\subsection{Implications}
The results of our study provide valuable insights that can help the SWSs community prioritize their efforts on the most critical challenges in SWSs development. Our analysis of SO and GitHub data reveals several key implications for both the development and utilization of these systems. Below, we outline how our findings can guide practitioners, researchers, and educators in enhancing the practice and learning of SWSs development.
\newline
\textbf{Implication for Practitioners: }To make the findings of this study actionable for practitioners of SWSs, we outline several evidence-based recommendations to improve system design, documentation quality, and development practice.
\begin{table}[htbp]
\centering
\caption{Characteristics of an Ideal SWS}
\label{tab:ideal_sws_characteristics}
\begin{tabularx}{\linewidth}{>{\hsize=0.7\hsize}X|>{\hsize=1.3\hsize}X}
\toprule
\textbf{Challenge Area} & \textbf{Design Consideration} \\
\midrule
Task Scheduling and Management Complexity & Simplify task and dependency specification with intuitive, high-level abstractions; automatically infer dependency graphs where possible. \\
\hline
Distributed Execution and Resource Management & Provide scalable, native support for distributed execution; include templates for configuring resources across different environments. \\
\hline
Containerization Challenges & Seamlessly integrate container support with automatic environment detection and detailed error validation. \\
\hline
Error Diagnosis and Debugging & Deliver actionable error messages with suggested resolutions and offer integrated workflow tracing tools. \\
\hline
Data Handling and File Management & Automate file staging, organization, and cleanup to reduce user burden. \\
\hline
Workflow Reusability and Modularity & Encourage modular workflow design and provide libraries of reusable components. \\
\hline
User Guidance and Documentation & Embed contextual documentation, usage hints, and annotated templates within the development environment. \\
\hline
Scalability and Optimization Support & Offer built-in profiling tools and optimization guidance for large-scale workflows. \\
\hline
Monitoring and Visualization & Provide real-time execution dashboards to visualize workflow progress and states. \\
\hline
Extensibility and Integration & Enable easy integration with external tools and data sources via standardized extension mechanisms. \\
\bottomrule
\end{tabularx}
\end{table}

First, based on the challenges identified in our analysis, we proposed a vision of an ideal SWS that minimizes common developer pain points. Table~\ref{tab:ideal_sws_characteristics} summarizes key design considerations that can help practitioners plan improvements. An ideal SWS would simplify task and dependency management through intuitive abstractions, provide native support for distributed execution with minimal configuration, seamlessly integrate containerization tools, deliver actionable and user-friendly error diagnostics, and offer built-in support for workflow modularity, monitoring, and extensibility. Focusing on these areas can significantly reduce user frustration and improve the adoption and usability of workflow systems across diverse scientific communities.

\begin{table}[htbp]
\centering
\caption{Documentation Best Practices Informed by Developer Challenges}
\label{tab:documentation_guidance}
\begin{tabularx}{\linewidth}{>{\hsize=0.6\hsize}X|>{\hsize=1.4\hsize}X}
\toprule
\textbf{Observed Challenge} & \textbf{Recommended Documentation Guidance} \\
\midrule
Task Scheduling Confusion & Provide detailed examples of workflow graphs, dependency patterns, and troubleshooting tips. \\
\hline
Distributed Execution Difficulties & Offer step-by-step setup guides and resource configuration templates for common environments. \\
\hline
Containerization Failures & Clearly document container usage, dependency handling, and troubleshooting strategies. \\
\hline
Troubleshooting Obscure Errors & Maintain an error message glossary with explanations and suggested fixes. \\
\hline
Data Handling Complexity & Provide best practices for file organization, staging, and naming conventions. \\
\hline
Onboarding New Users & Include beginner tutorials, quick-start guides, and annotated example workflows. \\
\hline
Extending and Customizing Workflows & Document extension points and APIs with modular, reusable examples. \\
\hline
Lack of Runtime Visibility & Explain available logging and visualization tools to support real-time workflow monitoring. \\
\hline
Scalability and Optimization Challenges & Provide profiling guidance and tuning strategies for large-scale execution. \\
\bottomrule
\end{tabularx}
\end{table}

Second, high-quality documentation plays a critical role in addressing many of the recurring challenges SWS users face. Table~\ref{tab:documentation_guidance} provides evidence-based recommendations for improving SWS documentation, directly informed by the issues frequently discussed on Stack Overflow and GitHub. Developers should ensure that documentation goes beyond static feature descriptions to address common error scenarios proactively, onboarding barriers, environment setup complexities, and workflow optimization strategies. Clear, example-driven, and error-aware documentation can significantly reduce support overhead and enhance user success.

Finally, SWS development teams can directly apply our methodology to continuously monitor and understand user challenges through community data mining. By systematically mining Stack Overflow, GitHub issues, and other community platforms, teams can extract relevant posts, preprocess the data following our approach, and apply topic modeling (e.g., BERTopic) to uncover dominant themes and emerging concerns. Frequency and trend analysis can reveal shifts in user difficulties over time, helping teams prioritize feature development, improve documentation, and better support their user communities. This lightweight and scalable approach provides an evidence-based mechanism for capturing real-world feedback, complementing traditional channels such as formal bug reports and direct user surveys. Regular community mining can serve as an early warning system for usability issues, evolving needs, and documentation gaps, ultimately informing more user-centered and sustainable SWS development.
\newline
\textbf{Implications for Researchers: }To facilitate future research in the domain of SWSs, we have shared our dataset along with a comprehensive replication package. While this paper presents a focused analysis of Stack Overflow and GitHub data, we envision that our dataset, particularly when integrated with complementary sources such as official community forums, Slack channels, mailing lists, and workflow-sharing platforms, can serve as a valuable foundation for a wide range of research endeavors. In particular, researchers in empirical software engineering, developer experience, and workflow automation may benefit from leveraging this resource to explore emerging trends, identify persistent challenges, and design targeted interventions. We, therefore, outline several promising directions for future work that can be pursued using our dataset in isolation or in conjunction with additional sources such as GitHub repositories, SWS documentation, community discussions, and system-level telemetry data. Some examples are listed below.

\begin{itemize}
    \item \emph{How do error types and troubleshooting strategies vary between graphical and script-based SWS platforms?} The motivation behind this question is to identify differences in usability, debugging complexity, and support mechanisms between graphical and script-based scientific workflow systems (SWS), with the goal of improving user support, tool development, and training resources tailored to each interface type.
    \newline
    \textbf{Approach: }To investigate how error types and troubleshooting strategies vary between graphical and script-based SWSs, researchers can select representative SWSs from both categories, for example, KNIME, Galaxy, and Taverna for graphical systems and Snakemake, Nextflow, and Apache Airflow for script-based ones. Then, they can collect data from GitHub and Stack Overflow following our approaches. After cleaning and preprocessing the textual data, researchers can categorize issues/posts into common error types such as syntax errors, dependency issues, runtime failures, and UI-related problems using manual coding and topic modeling techniques like BERTopic. They can also identify the troubleshooting strategies employed, such as community-suggested fixes, code snippets, documentation references, and GUI-based vs. command-line steps. A comparative analysis can be conducted to examine the distribution and frequency of error types, time to resolution, and the nature of support across both interface types. Finally, findings can be validated through manual review and summarized using visualizations and case-based insights to highlight key differences in user challenges and support needs between graphical and script-based SWS platforms.
    
    \item \textit{What are the most frequent problems SWS developers encounter when integrating container technologies (e.g., Docker, Singularity) into workflows?} Containerization plays a vital role in ensuring the reproducibility, portability, and scalability of scientific workflows. Investigating the challenges developers face when integrating container technologies such as Docker and Singularity can provide valuable insights into common pain points and technical barriers. This understanding can guide the development of more effective tools, enhance documentation, and shape best practices within the community. Moreover, it can help identify gaps in support, compatibility issues, and configuration complexities that may hinder seamless workflow development and deployment across diverse computational environments.
    \newline
    \textbf{Approach: }To identify the most frequent problems SWS developers encounter when integrating container technologies such as Docker and Singularity, researchers need to select a range of SWSs that actively support containerization, such as Nextflow, Snakemake, Galaxy, CWL, and Airflow. They can use our data or extract data following our approach from GitHub and Stack Overflow posts tagged with relevant SWS and container-related keywords (e.g., \emph{docker, singularity, container, image, runtime}). After cleaning and preprocessing the text data, they need to filter for posts specifically addressing container integration challenges. Using thematic analysis or topic modeling techniques such as the BERTopic we utilized in our work, researchers can categorize the issues into recurring themes such as image compatibility, dependency conflicts, volume mounting problems, and runtime configuration errors. They can analyze the frequency and impact of each problem type, considering metrics like resolution time and community engagement like we did in our paper. To ensure accuracy, a subset of categorized posts can be manually validated, and insights may be cross-checked with input from SWS developers. The findings can be used for practical insights to guide improvements in container support, tooling, and documentation within the SWS ecosystem.

    \item \textit{How do the types, complexity, and resolution patterns of developer challenges in SWS differ across support platforms such as Stack Overflow, GitHub, community forums, Slack channels, and workflow-sharing repositories?} SWSs are supported by a diverse ecosystem of communication and collaboration platforms, including public Q\&A sites (e.g., Stack Overflow), issue trackers (e.g., GitHub), community-maintained forums, real-time chat environments (e.g., Slack), and workflow-sharing repositories (e.g., WorkflowHub). Each of these platforms attracts different user groups and fosters distinct support-seeking behaviors shaped by their communication norms, technical affordances, and moderation policies. However, little is known about how developer challenges differ in type, complexity, and resolution across these platforms. Understanding these differences is essential for building a comprehensive view of developer pain points and for designing tailored support mechanisms that reflect the context-specific needs of SWS users. By comparing these platforms, researchers and tool developers can identify persistent usability gaps, uncover platform-specific blind spots, and inform the development of more effective documentation, learning resources, and automated support tools.
    \newline
    \textbf{Approach: }To investigate how the types, complexity, and resolution patterns of developer challenges in SWSs vary across support platforms, researchers can collect data from multiple sources, including Stack Overflow, GitHub repositories, official community forums (e.g., Galaxy Help, KNIME Hub), Slack or Gitter archives (where available), and workflow-sharing platforms such as WorkflowHub and nf-core. For each selected SWS, relevant posts, issues, and discussions can be extracted using platform-specific APIs or publicly available archives, along with associated metadata such as timestamps, user interactions, and resolution status. To ensure consistency across sources, all textual content should be preprocessed using a unified pipeline, removing noise, stopwords, and irrelevant markup, and applying lemmatization, following the preprocessing steps outlined in our study. Topic modeling techniques such as BERTopic can then be applied to extract themes from each platform and organize them under a shared taxonomy of developer challenges. The complexity of issues can be assessed using proxy indicators such as post length, number of replies, and technical specificity, while resolution patterns can be evaluated through metrics like response time, resolution status, and community engagement levels. A cross-platform comparative analysis can reveal thematic divergences, behavioral differences, and platform-specific support dynamics. Finally, these quantitative findings can be enriched with qualitative case comparisons to uncover persistent pain points and context-sensitive insights.
    \item \textit{Can we identify frequently asked questions or unresolved GitHub issues that highlight shortcomings in the official documentation or SWS onboarding processes?} Effective documentation and smooth onboarding are necessary for lowering the entry barrier to SWSs, especially given their inherent complexity and the diversity of their user base. However, developers and end-users often encounter unclear instructions, missing usage examples, or inconsistent behavior between documented and actual system behavior. Community-driven platforms such as Stack Overflow and GitHub provide a valuable lens through which these pain points surface organically. Frequently asked questions, recurring misconceptions and unresolved issues can serve as proxies for gaps in official documentation or friction in onboarding workflows. By systematically identifying and analyzing these patterns, researchers and tool maintainers can prioritize improvements in educational resources, user guides, and tool messaging. This, in turn, can enhance usability, reduce support overhead, and accelerate the adoption of SWSs by both novice and experienced users.
    \newline
    \textbf{Approach: }To address this question, researchers can adopt a methodology that integrates topic modeling, frequency analysis, and qualitative comparison with official documentation. Relevant posts from Stack Overflow and issues from GitHub repositories associated with selected SWSs (e.g., Galaxy, Nextflow, Snakemake, Airflow) can be collected or reused from our curated dataset, focusing specifically on frequently asked questions and unresolved discussions. Following our preprocessing pipeline, researchers can clean the textual data by removing noise, stopwords, and code snippets, and then apply lemmatization for consistency. Using BERTopic, dominant themes can be extracted from unresolved GitHub issues and high-frequency Stack Overflow posts. Particular attention should be given to recurring themes related to installation, configuration, initial setup, and tool integration, common areas of concern during onboarding. These extracted themes can then be systematically cross-referenced against official documentation to identify mismatches, missing examples, or unclear instructional content. Finally, a qualitative analysis of representative examples can help pinpoint concrete documentation gaps and usability pain points. This approach facilitates the identification of shortcomings in onboarding resources and offers actionable insights to enhance the documentation and learning experience for SWS users.
\end{itemize}

\begin{table}[htbp]
\centering
\caption{Additional Research Questions for Future Work}
\label{tab:future_rq}
\begin{tabularx}{\textwidth}{X|X}
\toprule
\textbf{Research Question} & \textbf{Relevance} \\
\midrule
Are certain types of questions (e.g., onboarding vs. optimization) more prevalent in certain SWS domains? & Useful for tailoring educational or support materials by domain. \\
\hline
Can unresolved issues and FAQs be automatically mapped to outdated or missing documentation? & Supports semi-automated documentation updating. \\
\hline
How do discussion patterns and resolution effectiveness vary across platforms (SO, GitHub, forums)? & Reveals platform-specific strengths in developer engagement. \\
\hline
Can user intent or expertise be inferred from question structure and used to personalize support? & Useful for recommender systems and intelligent agents in SWSs. \\
\bottomrule
\end{tabularx}
\end{table}

Our study provides not only key insights into the challenges developers face in SWSs but also a reusable dataset for broader empirical investigations. To support future research, we shared over 175,000 curated entries from Stack Overflow and GitHub, covering issues, pull requests, and developer Q\&A posts. This dataset enables various lines of inquiry that are not covered in this paper. For instance, researchers can explore how certain issues dominate specific domains (e.g., onboarding in bioinformatics vs. optimization in climate modeling) or investigate how unresolved posts and FAQs align with documentation gaps. Table \ref{tab:future_rq} outlines several such research questions and their potential impact on the community:

\begin{table}[htbp]
    \centering
    \caption{Evidence-Based Checklist for Teaching SWSs}
    \label{tab:sws_educator_checklist}
     \begin{tabularx}{\linewidth}{>{\hsize=0.7\hsize}X|>{\hsize=1.3\hsize}X}
        \toprule
        \textbf{Topic} & \textbf{Learning Objective} \\
        \midrule
        Workflow Creation and Scheduling & Teach students how to design, define, and schedule workflows using Directed Acyclic Graphs (DAGs) and manage task dependencies. \\
        \hline
        Distributed Task Management & Enable students to orchestrate distributed tasks, configure worker-queue architectures, and ensure fault tolerance across systems. \\
        \hline
        Data Structures and Operations & Develop students' ability to manipulate large datasets and dataframes efficiently, ensuring scalable and memory-optimized workflows. \\
        \hline
        Workflow Execution and Monitoring & Train students to execute workflows reliably, monitor job statuses, handle failures, and debug execution errors. \\
        \hline
        Containerization and Environment Management & Educate students on building, deploying, and troubleshooting containerized workflows to ensure reproducibility across different environments. \\
        \hline
        Rule and Dependency Management & Instruct students on defining input/output rules, using wildcards, and managing complex task dependencies within workflows. \\
        \hline
        Error Diagnosis and Troubleshooting & Equip students with systematic debugging strategies to identify and resolve common workflow execution and configuration issues. \\
        \hline
        Resource Configuration and Scalability & Prepare students to allocate resources efficiently and design workflows that can scale across high-performance or cloud-based environments. \\
        \hline
        Workflow Sharing and Reuse & Familiarize students with workflow repositories (e.g., WorkflowHub, nf-core) to promote reuse, reproducibility, and community standards. \\
        \hline
        Best Practices for Documentation & Encourage students to produce clear, comprehensive workflow documentation to enhance usability, reproducibility, and collaboration. \\
        \bottomrule
    \end{tabularx}
\end{table}

\textbf{Implications for Educators: }To support educators in preparing students to work effectively with SWSs, we provide an evidence-based checklist in Table \ref{tab:sws_educator_checklist}, derived from our empirical findings. This checklist highlights the common challenges faced by SWS developers and serves as a guide for structuring lesson plans around essential competencies. It outlines key topics that should be addressed along with the corresponding learning objectives students are expected to achieve. By integrating these elements into their teaching, educators can better equip students to manage the complexities of SWSs and succeed in real-world workflow development.

Educators can utilize this checklist to evaluate and refine their lesson plans, ensuring that course content aligns with the practical challenges developers face in real-world SWS development. By emphasizing these critical areas, instructors can strengthen students' technical preparedness, enhance their problem-solving skills, and promote best practices in the design and maintenance of scientific workflows.

Practitioners, developers, educators, and researchers can consider many aspects when deciding where to focus their efforts. Nevertheless, our study underscores the critical challenges and opportunities in developing and using SWSs. By addressing the identified issues and leveraging the insights from our analysis, stakeholders can enhance the effectiveness and impact of SWSs, ultimately advancing scientific research and innovation.

\subsection{Lesson Learned}

Throughout this study, we encountered several methodological and practical insights that can inform future research on mining developer discussions related to SWSs.
\begin{itemize}
    \item \textbf{Importance of Human-in-the-Loop Filtering: }Relying solely on automated filtering based on tags proved inadequate, as several SWSs share names with unrelated technologies or common terms (e.g., "Galaxy"). This highlighted the necessity of rigorous manual validation to ensure dataset accuracy. Human judgment played a crucial role in distinguishing truly relevant posts, underscoring the importance of domain expertise in empirical software engineering research.
    \item \textbf{Fragmentation of the SWS Ecosystem: }The heterogeneity of SWSs, ranging from GUI-based platforms like Galaxy to script-driven tools such as Nextflow and Snakemake, introduced significant challenges in applying consistent preprocessing and interpretation techniques. Each system uses distinct terminology, conventions, and usage patterns, which limited the overlap of topics and reduced the generalizability of findings across platforms. To address this, future studies could benefit from grouping SWSs based on interface style (e.g., graphical vs. DSL-based) or domain focus prior to analysis, enabling more targeted and meaningful comparisons.
    \item \textbf{Topic Modeling for Technical Corpora: }Though BERTopic performed well in uncovering latent topics, we observed that many topics still required semantic refinement and contextual interpretation. In highly technical domains like SWSs, relying solely on unsupervised models can yield ambiguous or overlapping clusters. Thus, we recommend combining topic modeling with iterative, expert-guided interpretation to ensure clarity and practical relevance.
    \item \textbf{Collaborative Review Strengthens Rigor: }The multi-author review and consensus-building approach not only ensured reliability of classification and topic interpretation but also helped us identify nuanced themes we may have overlooked individually.
    \item \textbf{Dataset Reusability and Reproducibility: } Maintaining a reproducible workflow for data extraction, cleaning, and topic modeling allowed us to easily re-run and extend our analyses. By sharing our data processing scripts and outputs, we hope to support future longitudinal or comparative studies in this domain.
    \item \textbf{Need for Actionable Synthesis: }Finally, we learned that simply presenting topic distributions or difficulty metrics is not sufficient. The real value lies in translating these findings into actionable guidance for researchers, educators, and practitioners. This realization shaped our extended implications and recommendations, aiming to make our work more impactful and useful to the SWS community.
\end{itemize}

\section{Threats to Validity}\label{threats-to-validity}
Threats to validity refer to factors or influences that can compromise the accuracy, reliability, or generalizability of study findings. Validity of research is concerned with the question of how the conclusions might be wrong, i.e., the relationship between conclusions and reality \cite{bean2007qualitative}. These threats can introduce errors, biases, or limitations that affect the validity and credibility of research results. Identifying and addressing these threats is essential to ensure the robustness and trustworthiness of research findings. In our study, we acknowledge the presence of the following potential threats.

\textbf{External Validity:} refers to the extent to which research findings can be generalized to broader populations or situations \cite{wohlin2012experimentation}. In this work, we analyzed Stack Overflow and GitHub repositories to understand the challenges of SWSs development. First, we used SO data to identify the challenges faced by developers of SWSs. However, this data might not encompass all the difficulties experienced by developers, as there could be subtle aspects rarely discussed. To address this limitation, we also examined prevalent issues and pull requests using GitHub, identifying common topics discussed there. Developers may also communicate their discussions through other media (e.g., mailing lists and SWSs forums). Developers working in proprietary environments, private organizations, or internal systems may encounter additional challenges that are not reflected in public discussions. Future work considering other data sources may complement our study. Furthermore, another issue arises from the limited selection of only 11 SWSs based on their high prevalence on Stack Overflow. This selection may not represent the broader population of SWSs, as it focuses only on the most discussed systems, potentially overlooking less popular or emerging SWSs that could exhibit different characteristics or trends. The findings derived from these 11 systems may not generalize to all SWSs, especially those that are less documented or used by smaller, specialized communities. Consequently, the conclusions drawn might reflect biases inherent in the selection process, limiting the applicability of the results to the broader landscape of scientific workflow systems.

Additionally, although our study covers SWSs used in multiple domains, including bioinformatics, genomics, cheminformatics, climate science, epidemiology, and machine learning, our findings are generalized across all systems. Specific domains may introduce unique challenges that are not fully captured in this study. For example, climate modeling workflows often require large-scale, distributed computing, whereas bioinformatics workflows emphasize data reproducibility and integration with specialized tools. A deeper domain-specific analysis would further strengthen the applicability of our findings.

\textbf{Internal Validity:} pertains to the causal relationship between a treatment and its outcome \cite{wohlin2012experimentation}. In this study, we utilized topic models (BERTopic) to group related SO posts, and GitHub issues and pull requests, assuming that similar textual content would be present within the same clusters. However, alternative methods could result in different clusters of posts and issue reports. To ensure the quality of the clusters, we manually reviewed the generated topics, merging similar ones if necessary and assigning meaningful labels to each cluster. Another potential concern is selecting the optimal number of topics. In BERTopic, though there is a parameter nr\_topics to control the number of topics, we controlled the number of topics through a cluster model. We used the HBDSCAN cluster model to perform this task. We improved the default representation of data using the Countvectorizer. It helps to ignore infrequent words and increase the n-gram range.

The subjectivity involved in labeling posts types poses a potential threat to the validity of our results. To mitigate this, we conducted three independent classifications and assessed the interrater agreement using the Cohen-Kappa test, which demonstrated a strong level of consistency among the annotators.

\textbf{Construct Validity:} involves the relationship between theory and observation \cite{wohlin2012experimentation}. A potential concern is the accuracy of labeling automatically generated topics, as the assigned names may not accurately capture the essence of the related posts, issues, and pull requests. To mitigate this threat, three authors independently reviewed the keywords and examined over 30 randomly selected posts/issues/pull requests for each topic. We then engaged in discussions to reach a consensus on a label that best represents the content of each topic.

One potential threat to construct validity arises from the inclusion of Apache Airflow in our analysis. Although Airflow is a widely recognized workflow orchestration tool, it is also used in industrial and business-oriented contexts. Consequently, some posts related to Airflow in our dataset may reflect challenges associated with general-purpose workflow automation rather than scientific workflows. While we believe that Airflow remains relevant given the increasing overlap between scientific data processing practices and industrial data engineering, we acknowledge that this broader usage context may introduce minor noise into our characterization of scientific workflow development challenges.
\newline
Our study relies on publicly available discussions from Stack Overflow and GitHub, which do not explicitly indicate the research domain or expertise levels of users. As a result, while we analyzed widely discussed challenges across different Scientific Workflow Systems (SWSs), we may not capture domain-specific nuances or variations in challenges based on user proficiency. Future studies could integrate metadata analysis or user profiling to better distinguish challenges faced by specific domains and varying levels of expert users.

Additionally, we evaluated the difficulty of SWSs topics by considering the percentage of unresolved posts, issues, and pull requests, as well as the median time to resolution. These metrics, while useful, may pose a threat to construct validity. Other potential indicators, such as the median time to the first answer, the number of answers per post (for Stack Overflow), and the number of comments, issue labels, review comments, and review cycles (for GitHub), could also be explored. However, the metrics we used are consistent with those applied in similar studies \citep{abdellatif2020challenges, ahmed2018concurrency, bajaj2014mining, nadi2016jumping, rosen2016mobile, yang2016security}.

\section{Conclusion \& Future Work}\label{conclusion}
In this study, we conducted a comprehensive empirical analysis to uncover the challenges faced by developers of SWSs. By mining and modeling over 35,000 Stack Overflow posts and 162,000 GitHub issues and pull requests, we identified recurring topics of concern, assessed their difficulty, and analyzed the nature of developer questions across platforms. Our investigation revealed 10 dominant topics on Stack Overflow and 13 on GitHub, with challenges in workflow creation, execution, and distributed task management being especially prevalent. Fine-grained topic decomposition further exposed critical pain points, such as DAG configuration issues, Kubernetes-based deployment failures, and cross-task data handling inconsistencies. We found that 'How'-type questions dominate developer queries, underscoring a widespread need for procedural guidance. Difficulty analysis indicated that topics involving large-scale orchestration, infrastructure integration, and API migrations are particularly hard to resolve, often with prolonged resolution times or high unresolved rates.

These findings carry important implications for tool designers, documentation authors, and the broader SWS research community. They suggest the need for more robust abstractions, improved deployment support, and clearer documentation practices to reduce friction in SWS development. Our methodology also provides a replicable, lightweight approach for continuously monitoring developer pain points using public data, offering actionable insights to improve usability, reliability, and maintainability in the evolution of SWSs.

Future research will explore the use of prompt-based large language models (LLMs) to assist in automated workflow creation, debugging, and documentation, potentially reducing the technical burden on developers and end-users. To broaden our understanding of real-world challenges in SWS development and usage, we also plan to expand our analysis beyond Stack Overflow and GitHub by mining discussions from platforms such as WorkflowHub, Slack communities, and other relevant developer forums.
In parallel, we aim to conduct interviews with researchers, workflow system developers, and research software engineers to identify critical pain points and design requirements. To complement these qualitative insights, we will deploy surveys targeting both novice and experienced SWS practitioners to gather quantitative data on usability barriers, perceived difficulties, and satisfaction with existing solutions. Furthermore, we will perform in-depth case studies across scientific domains such as bioinformatics, geosciences, and machine learning, examining workflow repositories and community discussions to surface domain-specific concerns.

\section{Declarations}
\subsection*{Funding }This research is supported in part by the Natural Sciences and Engineering Research Council of Canada (NSERC), and by the industry-stream NSERC CREATE in Software Analytics Research (SOAR)

\subsection*{Ethical approval } Not Applicable

\subsection*{Informed consent }Not Applicable

\subsection*{Author Contributions}
\begin{itemize}
    \item \textbf{Khairul Alam} led the conceptualization and design of the study, performed the data collection and preprocessing, conducted the topic modeling and empirical analysis, and drafted the manuscript.
    \item \textbf{Banani Roy }contributed to the study design, provided methodological guidance, reviewed and refined the data collection and topic analysis procedures, and critically revised the manuscript.
    \item \textbf{Chanchal K. Roy} supervised the overall research process, provided strategic direction for the study, contributed to the interpretation of results, refined topic analysis procedures, and helped with manuscript revision and finalization.
    \item \textbf{Kartik Mittal} assisted in data acquisition and analysis, contributed to reviewing and validating results, and participated in reviewing.
\end{itemize}

\subsection*{Data Availability Statements}
Our replication package can be found in our online appendix \citep{replicationpackage}.

\subsection*{Conflict of Interest}
We have no conflict of interest.
\subsection*{Clinical Trial Number in the manuscript }Not Applicable
\subsection*{Acknowledgments}
This research is supported in part by the Natural Sciences and Engineering Research Council of Canada (NSERC), and by the industry-stream NSERC CREATE in Software Analytics Research (SOAR).
\bibliography{sn-bibliography}


\begin{thebibliography}{164}
\ifx \bisbn   \undefined \def \bisbn  #1{ISBN #1}\fi
\ifx \binits  \undefined \def \binits#1{#1}\fi
\ifx \bauthor  \undefined \def \bauthor#1{#1}\fi
\ifx \batitle  \undefined \def \batitle#1{#1}\fi
\ifx \bjtitle  \undefined \def \bjtitle#1{#1}\fi
\ifx \bvolume  \undefined \def \bvolume#1{\textbf{#1}}\fi
\ifx \byear  \undefined \def \byear#1{#1}\fi
\ifx \bissue  \undefined \def \bissue#1{#1}\fi
\ifx \bfpage  \undefined \def \bfpage#1{#1}\fi
\ifx \blpage  \undefined \def \blpage #1{#1}\fi
\ifx \burl  \undefined \def \burl#1{\textsf{#1}}\fi
\ifx \doiurl  \undefined \def \doiurl#1{\url{https://doi.org/#1}}\fi
\ifx \betal  \undefined \def \betal{\textit{et al.}}\fi
\ifx \binstitute  \undefined \def \binstitute#1{#1}\fi
\ifx \binstitutionaled  \undefined \def \binstitutionaled#1{#1}\fi
\ifx \bctitle  \undefined \def \bctitle#1{#1}\fi
\ifx \beditor  \undefined \def \beditor#1{#1}\fi
\ifx \bpublisher  \undefined \def \bpublisher#1{#1}\fi
\ifx \bbtitle  \undefined \def \bbtitle#1{#1}\fi
\ifx \bedition  \undefined \def \bedition#1{#1}\fi
\ifx \bseriesno  \undefined \def \bseriesno#1{#1}\fi
\ifx \blocation  \undefined \def \blocation#1{#1}\fi
\ifx \bsertitle  \undefined \def \bsertitle#1{#1}\fi
\ifx \bsnm \undefined \def \bsnm#1{#1}\fi
\ifx \bsuffix \undefined \def \bsuffix#1{#1}\fi
\ifx \bparticle \undefined \def \bparticle#1{#1}\fi
\ifx \barticle \undefined \def \barticle#1{#1}\fi
\bibcommenthead
\ifx \bconfdate \undefined \def \bconfdate #1{#1}\fi
\ifx \botherref \undefined \def \botherref #1{#1}\fi
\ifx \url \undefined \def \url#1{\textsf{#1}}\fi
\ifx \bchapter \undefined \def \bchapter#1{#1}\fi
\ifx \bbook \undefined \def \bbook#1{#1}\fi
\ifx \bcomment \undefined \def \bcomment#1{#1}\fi
\ifx \oauthor \undefined \def \oauthor#1{#1}\fi
\ifx \citeauthoryear \undefined \def \citeauthoryear#1{#1}\fi
\ifx \endbibitem  \undefined \def \endbibitem {}\fi
\ifx \bconflocation  \undefined \def \bconflocation#1{#1}\fi
\ifx \arxivurl  \undefined \def \arxivurl#1{\textsf{#1}}\fi
\csname PreBibitemsHook\endcsname

\bibitem[\protect\citeauthoryear{Alam et~al.}{2023}]{10479414}
\begin{bchapter}
\bauthor{\bsnm{Alam}, \binits{K.}},
\bauthor{\bsnm{Roy}, \binits{B.}},
\bauthor{\bsnm{Serebrenik}, \binits{A.}}:
\bctitle{Reusability challenges of scientific workflows: A case study for galaxy}.
In: \bbtitle{2023 30th Asia-Pacific Software Engineering Conference (APSEC)},
pp. \bfpage{289}--\blpage{298}
(\byear{2023}).
\doiurl{10.1109/APSEC60848.2023.00039}
\end{bchapter}
\endbibitem

\bibitem[\protect\citeauthoryear{Liu et~al.}{2015}]{liu2015survey}
\begin{barticle}
\bauthor{\bsnm{Liu}, \binits{J.}},
\bauthor{\bsnm{Pacitti}, \binits{E.}},
\bauthor{\bsnm{Valduriez}, \binits{P.}},
\bauthor{\bsnm{Mattoso}, \binits{M.}}:
\batitle{A survey of data-intensive scientific workflow management}.
\bjtitle{Journal of Grid Computing}
\bvolume{13},
\bfpage{457}--\blpage{493}
(\byear{2015})
\end{barticle}
\endbibitem

\bibitem[\protect\citeauthoryear{Almeida et~al.}{2018}]{almeida2018modular}
\begin{bchapter}
\bauthor{\bsnm{Almeida}, \binits{J.R.}},
\bauthor{\bsnm{Ribeiro}, \binits{R.F.}},
\bauthor{\bsnm{Oliveira}, \binits{J.L.}}:
\bctitle{A modular workflow management framework.}
In: \bbtitle{HEALTHINF},
pp. \bfpage{414}--\blpage{421}
(\byear{2018})
\end{bchapter}
\endbibitem

\bibitem[\protect\citeauthoryear{Olabarriaga et~al.}{2014}]{olabarriaga2014scientific}
\begin{bchapter}
\bauthor{\bsnm{Olabarriaga}, \binits{S.}},
\bauthor{\bsnm{Pierantoni}, \binits{G.}},
\bauthor{\bsnm{Taffoni}, \binits{G.}},
\bauthor{\bsnm{Sciacca}, \binits{E.}},
\bauthor{\bsnm{Jaghoori}, \binits{M.}},
\bauthor{\bsnm{Korkhov}, \binits{V.}},
\bauthor{\bsnm{Castelli}, \binits{G.}},
\bauthor{\bsnm{Vuerli}, \binits{C.}},
\bauthor{\bsnm{Becciani}, \binits{U.}},
\bauthor{\bsnm{Carley}, \binits{E.}}, \betal:
\bctitle{Scientific workflow management--for whom?}
In: \bbtitle{2014 IEEE 10th International Conference on e-Science},
vol. \bseriesno{1},
pp. \bfpage{298}--\blpage{305}
(\byear{2014}).
\bcomment{IEEE}
\end{bchapter}
\endbibitem

\bibitem[\protect\citeauthoryear{Giardine et~al.}{2005}]{giardine2005galaxy}
\begin{barticle}
\bauthor{\bsnm{Giardine}, \binits{B.}},
\bauthor{\bsnm{Riemer}, \binits{C.}},
\bauthor{\bsnm{Hardison}, \binits{R.C.}},
\bauthor{\bsnm{Burhans}, \binits{R.}},
\bauthor{\bsnm{Elnitski}, \binits{L.}},
\bauthor{\bsnm{Shah}, \binits{P.}},
\bauthor{\bsnm{Zhang}, \binits{Y.}},
\bauthor{\bsnm{Blankenberg}, \binits{D.}},
\bauthor{\bsnm{Albert}, \binits{I.}},
\bauthor{\bsnm{Taylor}, \binits{J.}}, \betal:
\batitle{Galaxy: a platform for interactive large-scale genome analysis}.
\bjtitle{Genome research}
\bvolume{15}(\bissue{10}),
\bfpage{1451}--\blpage{1455}
(\byear{2005})
\end{barticle}
\endbibitem

\bibitem[\protect\citeauthoryear{Berthold et~al.}{2009}]{berthold2009knime}
\begin{barticle}
\bauthor{\bsnm{Berthold}, \binits{M.R.}},
\bauthor{\bsnm{Cebron}, \binits{N.}},
\bauthor{\bsnm{Dill}, \binits{F.}},
\bauthor{\bsnm{Gabriel}, \binits{T.R.}},
\bauthor{\bsnm{K{\"o}tter}, \binits{T.}},
\bauthor{\bsnm{Meinl}, \binits{T.}},
\bauthor{\bsnm{Ohl}, \binits{P.}},
\bauthor{\bsnm{Thiel}, \binits{K.}},
\bauthor{\bsnm{Wiswedel}, \binits{B.}}:
\batitle{Knime-the konstanz information miner: version 2.0 and beyond}.
\bjtitle{AcM SIGKDD explorations Newsletter}
\bvolume{11}(\bissue{1}),
\bfpage{26}--\blpage{31}
(\byear{2009})
\end{barticle}
\endbibitem

\bibitem[\protect\citeauthoryear{K{\"o}ster and Rahmann}{2012}]{koster2012snakemake}
\begin{barticle}
\bauthor{\bsnm{K{\"o}ster}, \binits{J.}},
\bauthor{\bsnm{Rahmann}, \binits{S.}}:
\batitle{Snakemake—a scalable bioinformatics workflow engine}.
\bjtitle{Bioinformatics}
\bvolume{28}(\bissue{19}),
\bfpage{2520}--\blpage{2522}
(\byear{2012})
\end{barticle}
\endbibitem

\bibitem[\protect\citeauthoryear{Di~Tommaso et~al.}{2017}]{di2017nextflow}
\begin{barticle}
\bauthor{\bsnm{Di~Tommaso}, \binits{P.}},
\bauthor{\bsnm{Chatzou}, \binits{M.}},
\bauthor{\bsnm{Floden}, \binits{E.W.}},
\bauthor{\bsnm{Barja}, \binits{P.P.}},
\bauthor{\bsnm{Palumbo}, \binits{E.}},
\bauthor{\bsnm{Notredame}, \binits{C.}}:
\batitle{Nextflow enables reproducible computational workflows}.
\bjtitle{Nature biotechnology}
\bvolume{35}(\bissue{4}),
\bfpage{316}--\blpage{319}
(\byear{2017})
\end{barticle}
\endbibitem

\bibitem[\protect\citeauthoryear{McPhillips et~al.}{2009}]{mcphillips2009scientific}
\begin{barticle}
\bauthor{\bsnm{McPhillips}, \binits{T.}},
\bauthor{\bsnm{Bowers}, \binits{S.}},
\bauthor{\bsnm{Zinn}, \binits{D.}},
\bauthor{\bsnm{Lud{\"a}scher}, \binits{B.}}:
\batitle{Scientific workflow design for mere mortals}.
\bjtitle{FGCS}
\bvolume{25},
\bfpage{541}--\blpage{551}
(\byear{2009})
\end{barticle}
\endbibitem

\bibitem[\protect\citeauthoryear{Bowers}{2012}]{bowers2012scientific}
\begin{barticle}
\bauthor{\bsnm{Bowers}, \binits{S.}}:
\batitle{Scientific workflow, provenance, and data modeling challenges and approaches}.
\bjtitle{JoDS}
\bvolume{1},
\bfpage{19}--\blpage{30}
(\byear{2012})
\end{barticle}
\endbibitem

\bibitem[\protect\citeauthoryear{Lin et~al.}{2009}]{lin2009reference}
\begin{barticle}
\bauthor{\bsnm{Lin}, \binits{C.}},
\bauthor{\bsnm{Lu}, \binits{S.}},
\bauthor{\bsnm{Fei}, \binits{X.}},
\bauthor{\bsnm{Chebotko}, \binits{A.}},
\bauthor{\bsnm{Pai}, \binits{D.}},
\bauthor{\bsnm{Lai}, \binits{Z.}},
\bauthor{\bsnm{Fotouhi}, \binits{F.}},
\bauthor{\bsnm{Hua}, \binits{J.}}:
\batitle{A reference architecture for scientific workflow management systems and the view soa solution}.
\bjtitle{IEEE TSC}
\bvolume{2},
\bfpage{79}--\blpage{92}
(\byear{2009})
\end{barticle}
\endbibitem

\bibitem[\protect\citeauthoryear{Deelman et~al.}{2019}]{deelman2019role}
\begin{barticle}
\bauthor{\bsnm{Deelman}, \binits{E.}},
\bauthor{\bsnm{Mandal}, \binits{A.}},
\bauthor{\bsnm{Jiang}, \binits{M.}},
\bauthor{\bsnm{Sakellariou}, \binits{R.}}:
\batitle{The role of machine learning in scientific workflows}.
\bjtitle{The International Journal of High Performance Computing Applications}
\bvolume{33}(\bissue{6}),
\bfpage{1128}--\blpage{1139}
(\byear{2019})
\end{barticle}
\endbibitem

\bibitem[\protect\citeauthoryear{Marozzo et~al.}{2016}]{marozzo2016workflow}
\begin{barticle}
\bauthor{\bsnm{Marozzo}, \binits{F.}},
\bauthor{\bsnm{Talia}, \binits{D.}},
\bauthor{\bsnm{Trunfio}, \binits{P.}}:
\batitle{A workflow management system for scalable data mining on clouds}.
\bjtitle{IEEE Transactions on Services Computing}
\bvolume{11}(\bissue{3}),
\bfpage{480}--\blpage{492}
(\byear{2016})
\end{barticle}
\endbibitem

\bibitem[\protect\citeauthoryear{Oinn et~al.}{2004}]{oinn2004taverna}
\begin{barticle}
\bauthor{\bsnm{Oinn}, \binits{T.}},
\bauthor{\bsnm{Addis}, \binits{M.}},
\bauthor{\bsnm{Ferris}, \binits{J.}},
\bauthor{\bsnm{Marvin}, \binits{D.}},
\bauthor{\bsnm{Senger}, \binits{M.}},
\bauthor{\bsnm{Greenwood}, \binits{M.}},
\bauthor{\bsnm{Carver}, \binits{T.}},
\bauthor{\bsnm{Glover}, \binits{K.}},
\bauthor{\bsnm{Pocock}, \binits{M.R.}},
\bauthor{\bsnm{Wipat}, \binits{A.}}, \betal:
\batitle{Taverna: a tool for the composition and enactment of bioinformatics workflows}.
\bjtitle{Bioinformatics}
\bvolume{20}(\bissue{17}),
\bfpage{3045}--\blpage{3054}
(\byear{2004})
\end{barticle}
\endbibitem

\bibitem[\protect\citeauthoryear{Fern{\'a}ndez et~al.}{2013}]{fernandez2013chemistry}
\begin{barticle}
\bauthor{\bsnm{Fern{\'a}ndez}, \binits{H.}},
\bauthor{\bsnm{Tedeschi}, \binits{C.}},
\bauthor{\bsnm{Priol}, \binits{T.}}:
\batitle{A chemistry-inspired workflow management system for decentralizing workflow execution}.
\bjtitle{IEEE Transactions on Services Computing}
\bvolume{9}(\bissue{2}),
\bfpage{213}--\blpage{226}
(\byear{2013})
\end{barticle}
\endbibitem

\bibitem[\protect\citeauthoryear{Li et~al.}{2015}]{li2015enabling}
\begin{barticle}
\bauthor{\bsnm{Li}, \binits{Z.}},
\bauthor{\bsnm{Yang}, \binits{C.}},
\bauthor{\bsnm{Jin}, \binits{B.}},
\bauthor{\bsnm{Yu}, \binits{M.}},
\bauthor{\bsnm{Liu}, \binits{K.}},
\bauthor{\bsnm{Sun}, \binits{M.}},
\bauthor{\bsnm{Zhan}, \binits{M.}}:
\batitle{Enabling big geoscience data analytics with a cloud-based, mapreduce-enabled and service-oriented workflow framework}.
\bjtitle{PloS one}
\bvolume{10}(\bissue{3}),
\bfpage{0116781}
(\byear{2015})
\end{barticle}
\endbibitem

\bibitem[\protect\citeauthoryear{Hendrix et~al.}{2016}]{hendrix2016tigres}
\begin{bchapter}
\bauthor{\bsnm{Hendrix}, \binits{V.}},
\bauthor{\bsnm{Fox}, \binits{J.}},
\bauthor{\bsnm{Ghoshal}, \binits{D.}},
\bauthor{\bsnm{Ramakrishnan}, \binits{L.}}:
\bctitle{Tigres workflow library: Supporting scientific pipelines on hpc systems}.
In: \bbtitle{2016 16th IEEE/ACM International Symposium on Cluster, Cloud and Grid Computing (CCGrid)},
pp. \bfpage{146}--\blpage{155}
(\byear{2016}).
\bcomment{IEEE}
\end{bchapter}
\endbibitem

\bibitem[\protect\citeauthoryear{Mamykina et~al.}{2011}]{mamykina2011design}
\begin{bchapter}
\bauthor{\bsnm{Mamykina}, \binits{L.}},
\bauthor{\bsnm{Manoim}, \binits{B.}},
\bauthor{\bsnm{Mittal}, \binits{M.}},
\bauthor{\bsnm{Hripcsak}, \binits{G.}},
\bauthor{\bsnm{Hartmann}, \binits{B.}}:
\bctitle{Design lessons from the fastest q\&a site in the west}.
In: \bbtitle{Proceedings of the SIGCHI Conference on Human Factors in Computing Systems},
pp. \bfpage{2857}--\blpage{2866}
(\byear{2011})
\end{bchapter}
\endbibitem

\bibitem[\protect\citeauthoryear{Dabbish et~al.}{2012}]{dabbish2012social}
\begin{bchapter}
\bauthor{\bsnm{Dabbish}, \binits{L.}},
\bauthor{\bsnm{Stuart}, \binits{C.}},
\bauthor{\bsnm{Tsay}, \binits{J.}},
\bauthor{\bsnm{Herbsleb}, \binits{J.}}:
\bctitle{Social coding in github: transparency and collaboration in an open software repository}.
In: \bbtitle{Proceedings of the ACM 2012 Conference on Computer Supported Cooperative Work},
pp. \bfpage{1277}--\blpage{1286}
(\byear{2012})
\end{bchapter}
\endbibitem

\bibitem[\protect\citeauthoryear{Mork et~al.}{2015}]{mork2015contemporary}
\begin{bchapter}
\bauthor{\bsnm{Mork}, \binits{R.}},
\bauthor{\bsnm{Martin}, \binits{P.}},
\bauthor{\bsnm{Zhao}, \binits{Z.}}:
\bctitle{Contemporary challenges for data-intensive scientific workflow management systems}.
In: \bbtitle{Proceedings of the 10th Workshop on Workflows in Support of Large-Scale Science},
pp. \bfpage{1}--\blpage{11}
(\byear{2015})
\end{bchapter}
\endbibitem

\bibitem[\protect\citeauthoryear{Rosen and Shihab}{2016}]{rosen2016mobile}
\begin{botherref}
\oauthor{\bsnm{Rosen}, \binits{C.}},
\oauthor{\bsnm{Shihab}, \binits{E.}}:
What are mobile developers asking about? a large scale study using stack overflow.
ESE
\textbf{21}
(2016)
\end{botherref}
\endbibitem

\bibitem[\protect\citeauthoryear{Abdellatif et~al.}{2020}]{abdellatif2020challenges}
\begin{bchapter}
\bauthor{\bsnm{Abdellatif}, \binits{A.}},
\bauthor{\bsnm{Costa}, \binits{D.}},
\bauthor{\bsnm{Badran}, \binits{K.}},
\bauthor{\bsnm{Abdalkareem}, \binits{R.}},
\bauthor{\bsnm{Shihab}, \binits{E.}}:
\bctitle{Challenges in chatbot development: A study of stack overflow posts}.
In: \bbtitle{MSR},
pp. \bfpage{174}--\blpage{185}
(\byear{2020})
\end{bchapter}
\endbibitem

\bibitem[\protect\citeauthoryear{Treude et~al.}{2011}]{treude2011programmers}
\begin{bchapter}
\bauthor{\bsnm{Treude}, \binits{C.}},
\bauthor{\bsnm{Barzilay}, \binits{O.}},
\bauthor{\bsnm{Storey}, \binits{M.-A.}}:
\bctitle{How do programmers ask and answer questions on the web?(nier track)}.
In: \bbtitle{Proceedings of the 33rd International Conference on Software Engineering},
pp. \bfpage{804}--\blpage{807}
(\byear{2011})
\end{bchapter}
\endbibitem

\bibitem[\protect\citeauthoryear{Bagherzadeh and Khatchadourian}{2019}]{bagherzadeh2019going}
\begin{bchapter}
\bauthor{\bsnm{Bagherzadeh}, \binits{M.}},
\bauthor{\bsnm{Khatchadourian}, \binits{R.}}:
\bctitle{Going big: a large-scale study on what big data developers ask}.
In: \bbtitle{Proceedings of the 2019 27th ACM Joint Meeting on European Software Engineering Conference and Symposium on the Foundations of Software Engineering},
pp. \bfpage{432}--\blpage{442}
(\byear{2019})
\end{bchapter}
\endbibitem

\bibitem[\protect\citeauthoryear{Yang et~al.}{2016}]{yang2016security}
\begin{barticle}
\bauthor{\bsnm{Yang}, \binits{X.-L.}},
\bauthor{\bsnm{Lo}, \binits{D.}},
\bauthor{\bsnm{Xia}, \binits{X.}},
\bauthor{\bsnm{Wan}, \binits{Z.-Y.}},
\bauthor{\bsnm{Sun}, \binits{J.-L.}}:
\batitle{What security questions do developers ask? a large-scale study of stack overflow posts}.
\bjtitle{JCST}
\bvolume{31},
\bfpage{910}--\blpage{924}
(\byear{2016})
\end{barticle}
\endbibitem

\bibitem[\protect\citeauthoryear{Li et~al.}{2021}]{li2021understanding}
\begin{bchapter}
\bauthor{\bsnm{Li}, \binits{H.}},
\bauthor{\bsnm{Khomh}, \binits{F.}},
\bauthor{\bsnm{Openja}, \binits{M.}}, \betal:
\bctitle{Understanding quantum software engineering challenges an empirical study on stack exchange forums and github issues}.
In: \bbtitle{ICSME},
pp. \bfpage{343}--\blpage{354}.
\bpublisher{IEEE}, \blocation{???}
(\byear{2021}).
\bcomment{IEEE}
\end{bchapter}
\endbibitem

\bibitem[\protect\citeauthoryear{Scoccia et~al.}{2021}]{scoccia2021challenges}
\begin{bchapter}
\bauthor{\bsnm{Scoccia}, \binits{G.L.}},
\bauthor{\bsnm{Migliarini}, \binits{P.}},
\bauthor{\bsnm{Autili}, \binits{M.}}:
\bctitle{Challenges in developing desktop web apps: a study of stack overflow and github}.
In: \bbtitle{2021 IEEE/ACM 18th International Conference on Mining Software Repositories (MSR)},
pp. \bfpage{271}--\blpage{282}
(\byear{2021}).
\bcomment{IEEE}
\end{bchapter}
\endbibitem

\bibitem[\protect\citeauthoryear{Atkinson et~al.}{2017}]{atkinson2017scientific}
\begin{botherref}
\oauthor{\bsnm{Atkinson}, \binits{M.}},
\oauthor{\bsnm{Gesing}, \binits{S.}},
\oauthor{\bsnm{Montagnat}, \binits{J.}},
\oauthor{\bsnm{Taylor}, \binits{I.}}:
Scientific workflows: Past, present and future.
Elsevier
(2017)
\end{botherref}
\endbibitem

\bibitem[\protect\citeauthoryear{M{\"o}lder et~al.}{2021}]{molder2021sustainable}
\begin{botherref}
\oauthor{\bsnm{M{\"o}lder}, \binits{F.}},
\oauthor{\bsnm{Jablonski}, \binits{K.P.}},
\oauthor{\bsnm{Letcher}, \binits{B.}},
\oauthor{\bsnm{Hall}, \binits{M.B.}},
\oauthor{\bsnm{Tomkins-Tinch}, \binits{C.H.}},
\oauthor{\bsnm{Sochat}, \binits{V.}},
\oauthor{\bsnm{Forster}, \binits{J.}},
\oauthor{\bsnm{Lee}, \binits{S.}},
\oauthor{\bsnm{Twardziok}, \binits{S.O.}},
\oauthor{\bsnm{Kanitz}, \binits{A.}}, et al.:
Sustainable data analysis with snakemake.
F1000Research
\textbf{10}
(2021)
\end{botherref}
\endbibitem

\bibitem[\protect\citeauthoryear{Kluge et~al.}{2020}]{kluge2020watchdog}
\begin{barticle}
\bauthor{\bsnm{Kluge}, \binits{M.}},
\bauthor{\bsnm{Friedl}, \binits{M.-S.}},
\bauthor{\bsnm{Menzel}, \binits{A.L.}},
\bauthor{\bsnm{Friedel}, \binits{C.C.}}:
\batitle{Watchdog 2.0: New developments for reusability, reproducibility, and workflow execution}.
\bjtitle{GigaScience}
\bvolume{9}(\bissue{6}),
\bfpage{068}
(\byear{2020})
\end{barticle}
\endbibitem

\bibitem[\protect\citeauthoryear{Cervera et~al.}{2019}]{cervera2019anduril}
\begin{barticle}
\bauthor{\bsnm{Cervera}, \binits{A.}},
\bauthor{\bsnm{Rantanen}, \binits{V.}},
\bauthor{\bsnm{Ovaska}, \binits{K.}},
\bauthor{\bsnm{Laakso}, \binits{M.}},
\bauthor{\bsnm{Nunez-Fontarnau}, \binits{J.}},
\bauthor{\bsnm{Alkodsi}, \binits{A.}},
\bauthor{\bsnm{Casado}, \binits{J.}},
\bauthor{\bsnm{Facciotto}, \binits{C.}},
\bauthor{\bsnm{H{\"a}kkinen}, \binits{A.}},
\bauthor{\bsnm{Louhimo}, \binits{R.}}, \betal:
\batitle{Anduril 2: upgraded large-scale data integration framework}.
\bjtitle{Bioinformatics}
\bvolume{35}(\bissue{19}),
\bfpage{3815}--\blpage{3817}
(\byear{2019})
\end{barticle}
\endbibitem

\bibitem[\protect\citeauthoryear{Salim et~al.}{2019}]{salim2019balsam}
\begin{botherref}
\oauthor{\bsnm{Salim}, \binits{M.A.}},
\oauthor{\bsnm{Uram}, \binits{T.D.}},
\oauthor{\bsnm{Childers}, \binits{J.T.}},
\oauthor{\bsnm{Balaprakash}, \binits{P.}},
\oauthor{\bsnm{Vishwanath}, \binits{V.}},
\oauthor{\bsnm{Papka}, \binits{M.E.}}:
Balsam: Automated scheduling and execution of dynamic, data-intensive hpc workflows.
arXiv preprint arXiv:1909.08704
(2019)
\end{botherref}
\endbibitem

\bibitem[\protect\citeauthoryear{Lampa et~al.}{2019}]{lampa2019scipipe}
\begin{barticle}
\bauthor{\bsnm{Lampa}, \binits{S.}},
\bauthor{\bsnm{Dahl{\"o}}, \binits{M.}},
\bauthor{\bsnm{Alvarsson}, \binits{J.}},
\bauthor{\bsnm{Spjuth}, \binits{O.}}:
\batitle{Scipipe: A workflow library for agile development of complex and dynamic bioinformatics pipelines}.
\bjtitle{GigaScience}
\bvolume{8}(\bissue{5}),
\bfpage{044}
(\byear{2019})
\end{barticle}
\endbibitem

\bibitem[\protect\citeauthoryear{Pal and Przytycka}{2020}]{pal2020bioinformatics}
\begin{barticle}
\bauthor{\bsnm{Pal}, \binits{S.}},
\bauthor{\bsnm{Przytycka}, \binits{T.M.}}:
\batitle{Bioinformatics pipeline using judi: just do it!}
\bjtitle{Bioinformatics}
\bvolume{36}(\bissue{8}),
\bfpage{2572}--\blpage{2574}
(\byear{2020})
\end{barticle}
\endbibitem

\bibitem[\protect\citeauthoryear{Sadedin et~al.}{2012}]{sadedin2012bpipe}
\begin{barticle}
\bauthor{\bsnm{Sadedin}, \binits{S.P.}},
\bauthor{\bsnm{Pope}, \binits{B.}},
\bauthor{\bsnm{Oshlack}, \binits{A.}}:
\batitle{Bpipe: a tool for running and managing bioinformatics pipelines}.
\bjtitle{Bioinformatics}
\bvolume{28}(\bissue{11}),
\bfpage{1525}--\blpage{1526}
(\byear{2012})
\end{barticle}
\endbibitem

\bibitem[\protect\citeauthoryear{Ewels et~al.}{2016}]{ewels2016cluster}
\begin{botherref}
\oauthor{\bsnm{Ewels}, \binits{P.}},
\oauthor{\bsnm{Krueger}, \binits{F.}},
\oauthor{\bsnm{K{\"a}ller}, \binits{M.}},
\oauthor{\bsnm{Andrews}, \binits{S.}}:
Cluster flow: A user-friendly bioinformatics workflow tool.
F1000Research
\textbf{5}
(2016)
\end{botherref}
\endbibitem

\bibitem[\protect\citeauthoryear{Cingolani et~al.}{2015}]{cingolani2015bigdatascript}
\begin{barticle}
\bauthor{\bsnm{Cingolani}, \binits{P.}},
\bauthor{\bsnm{Sladek}, \binits{R.}},
\bauthor{\bsnm{Blanchette}, \binits{M.}}:
\batitle{Bigdatascript: a scripting language for data pipelines}.
\bjtitle{Bioinformatics}
\bvolume{31}(\bissue{1}),
\bfpage{10}--\blpage{16}
(\byear{2015})
\end{barticle}
\endbibitem

\bibitem[\protect\citeauthoryear{Jimenez et~al.}{2017}]{jimenez2017popper}
\begin{bchapter}
\bauthor{\bsnm{Jimenez}, \binits{I.}},
\bauthor{\bsnm{Sevilla}, \binits{M.}},
\bauthor{\bsnm{Watkins}, \binits{N.}},
\bauthor{\bsnm{Maltzahn}, \binits{C.}},
\bauthor{\bsnm{Lofstead}, \binits{J.}},
\bauthor{\bsnm{Mohror}, \binits{K.}},
\bauthor{\bsnm{Arpaci-Dusseau}, \binits{A.}},
\bauthor{\bsnm{Arpaci-Dusseau}, \binits{R.}}:
\bctitle{The popper convention: Making reproducible systems evaluation practical}.
In: \bbtitle{2017 Ieee International Parallel and Distributed Processing Symposium Workshops (ipdpsw)},
pp. \bfpage{1561}--\blpage{1570}
(\byear{2017}).
\bcomment{IEEE}
\end{bchapter}
\endbibitem

\bibitem[\protect\citeauthoryear{Ben-Kiki et~al.}{2009}]{ben2009yaml}
\begin{botherref}
\oauthor{\bsnm{Ben-Kiki}, \binits{O.}},
\oauthor{\bsnm{Evans}, \binits{C.}},
\oauthor{\bsnm{Ingerson}, \binits{B.}}:
Yaml ain’t markup language (yaml™) version 1.1.
Working Draft 2008
\textbf{5}(11)
(2009)
\end{botherref}
\endbibitem

\bibitem[\protect\citeauthoryear{Amstutz et~al.}{2016}]{amstutz2016common}
\begin{botherref}
\oauthor{\bsnm{Amstutz}, \binits{P.}},
\oauthor{\bsnm{Crusoe}, \binits{M.R.}},
\oauthor{\bsnm{Tijani{\'c}}, \binits{N.}},
\oauthor{\bsnm{Chapman}, \binits{B.}},
\oauthor{\bsnm{Chilton}, \binits{J.}},
\oauthor{\bsnm{Heuer}, \binits{M.}},
\oauthor{\bsnm{Kartashov}, \binits{A.}},
\oauthor{\bsnm{Leehr}, \binits{D.}},
\oauthor{\bsnm{M{\'e}nager}, \binits{H.}},
\oauthor{\bsnm{Nedeljkovich}, \binits{M.}}, et al.:
Common workflow language, v1. 0
(2016)
\end{botherref}
\endbibitem

\bibitem[\protect\citeauthoryear{Voss et~al.}{}]{vossfull}
\begin{botherref}
\oauthor{\bsnm{Voss}, \binits{K.}},
\oauthor{\bsnm{Gentry}, \binits{J.}},
\oauthor{\bsnm{Auwera}, \binits{G.}}:
Full-stack genomics pipelining with GATK4+ WDL+ Cromwell. F1000Research 2017
\end{botherref}
\endbibitem

\bibitem[\protect\citeauthoryear{Vivian et~al.}{2017}]{vivian2017toil}
\begin{barticle}
\bauthor{\bsnm{Vivian}, \binits{J.}},
\bauthor{\bsnm{Rao}, \binits{A.A.}},
\bauthor{\bsnm{Nothaft}, \binits{F.A.}},
\bauthor{\bsnm{Ketchum}, \binits{C.}},
\bauthor{\bsnm{Armstrong}, \binits{J.}},
\bauthor{\bsnm{Novak}, \binits{A.}},
\bauthor{\bsnm{Pfeil}, \binits{J.}},
\bauthor{\bsnm{Narkizian}, \binits{J.}},
\bauthor{\bsnm{Deran}, \binits{A.D.}},
\bauthor{\bsnm{Musselman-Brown}, \binits{A.}}, \betal:
\batitle{Toil enables reproducible, open source, big biomedical data analyses}.
\bjtitle{Nature biotechnology}
\bvolume{35}(\bissue{4}),
\bfpage{314}--\blpage{316}
(\byear{2017})
\end{barticle}
\endbibitem

\bibitem[\protect\citeauthoryear{Santana-Perez and P{\'e}rez-Hern{\'a}ndez}{2015}]{santana2015towards}
\begin{barticle}
\bauthor{\bsnm{Santana-Perez}, \binits{I.}},
\bauthor{\bsnm{P{\'e}rez-Hern{\'a}ndez}, \binits{M.S.}}:
\batitle{Towards reproducibility in scientific workflows: An infrastructure-based approach}.
\bjtitle{Scientific Programming}
\bvolume{2015}(\bissue{1}),
\bfpage{243180}
(\byear{2015})
\end{barticle}
\endbibitem

\bibitem[\protect\citeauthoryear{Chirigati et~al.}{2016}]{chirigati2016reprozip}
\begin{bchapter}
\bauthor{\bsnm{Chirigati}, \binits{F.}},
\bauthor{\bsnm{Rampin}, \binits{R.}},
\bauthor{\bsnm{Shasha}, \binits{D.}},
\bauthor{\bsnm{Freire}, \binits{J.}}:
\bctitle{Reprozip: Computational reproducibility with ease}.
In: \bbtitle{Proceedings of the 2016 International Conference on Management of Data},
pp. \bfpage{2085}--\blpage{2088}
(\byear{2016})
\end{bchapter}
\endbibitem

\bibitem[\protect\citeauthoryear{Garijo et~al.}{2017}]{garijo2017abstract}
\begin{barticle}
\bauthor{\bsnm{Garijo}, \binits{D.}},
\bauthor{\bsnm{Gil}, \binits{Y.}},
\bauthor{\bsnm{Corcho}, \binits{O.}}:
\batitle{Abstract, link, publish, exploit: An end to end framework for workflow sharing}.
\bjtitle{Future Generation Computer Systems}
\bvolume{75},
\bfpage{271}--\blpage{283}
(\byear{2017})
\end{barticle}
\endbibitem

\bibitem[\protect\citeauthoryear{}{2022}]{galaxy2022galaxy}
\begin{botherref}
The galaxy platform for accessible, reproducible and collaborative biomedical analyses: 2022 update.
Nucleic Acids Research
\textbf{50}(W1),
345--351
(2022)
\end{botherref}
\endbibitem

\bibitem[\protect\citeauthoryear{Jagla et~al.}{2011}]{jagla2011extending}
\begin{barticle}
\bauthor{\bsnm{Jagla}, \binits{B.}},
\bauthor{\bsnm{Wiswedel}, \binits{B.}},
\bauthor{\bsnm{Copp{\'e}e}, \binits{J.-Y.}}:
\batitle{Extending knime for next-generation sequencing data analysis}.
\bjtitle{Bioinformatics}
\bvolume{27}(\bissue{20}),
\bfpage{2907}--\blpage{2909}
(\byear{2011})
\end{barticle}
\endbibitem

\bibitem[\protect\citeauthoryear{Kiran et~al.}{2023}]{kiran2023criteria}
\begin{barticle}
\bauthor{\bsnm{Kiran}, \binits{A.D.}},
\bauthor{\bsnm{Ay}, \binits{M.C.}},
\bauthor{\bsnm{Allmer}, \binits{J.}}:
\batitle{Criteria for the evaluation of workflow management systems for scientific data analysis}.
\bjtitle{Journal of Bioinformatics and Systems Biology}
\bvolume{6},
\bfpage{121}--\blpage{133}
(\byear{2023})
\end{barticle}
\endbibitem

\bibitem[\protect\citeauthoryear{Sethi and Gil}{2017}]{sethi2017scientific}
\begin{barticle}
\bauthor{\bsnm{Sethi}, \binits{R.J.}},
\bauthor{\bsnm{Gil}, \binits{Y.}}:
\batitle{Scientific workflows in data analysis: Bridging expertise across multiple domains}.
\bjtitle{Future Generation Computer Systems}
\bvolume{75},
\bfpage{256}--\blpage{270}
(\byear{2017})
\end{barticle}
\endbibitem

\bibitem[\protect\citeauthoryear{Kotliar et~al.}{2019}]{kotliar2019cwl}
\begin{barticle}
\bauthor{\bsnm{Kotliar}, \binits{M.}},
\bauthor{\bsnm{Kartashov}, \binits{A.V.}},
\bauthor{\bsnm{Barski}, \binits{A.}}:
\batitle{Cwl-airflow: a lightweight pipeline manager supporting common workflow language}.
\bjtitle{GigaScience}
\bvolume{8}(\bissue{7}),
\bfpage{084}
(\byear{2019})
\end{barticle}
\endbibitem

\bibitem[\protect\citeauthoryear{Jain et~al.}{2015}]{jain2015fireworks}
\begin{barticle}
\bauthor{\bsnm{Jain}, \binits{A.}},
\bauthor{\bsnm{Ong}, \binits{S.P.}},
\bauthor{\bsnm{Chen}, \binits{W.}},
\bauthor{\bsnm{Medasani}, \binits{B.}},
\bauthor{\bsnm{Qu}, \binits{X.}},
\bauthor{\bsnm{Kocher}, \binits{M.}},
\bauthor{\bsnm{Brafman}, \binits{M.}},
\bauthor{\bsnm{Petretto}, \binits{G.}},
\bauthor{\bsnm{Rignanese}, \binits{G.-M.}},
\bauthor{\bsnm{Hautier}, \binits{G.}}, \betal:
\batitle{Fireworks: a dynamic workflow system designed for high-throughput applications}.
\bjtitle{Concurrency and Computation: Practice and Experience}
\bvolume{27}(\bissue{17}),
\bfpage{5037}--\blpage{5059}
(\byear{2015})
\end{barticle}
\endbibitem

\bibitem[\protect\citeauthoryear{Ahmad et~al.}{2021}]{ahmad2021scientific}
\begin{barticle}
\bauthor{\bsnm{Ahmad}, \binits{Z.}},
\bauthor{\bsnm{Jehangiri}, \binits{A.I.}},
\bauthor{\bsnm{Ala'anzy}, \binits{M.A.}},
\bauthor{\bsnm{Othman}, \binits{M.}},
\bauthor{\bsnm{Latip}, \binits{R.}},
\bauthor{\bsnm{Zaman}, \binits{S.K.U.}},
\bauthor{\bsnm{Umar}, \binits{A.I.}}:
\batitle{Scientific workflows management and scheduling in cloud computing: taxonomy, prospects, and challenges}.
\bjtitle{IEEE Access}
\bvolume{9},
\bfpage{53491}--\blpage{53508}
(\byear{2021})
\end{barticle}
\endbibitem

\bibitem[\protect\citeauthoryear{Alam et~al.}{2025}]{replicationpackage}
\begin{botherref}
\oauthor{\bsnm{Alam}, \binits{K.}},
\oauthor{\bsnm{Roy}, \binits{B.}},
\oauthor{\bsnm{Roy}, \binits{C.}},
\oauthor{\bsnm{Mittal}, \binits{K.}}:
{Artifact of the paper "An Empirical Investigation on the Challenges in Scientific Workflow Systems Development."}.
Online; last accessed May, 2025
(2025).
\url{https://zenodo.org/records/15454588}
\end{botherref}
\endbibitem

\bibitem[\protect\citeauthoryear{Zhao et~al.}{2008}]{zhao2008scientific}
\begin{bchapter}
\bauthor{\bsnm{Zhao}, \binits{Y.}},
\bauthor{\bsnm{Raicu}, \binits{I.}},
\bauthor{\bsnm{Foster}, \binits{I.}}:
\bctitle{Scientific workflow systems for 21st century, new bottle or new wine?}
In: \bbtitle{2008 IEEE Congress on Services-Part I},
pp. \bfpage{467}--\blpage{471}
(\byear{2008}).
\bcomment{IEEE}
\end{bchapter}
\endbibitem

\bibitem[\protect\citeauthoryear{Zhao et~al.}{2011}]{zhao2011opportunities}
\begin{bchapter}
\bauthor{\bsnm{Zhao}, \binits{Y.}},
\bauthor{\bsnm{Fei}, \binits{X.}},
\bauthor{\bsnm{Raicu}, \binits{I.}},
\bauthor{\bsnm{Lu}, \binits{S.}}:
\bctitle{Opportunities and challenges in running scientific workflows on the cloud}.
In: \bbtitle{CyberC}
(\byear{2011}).
\bcomment{IEEE}
\end{bchapter}
\endbibitem

\bibitem[\protect\citeauthoryear{Davidson and Freire}{2008}]{davidson2008provenance}
\begin{bchapter}
\bauthor{\bsnm{Davidson}, \binits{S.B.}},
\bauthor{\bsnm{Freire}, \binits{J.}}:
\bctitle{Provenance and scientific workflows: challenges and opportunities}.
In: \bbtitle{ACM SIGMOD},
pp. \bfpage{1345}--\blpage{1350}
(\byear{2008})
\end{bchapter}
\endbibitem

\bibitem[\protect\citeauthoryear{Gil et~al.}{2007}]{gil2007examining}
\begin{barticle}
\bauthor{\bsnm{Gil}, \binits{Y.}},
\bauthor{\bsnm{Deelman}, \binits{E.}},
\bauthor{\bsnm{Ellisman}, \binits{M.}},
\bauthor{\bsnm{Fahringer}, \binits{T.}},
\bauthor{\bsnm{Fox}, \binits{G.}},
\bauthor{\bsnm{Gannon}, \binits{D.}},
\bauthor{\bsnm{Goble}, \binits{C.}},
\bauthor{\bsnm{Livny}, \binits{M.}},
\bauthor{\bsnm{Moreau}, \binits{L.}},
\bauthor{\bsnm{Myers}, \binits{J.}}:
\batitle{Examining the challenges of scientific workflows}.
\bjtitle{Computer}
\bvolume{40}(\bissue{12}),
\bfpage{24}--\blpage{32}
(\byear{2007})
\end{barticle}
\endbibitem

\bibitem[\protect\citeauthoryear{Gu et~al.}{2023}]{gu2023plan}
\begin{bchapter}
\bauthor{\bsnm{Gu}, \binits{Y.}},
\bauthor{\bsnm{Cao}, \binits{J.}},
\bauthor{\bsnm{Guo}, \binits{Y.}},
\bauthor{\bsnm{Qian}, \binits{S.}},
\bauthor{\bsnm{Guan}, \binits{W.}}:
\bctitle{Plan, generate and match: Scientific workflow recommendation with large language models}.
In: \bbtitle{International Conference on Service-Oriented Computing},
pp. \bfpage{86}--\blpage{102}
(\byear{2023}).
\bcomment{Springer}
\end{bchapter}
\endbibitem

\bibitem[\protect\citeauthoryear{S{\"a}nger et~al.}{2024}]{sanger2024qualitative}
\begin{barticle}
\bauthor{\bsnm{S{\"a}nger}, \binits{M.}},
\bauthor{\bsnm{De~Mecquenem}, \binits{N.}},
\bauthor{\bsnm{Lewi{\'n}ska}, \binits{K.E.}},
\bauthor{\bsnm{Bountris}, \binits{V.}},
\bauthor{\bsnm{Lehmann}, \binits{F.}},
\bauthor{\bsnm{Leser}, \binits{U.}},
\bauthor{\bsnm{Kosch}, \binits{T.}}:
\batitle{A qualitative assessment of using chatgpt as large language model for scientific workflow development}.
\bjtitle{GigaScience}
\bvolume{13},
\bfpage{030}
(\byear{2024})
\end{barticle}
\endbibitem

\bibitem[\protect\citeauthoryear{Procko et~al.}{2023}]{procko2023towards}
\begin{botherref}
\oauthor{\bsnm{Procko}, \binits{T.}},
\oauthor{\bsnm{Davidoff}, \binits{A.}},
\oauthor{\bsnm{Elvira}, \binits{T.}},
\oauthor{\bsnm{Ochoa}, \binits{O.}}:
Towards improved scientific knowledge proliferation: Leveraging large language models on the traditional scientific writing workflow.
Available at SSRN 4594836
(2023)
\end{botherref}
\endbibitem

\bibitem[\protect\citeauthoryear{Liew et~al.}{2016}]{liew2016scientific}
\begin{barticle}
\bauthor{\bsnm{Liew}, \binits{C.S.}},
\bauthor{\bsnm{Atkinson}, \binits{M.P.}},
\bauthor{\bsnm{Galea}, \binits{M.}},
\bauthor{\bsnm{Ang}, \binits{T.F.}},
\bauthor{\bsnm{Martin}, \binits{P.}},
\bauthor{\bsnm{Hemert}, \binits{J.I.V.}}:
\batitle{Scientific workflows: moving across paradigms}.
\bjtitle{ACM Computing Surveys (CSUR)}
\bvolume{49}(\bissue{4}),
\bfpage{1}--\blpage{39}
(\byear{2016})
\end{barticle}
\endbibitem

\bibitem[\protect\citeauthoryear{Barker and Van~Hemert}{2007}]{barker2007scientific}
\begin{bchapter}
\bauthor{\bsnm{Barker}, \binits{A.}},
\bauthor{\bsnm{Van~Hemert}, \binits{J.}}:
\bctitle{Scientific workflow: a survey and research directions}.
In: \bbtitle{PPAM},
pp. \bfpage{746}--\blpage{753}
(\byear{2007}).
\bcomment{Springer}
\end{bchapter}
\endbibitem

\bibitem[\protect\citeauthoryear{}{2024}]{workflowhub}
\begin{botherref}
{WorkflowHub Community}.
Online; last accessed January, 2024
(2024).
\url{https://workflowhub.eu/workflows}
\end{botherref}
\endbibitem

\bibitem[\protect\citeauthoryear{}{2024}]{myexperimentrepo}
\begin{botherref}
{myExperiment Workflows}.
Online; last accessed January, 2024
(2024).
\url{https://www.myexperiment.org/workflows}
\end{botherref}
\endbibitem

\bibitem[\protect\citeauthoryear{}{2024}]{snakemakerepo}
\begin{botherref}
{SnakeMake Workflows}.
Online; last accessed January, 2024
(2024).
\url{https://nf-co.re/}
\end{botherref}
\endbibitem

\bibitem[\protect\citeauthoryear{Jelodar et~al.}{2019}]{jelodar2019latent}
\begin{barticle}
\bauthor{\bsnm{Jelodar}, \binits{H.}},
\bauthor{\bsnm{Wang}, \binits{Y.}},
\bauthor{\bsnm{Yuan}, \binits{C.}},
\bauthor{\bsnm{Feng}, \binits{X.}},
\bauthor{\bsnm{Jiang}, \binits{X.}},
\bauthor{\bsnm{Li}, \binits{Y.}},
\bauthor{\bsnm{Zhao}, \binits{L.}}:
\batitle{Latent dirichlet allocation (lda) and topic modeling: models, applications, a survey}.
\bjtitle{Multimedia Tools and Applications}
\bvolume{78},
\bfpage{15169}--\blpage{15211}
(\byear{2019})
\end{barticle}
\endbibitem

\bibitem[\protect\citeauthoryear{Wang et~al.}{2023}]{wang2023identifying}
\begin{botherref}
\oauthor{\bsnm{Wang}, \binits{Z.}},
\oauthor{\bsnm{Chen}, \binits{J.}},
\oauthor{\bsnm{Chen}, \binits{J.}},
\oauthor{\bsnm{Chen}, \binits{H.}}:
Identifying interdisciplinary topics and their evolution based on bertopic.
Scientometrics,
1--26
(2023)
\end{botherref}
\endbibitem

\bibitem[\protect\citeauthoryear{Yau et~al.}{2014}]{yau2014clustering}
\begin{barticle}
\bauthor{\bsnm{Yau}, \binits{C.-K.}},
\bauthor{\bsnm{Porter}, \binits{A.}},
\bauthor{\bsnm{Newman}, \binits{N.}},
\bauthor{\bsnm{Suominen}, \binits{A.}}:
\batitle{Clustering scientific documents with topic modeling}.
\bjtitle{Scientometrics}
\bvolume{100},
\bfpage{767}--\blpage{786}
(\byear{2014})
\end{barticle}
\endbibitem

\bibitem[\protect\citeauthoryear{Yi and Allan}{2009}]{yi2009comparative}
\begin{bchapter}
\bauthor{\bsnm{Yi}, \binits{X.}},
\bauthor{\bsnm{Allan}, \binits{J.}}:
\bctitle{A comparative study of utilizing topic models for information retrieval}.
In: \bbtitle{Advances in Information Retrieval: 31th European Conference on IR Research, ECIR 2009, Toulouse, France, April 6-9, 2009. Proceedings 31},
pp. \bfpage{29}--\blpage{41}
(\byear{2009}).
\bcomment{Springer}
\end{bchapter}
\endbibitem

\bibitem[\protect\citeauthoryear{Luostarinen and Kohonen}{2013}]{luostarinen2013using}
\begin{bchapter}
\bauthor{\bsnm{Luostarinen}, \binits{T.}},
\bauthor{\bsnm{Kohonen}, \binits{O.}}:
\bctitle{Using topic models in content-based news recommender systems}.
In: \bbtitle{Proceedings of the 19th Nordic Conference of Computational Linguistics (NODALIDA 2013)},
pp. \bfpage{239}--\blpage{251}
(\byear{2013})
\end{bchapter}
\endbibitem

\bibitem[\protect\citeauthoryear{Eidelman et~al.}{2012}]{eidelman2012topic}
\begin{bchapter}
\bauthor{\bsnm{Eidelman}, \binits{V.}},
\bauthor{\bsnm{Boyd-Graber}, \binits{J.}},
\bauthor{\bsnm{Resnik}, \binits{P.}}:
\bctitle{Topic models for dynamic translation model adaptation}.
In: \bbtitle{Proceedings of the 50th Annual Meeting of the Association for Computational Linguistics (Volume 2: Short Papers)},
pp. \bfpage{115}--\blpage{119}
(\byear{2012})
\end{bchapter}
\endbibitem

\bibitem[\protect\citeauthoryear{Belwal et~al.}{2023}]{belwal2023extractive}
\begin{barticle}
\bauthor{\bsnm{Belwal}, \binits{R.C.}},
\bauthor{\bsnm{Rai}, \binits{S.}},
\bauthor{\bsnm{Gupta}, \binits{A.}}:
\batitle{Extractive text summarization using clustering-based topic modeling}.
\bjtitle{Soft Computing}
\bvolume{27}(\bissue{7}),
\bfpage{3965}--\blpage{3982}
(\byear{2023})
\end{barticle}
\endbibitem

\bibitem[\protect\citeauthoryear{Asuncion et~al.}{2010}]{asuncion2010software}
\begin{bchapter}
\bauthor{\bsnm{Asuncion}, \binits{H.U.}},
\bauthor{\bsnm{Asuncion}, \binits{A.U.}},
\bauthor{\bsnm{Taylor}, \binits{R.N.}}:
\bctitle{Software traceability with topic modeling}.
In: \bbtitle{Proceedings of the 32nd ACM/IEEE International Conference on Software Engineering-Volume 1},
pp. \bfpage{95}--\blpage{104}
(\byear{2010})
\end{bchapter}
\endbibitem

\bibitem[\protect\citeauthoryear{Bagheri et~al.}{2014}]{bagheri2014adm}
\begin{barticle}
\bauthor{\bsnm{Bagheri}, \binits{A.}},
\bauthor{\bsnm{Saraee}, \binits{M.}},
\bauthor{\bsnm{De~Jong}, \binits{F.}}:
\batitle{Adm-lda: An aspect detection model based on topic modelling using the structure of review sentences}.
\bjtitle{Journal of Information Science}
\bvolume{40}(\bissue{5}),
\bfpage{621}--\blpage{636}
(\byear{2014})
\end{barticle}
\endbibitem

\bibitem[\protect\citeauthoryear{Jo and Oh}{2011}]{jo2011aspect}
\begin{bchapter}
\bauthor{\bsnm{Jo}, \binits{Y.}},
\bauthor{\bsnm{Oh}, \binits{A.H.}}:
\bctitle{Aspect and sentiment unification model for online review analysis}.
In: \bbtitle{Proceedings of the Fourth ACM International Conference on Web Search and Data Mining},
pp. \bfpage{815}--\blpage{824}
(\byear{2011})
\end{bchapter}
\endbibitem

\bibitem[\protect\citeauthoryear{Zhai et~al.}{2011}]{zhai2011constrained}
\begin{bchapter}
\bauthor{\bsnm{Zhai}, \binits{Z.}},
\bauthor{\bsnm{Liu}, \binits{B.}},
\bauthor{\bsnm{Xu}, \binits{H.}},
\bauthor{\bsnm{Jia}, \binits{P.}}:
\bctitle{Constrained lda for grouping product features in opinion mining}.
In: \bbtitle{Advances in Knowledge Discovery and Data Mining: 15th Pacific-Asia Conference, PAKDD 2011, Shenzhen, China, May 24-27, 2011, Proceedings, Part I 15},
pp. \bfpage{448}--\blpage{459}
(\byear{2011}).
\bcomment{Springer}
\end{bchapter}
\endbibitem

\bibitem[\protect\citeauthoryear{Chen et~al.}{2012}]{chen2012explaining}
\begin{bchapter}
\bauthor{\bsnm{Chen}, \binits{T.-H.}},
\bauthor{\bsnm{Thomas}, \binits{S.W.}},
\bauthor{\bsnm{Nagappan}, \binits{M.}},
\bauthor{\bsnm{Hassan}, \binits{A.E.}}:
\bctitle{Explaining software defects using topic models}.
In: \bbtitle{2012 9th IEEE Working Conference on Mining Software Repositories (MSR)},
pp. \bfpage{189}--\blpage{198}
(\byear{2012}).
\bcomment{IEEE}
\end{bchapter}
\endbibitem

\bibitem[\protect\citeauthoryear{Chen et~al.}{2016}]{chen2016survey}
\begin{barticle}
\bauthor{\bsnm{Chen}, \binits{T.-H.}},
\bauthor{\bsnm{Thomas}, \binits{S.W.}},
\bauthor{\bsnm{Hassan}, \binits{A.E.}}:
\batitle{A survey on the use of topic models when mining software repositories}.
\bjtitle{Empirical Software Engineering}
\bvolume{21},
\bfpage{1843}--\blpage{1919}
(\byear{2016})
\end{barticle}
\endbibitem

\bibitem[\protect\citeauthoryear{Thomas}{2011}]{thomas2011mining}
\begin{bchapter}
\bauthor{\bsnm{Thomas}, \binits{S.W.}}:
\bctitle{Mining software repositories using topic models}.
In: \bbtitle{Proceedings of the 33rd International Conference on Software Engineering},
pp. \bfpage{1138}--\blpage{1139}
(\byear{2011})
\end{bchapter}
\endbibitem

\bibitem[\protect\citeauthoryear{Thomas et~al.}{2011}]{thomas2011modeling}
\begin{bchapter}
\bauthor{\bsnm{Thomas}, \binits{S.W.}},
\bauthor{\bsnm{Adams}, \binits{B.}},
\bauthor{\bsnm{Hassan}, \binits{A.E.}},
\bauthor{\bsnm{Blostein}, \binits{D.}}:
\bctitle{Modeling the evolution of topics in source code histories}.
In: \bbtitle{Proceedings of the 8th Working Conference on Mining Software Repositories},
pp. \bfpage{173}--\blpage{182}
(\byear{2011})
\end{bchapter}
\endbibitem

\bibitem[\protect\citeauthoryear{Tian et~al.}{2009}]{tian2009using}
\begin{bchapter}
\bauthor{\bsnm{Tian}, \binits{K.}},
\bauthor{\bsnm{Revelle}, \binits{M.}},
\bauthor{\bsnm{Poshyvanyk}, \binits{D.}}:
\bctitle{Using latent dirichlet allocation for automatic categorization of software}.
In: \bbtitle{2009 6th IEEE International Working Conference on Mining Software Repositories},
pp. \bfpage{163}--\blpage{166}
(\byear{2009}).
\bcomment{IEEE}
\end{bchapter}
\endbibitem

\bibitem[\protect\citeauthoryear{Gethers and Poshyvanyk}{2010}]{gethers2010using}
\begin{bchapter}
\bauthor{\bsnm{Gethers}, \binits{M.}},
\bauthor{\bsnm{Poshyvanyk}, \binits{D.}}:
\bctitle{Using relational topic models to capture coupling among classes in object-oriented software systems}.
In: \bbtitle{2010 IEEE International Conference on Software Maintenance},
pp. \bfpage{1}--\blpage{10}
(\byear{2010}).
\bcomment{IEEE}
\end{bchapter}
\endbibitem

\bibitem[\protect\citeauthoryear{Linstead et~al.}{2007}]{linstead2007mining}
\begin{bchapter}
\bauthor{\bsnm{Linstead}, \binits{E.}},
\bauthor{\bsnm{Rigor}, \binits{P.}},
\bauthor{\bsnm{Bajracharya}, \binits{S.}},
\bauthor{\bsnm{Lopes}, \binits{C.}},
\bauthor{\bsnm{Baldi}, \binits{P.}}:
\bctitle{Mining concepts from code with probabilistic topic models}.
In: \bbtitle{Proceedings of the 22nd IEEE/ACM International Conference on Automated Software Engineering},
pp. \bfpage{461}--\blpage{464}
(\byear{2007})
\end{bchapter}
\endbibitem

\bibitem[\protect\citeauthoryear{Linstead et~al.}{2008}]{linstead2008application}
\begin{bchapter}
\bauthor{\bsnm{Linstead}, \binits{E.}},
\bauthor{\bsnm{Lopes}, \binits{C.}},
\bauthor{\bsnm{Baldi}, \binits{P.}}:
\bctitle{An application of latent dirichlet allocation to analyzing software evolution}.
In: \bbtitle{2008 Seventh International Conference on Machine Learning and Applications},
pp. \bfpage{813}--\blpage{818}
(\byear{2008}).
\bcomment{IEEE}
\end{bchapter}
\endbibitem

\bibitem[\protect\citeauthoryear{Lukins et~al.}{2010}]{lukins2010bug}
\begin{barticle}
\bauthor{\bsnm{Lukins}, \binits{S.K.}},
\bauthor{\bsnm{Kraft}, \binits{N.A.}},
\bauthor{\bsnm{Etzkorn}, \binits{L.H.}}:
\batitle{Bug localization using latent dirichlet allocation}.
\bjtitle{Information and Software Technology}
\bvolume{52}(\bissue{9}),
\bfpage{972}--\blpage{990}
(\byear{2010})
\end{barticle}
\endbibitem

\bibitem[\protect\citeauthoryear{Savage et~al.}{2010}]{savage2010topic}
\begin{bchapter}
\bauthor{\bsnm{Savage}, \binits{T.}},
\bauthor{\bsnm{Dit}, \binits{B.}},
\bauthor{\bsnm{Gethers}, \binits{M.}},
\bauthor{\bsnm{Poshyvanyk}, \binits{D.}}:
\bctitle{Topic xp: Exploring topics in source code using latent dirichlet allocation}.
In: \bbtitle{2010 IEEE International Conference on Software Maintenance},
pp. \bfpage{1}--\blpage{6}
(\byear{2010}).
\bcomment{IEEE}
\end{bchapter}
\endbibitem

\bibitem[\protect\citeauthoryear{Gokcimen and Das}{2024}]{10527342}
\begin{bchapter}
\bauthor{\bsnm{Gokcimen}, \binits{T.}},
\bauthor{\bsnm{Das}, \binits{B.}}:
\bctitle{Topic modelling using bertopic for robust spam detection}.
In: \bbtitle{2024 12th International Symposium on Digital Forensics and Security (ISDFS)},
pp. \bfpage{1}--\blpage{5}
(\byear{2024}).
\doiurl{10.1109/ISDFS60797.2024.10527342}
\end{bchapter}
\endbibitem

\bibitem[\protect\citeauthoryear{Cheng and Shen}{2016}]{cheng2016effective}
\begin{barticle}
\bauthor{\bsnm{Cheng}, \binits{Z.}},
\bauthor{\bsnm{Shen}, \binits{J.}}:
\batitle{On effective location-aware music recommendation}.
\bjtitle{ACM Transactions on Information Systems (TOIS)}
\bvolume{34}(\bissue{2}),
\bfpage{1}--\blpage{32}
(\byear{2016})
\end{barticle}
\endbibitem

\bibitem[\protect\citeauthoryear{Kim and Shim}{2014}]{kim2014twilite}
\begin{barticle}
\bauthor{\bsnm{Kim}, \binits{Y.}},
\bauthor{\bsnm{Shim}, \binits{K.}}:
\batitle{Twilite: A recommendation system for twitter using a probabilistic model based on latent dirichlet allocation}.
\bjtitle{Information Systems}
\bvolume{42},
\bfpage{59}--\blpage{77}
(\byear{2014})
\end{barticle}
\endbibitem

\bibitem[\protect\citeauthoryear{Lu and Lee}{2015}]{lu2015twitter}
\begin{barticle}
\bauthor{\bsnm{Lu}, \binits{H.-M.}},
\bauthor{\bsnm{Lee}, \binits{C.-H.}}:
\batitle{A twitter hashtag recommendation model that accommodates for temporal clustering effects}.
\bjtitle{IEEE Intelligent Systems}
\bvolume{30}(\bissue{3}),
\bfpage{18}--\blpage{25}
(\byear{2015})
\end{barticle}
\endbibitem

\bibitem[\protect\citeauthoryear{Zhao et~al.}{2016}]{zhao2016personalized}
\begin{barticle}
\bauthor{\bsnm{Zhao}, \binits{F.}},
\bauthor{\bsnm{Zhu}, \binits{Y.}},
\bauthor{\bsnm{Jin}, \binits{H.}},
\bauthor{\bsnm{Yang}, \binits{L.T.}}:
\batitle{A personalized hashtag recommendation approach using lda-based topic model in microblog environment}.
\bjtitle{Future Generation Computer Systems}
\bvolume{65},
\bfpage{196}--\blpage{206}
(\byear{2016})
\end{barticle}
\endbibitem

\bibitem[\protect\citeauthoryear{Zoghbi et~al.}{2016}]{zoghbi2016latent}
\begin{barticle}
\bauthor{\bsnm{Zoghbi}, \binits{S.}},
\bauthor{\bsnm{Vuli{\'c}}, \binits{I.}},
\bauthor{\bsnm{Moens}, \binits{M.-F.}}:
\batitle{Latent dirichlet allocation for linking user-generated content and e-commerce data}.
\bjtitle{Information Sciences}
\bvolume{367},
\bfpage{573}--\blpage{599}
(\byear{2016})
\end{barticle}
\endbibitem

\bibitem[\protect\citeauthoryear{Grootendorst}{2022}]{grootendorst2022bertopic}
\begin{botherref}
\oauthor{\bsnm{Grootendorst}, \binits{M.}}:
Bertopic: Neural topic modeling with a class-based tf-idf procedure.
arXiv preprint arXiv:2203.05794
(2022)
\end{botherref}
\endbibitem

\bibitem[\protect\citeauthoryear{Blei et~al.}{2003}]{blei2003latent}
\begin{barticle}
\bauthor{\bsnm{Blei}, \binits{D.M.}},
\bauthor{\bsnm{Ng}, \binits{A.Y.}},
\bauthor{\bsnm{Jordan}, \binits{M.I.}}:
\batitle{Latent dirichlet allocation}.
\bjtitle{Journal of machine Learning research}
\bvolume{3}(\bissue{Jan}),
\bfpage{993}--\blpage{1022}
(\byear{2003})
\end{barticle}
\endbibitem

\bibitem[\protect\citeauthoryear{Lee and Seung}{2000}]{lee2000algorithms}
\begin{botherref}
\oauthor{\bsnm{Lee}, \binits{D.}},
\oauthor{\bsnm{Seung}, \binits{H.S.}}:
Algorithms for non-negative matrix factorization.
Advances in neural information processing systems
\textbf{13}
(2000)
\end{botherref}
\endbibitem

\bibitem[\protect\citeauthoryear{Landauer et~al.}{1998}]{landauer1998introduction}
\begin{barticle}
\bauthor{\bsnm{Landauer}, \binits{T.K.}},
\bauthor{\bsnm{Foltz}, \binits{P.W.}},
\bauthor{\bsnm{Laham}, \binits{D.}}:
\batitle{An introduction to latent semantic analysis}.
\bjtitle{Discourse processes}
\bvolume{25}(\bissue{2-3}),
\bfpage{259}--\blpage{284}
(\byear{1998})
\end{barticle}
\endbibitem

\bibitem[\protect\citeauthoryear{Blei and Lafferty}{2006a}]{blei2006correlated}
\begin{barticle}
\bauthor{\bsnm{Blei}, \binits{D.}},
\bauthor{\bsnm{Lafferty}, \binits{J.}}:
\batitle{Correlated topic models}.
\bjtitle{Advances in neural information processing systems}
\bvolume{18},
\bfpage{147}
(\byear{2006})
\end{barticle}
\endbibitem

\bibitem[\protect\citeauthoryear{Blei and Lafferty}{2006b}]{blei2006dynamic}
\begin{bchapter}
\bauthor{\bsnm{Blei}, \binits{D.M.}},
\bauthor{\bsnm{Lafferty}, \binits{J.D.}}:
\bctitle{Dynamic topic models}.
In: \bbtitle{Proceedings of the 23rd International Conference on Machine Learning},
pp. \bfpage{113}--\blpage{120}
(\byear{2006})
\end{bchapter}
\endbibitem

\bibitem[\protect\citeauthoryear{Yan et~al.}{2013}]{yan2013biterm}
\begin{bchapter}
\bauthor{\bsnm{Yan}, \binits{X.}},
\bauthor{\bsnm{Guo}, \binits{J.}},
\bauthor{\bsnm{Lan}, \binits{Y.}},
\bauthor{\bsnm{Cheng}, \binits{X.}}:
\bctitle{A biterm topic model for short texts}.
In: \bbtitle{Proceedings of the 22nd International Conference on World Wide Web},
pp. \bfpage{1445}--\blpage{1456}
(\byear{2013})
\end{bchapter}
\endbibitem

\bibitem[\protect\citeauthoryear{Parlina and Maryati}{2023}]{10285737}
\begin{bchapter}
\bauthor{\bsnm{Parlina}, \binits{A.}},
\bauthor{\bsnm{Maryati}, \binits{I.}}:
\bctitle{Leveraging bertopic for the analysis of scientific papers on seaweed}.
In: \bbtitle{2023 International Conference on Computer, Control, Informatics and Its Applications (IC3INA)},
pp. \bfpage{279}--\blpage{283}
(\byear{2023}).
\doiurl{10.1109/IC3INA60834.2023.10285737}
\end{bchapter}
\endbibitem

\bibitem[\protect\citeauthoryear{Kang et~al.}{2023}]{10487248}
\begin{bchapter}
\bauthor{\bsnm{Kang}, \binits{W.}},
\bauthor{\bsnm{Kim}, \binits{Y.}},
\bauthor{\bsnm{Kim}, \binits{H.}},
\bauthor{\bsnm{Lee}, \binits{J.}}:
\bctitle{An analysis of research trends on language model using bertopic}.
In: \bbtitle{2023 Congress in Computer Science, Computer Engineering, and Applied Computing (CSCE)},
pp. \bfpage{168}--\blpage{172}
(\byear{2023}).
\doiurl{10.1109/CSCE60160.2023.00032}
\end{bchapter}
\endbibitem

\bibitem[\protect\citeauthoryear{Doi et~al.}{2024}]{doi2024topic}
\begin{bchapter}
\bauthor{\bsnm{Doi}, \binits{T.}},
\bauthor{\bsnm{Isonuma}, \binits{M.}},
\bauthor{\bsnm{Yanaka}, \binits{H.}}:
\bctitle{Topic modeling for short texts with large language models}.
In: \bbtitle{Proceedings of the 62nd Annual Meeting of the Association for Computational Linguistics (Volume 4: Student Research Workshop)},
pp. \bfpage{21}--\blpage{33}
(\byear{2024})
\end{bchapter}
\endbibitem

\bibitem[\protect\citeauthoryear{Mankolli}{2022}]{9854488}
\begin{bchapter}
\bauthor{\bsnm{Mankolli}, \binits{E.}}:
\bctitle{Reducing the complexity of candidate selection using natural language processing}.
In: \bbtitle{2022 29th International Conference on Systems, Signals and Image Processing (IWSSIP)},
vol. \bseriesno{CFP2255E-ART},
pp. \bfpage{1}--\blpage{4}
(\byear{2022}).
\doiurl{10.1109/IWSSIP55020.2022.9854488}
\end{bchapter}
\endbibitem

\bibitem[\protect\citeauthoryear{Atzeni et~al.}{2022}]{atzeni2022systematic}
\begin{barticle}
\bauthor{\bsnm{Atzeni}, \binits{D.}},
\bauthor{\bsnm{Bacciu}, \binits{D.}},
\bauthor{\bsnm{Mazzei}, \binits{D.}},
\bauthor{\bsnm{Prencipe}, \binits{G.}}:
\batitle{A systematic review of wi-fi and machine learning integration with topic modeling techniques}.
\bjtitle{Sensors}
\bvolume{22}(\bissue{13}),
\bfpage{4925}
(\byear{2022})
\end{barticle}
\endbibitem

\bibitem[\protect\citeauthoryear{Nie et~al.}{2017}]{nie2017data}
\begin{barticle}
\bauthor{\bsnm{Nie}, \binits{L.}},
\bauthor{\bsnm{Wei}, \binits{X.}},
\bauthor{\bsnm{Zhang}, \binits{D.}},
\bauthor{\bsnm{Wang}, \binits{X.}},
\bauthor{\bsnm{Gao}, \binits{Z.}},
\bauthor{\bsnm{Yang}, \binits{Y.}}:
\batitle{Data-driven answer selection in community qa systems}.
\bjtitle{IEEE transactions on knowledge and data engineering}
\bvolume{29}(\bissue{6}),
\bfpage{1186}--\blpage{1198}
(\byear{2017})
\end{barticle}
\endbibitem

\bibitem[\protect\citeauthoryear{Fiscus and Doddington}{2002}]{fiscus2002topic}
\begin{bchapter}
\bauthor{\bsnm{Fiscus}, \binits{J.G.}},
\bauthor{\bsnm{Doddington}, \binits{G.R.}}:
\bctitle{Topic detection and tracking evaluation overview}.
In: \bbtitle{Topic Detection and Tracking: Event-based Information Organization},
pp. \bfpage{17}--\blpage{31}.
\bpublisher{Springer}, \blocation{???}
(\byear{2002})
\end{bchapter}
\endbibitem

\bibitem[\protect\citeauthoryear{Barbosa}{2020}]{barbosa2020software}
\begin{bchapter}
\bauthor{\bsnm{Barbosa}, \binits{L.S.}}:
\bctitle{Software engineering for'quantum advantage'}.
In: \bbtitle{Proceedings of the IEEE/ACM 42nd International Conference on Software Engineering Workshops},
pp. \bfpage{427}--\blpage{429}
(\byear{2020})
\end{bchapter}
\endbibitem

\bibitem[\protect\citeauthoryear{Wang et~al.}{2013}]{wang2013empirical}
\begin{bchapter}
\bauthor{\bsnm{Wang}, \binits{S.}},
\bauthor{\bsnm{Lo}, \binits{D.}},
\bauthor{\bsnm{Jiang}, \binits{L.}}:
\bctitle{An empirical study on developer interactions in stackoverflow}.
In: \bbtitle{Proceedings of the 28th Annual ACM Symposium on Applied Computing},
pp. \bfpage{1019}--\blpage{1024}
(\byear{2013})
\end{bchapter}
\endbibitem

\bibitem[\protect\citeauthoryear{Chen et~al.}{2019}]{chen2019modeling}
\begin{barticle}
\bauthor{\bsnm{Chen}, \binits{H.}},
\bauthor{\bsnm{Coogle}, \binits{J.}},
\bauthor{\bsnm{Damevski}, \binits{K.}}:
\batitle{Modeling stack overflow tags and topics as a hierarchy of concepts}.
\bjtitle{Journal of Systems and Software}
\bvolume{156},
\bfpage{283}--\blpage{299}
(\byear{2019})
\end{barticle}
\endbibitem

\bibitem[\protect\citeauthoryear{Allamanis and Sutton}{2013}]{allamanis2013and}
\begin{bchapter}
\bauthor{\bsnm{Allamanis}, \binits{M.}},
\bauthor{\bsnm{Sutton}, \binits{C.}}:
\bctitle{Why, when, and what: analyzing stack overflow questions by topic, type, and code}.
In: \bbtitle{Mining Software Repositories (MSR)},
pp. \bfpage{53}--\blpage{56}
(\byear{2013}).
\bcomment{IEEE}
\end{bchapter}
\endbibitem

\bibitem[\protect\citeauthoryear{Alshangiti et~al.}{2019}]{alshangiti2019developing}
\begin{bchapter}
\bauthor{\bsnm{Alshangiti}, \binits{M.}},
\bauthor{\bsnm{Sapkota}, \binits{H.}},
\bauthor{\bsnm{Murukannaiah}, \binits{P.K.}},
\bauthor{\bsnm{Liu}, \binits{X.}},
\bauthor{\bsnm{Yu}, \binits{Q.}}:
\bctitle{Why is developing machine learning applications challenging? a study on stack overflow posts}.
In: \bbtitle{International Symposium on (ESEM)},
pp. \bfpage{1}--\blpage{11}
(\byear{2019}).
\bcomment{IEEE}
\end{bchapter}
\endbibitem

\bibitem[\protect\citeauthoryear{Ahmed and Bagherzadeh}{2018}]{ahmed2018concurrency}
\begin{bchapter}
\bauthor{\bsnm{Ahmed}, \binits{S.}},
\bauthor{\bsnm{Bagherzadeh}, \binits{M.}}:
\bctitle{What do concurrency developers ask about? a large-scale study using stack overflow}.
In: \bbtitle{Proceedings of the 12th ACM/IEEE International Symposium on Empirical Software Engineering and Measurement},
pp. \bfpage{1}--\blpage{10}
(\byear{2018})
\end{bchapter}
\endbibitem

\bibitem[\protect\citeauthoryear{Mi et~al.}{2023}]{10371495}
\begin{bchapter}
\bauthor{\bsnm{Mi}, \binits{Q.}},
\bauthor{\bsnm{Bao}, \binits{Q.}},
\bauthor{\bsnm{Cui}, \binits{L.}}:
\bctitle{Identifying topics and trends in devops: A study of stack overflow posts}.
In: \bbtitle{2023 49th Euromicro Conference on Software Engineering and Advanced Applications (SEAA)},
pp. \bfpage{402}--\blpage{409}
(\byear{2023}).
\doiurl{10.1109/SEAA60479.2023.00067}
\end{bchapter}
\endbibitem

\bibitem[\protect\citeauthoryear{Suwonchoochit and Senivongse}{2021}]{9509047}
\begin{bchapter}
\bauthor{\bsnm{Suwonchoochit}, \binits{N.}},
\bauthor{\bsnm{Senivongse}, \binits{T.}}:
\bctitle{Classification of database technology problems on stack overflow}.
In: \bbtitle{2021 IEEE/ACIS 19th International Conference on Software Engineering Research, Management and Applications (SERA)},
pp. \bfpage{21}--\blpage{26}
(\byear{2021}).
\doiurl{10.1109/SERA51205.2021.9509047}
\end{bchapter}
\endbibitem

\bibitem[\protect\citeauthoryear{Zou et~al.}{2017}]{zou2017towards}
\begin{barticle}
\bauthor{\bsnm{Zou}, \binits{J.}},
\bauthor{\bsnm{Xu}, \binits{L.}},
\bauthor{\bsnm{Yang}, \binits{M.}},
\bauthor{\bsnm{Zhang}, \binits{X.}},
\bauthor{\bsnm{Yang}, \binits{D.}}:
\batitle{Towards comprehending the non-functional requirements through developers’ eyes: An exploration of stack overflow using topic analysis}.
\bjtitle{Information and Software Technology}
\bvolume{84},
\bfpage{19}--\blpage{32}
(\byear{2017})
\end{barticle}
\endbibitem

\bibitem[\protect\citeauthoryear{Dhasade et~al.}{2020}]{dhasade2020towards}
\begin{bchapter}
\bauthor{\bsnm{Dhasade}, \binits{A.B.}},
\bauthor{\bsnm{Venigalla}, \binits{A.S.M.}},
\bauthor{\bsnm{Chimalakonda}, \binits{S.}}:
\bctitle{Towards prioritizing github issues}.
In: \bbtitle{Proceedings of the 13th Innovations in Software Engineering Conference on Formerly Known as India Software Engineering Conference},
pp. \bfpage{1}--\blpage{5}
(\byear{2020})
\end{bchapter}
\endbibitem

\bibitem[\protect\citeauthoryear{Jokhio}{2021}]{jokhio2021mining}
\begin{botherref}
\oauthor{\bsnm{Jokhio}, \binits{M.}}:
Mining github issues for bugs, feature requests and questions.
''
(2021)
\end{botherref}
\endbibitem

\bibitem[\protect\citeauthoryear{Campbell et~al.}{2015}]{campbell2015latent}
\begin{bchapter}
\bauthor{\bsnm{Campbell}, \binits{J.C.}},
\bauthor{\bsnm{Hindle}, \binits{A.}},
\bauthor{\bsnm{Stroulia}, \binits{E.}}:
\bctitle{Latent dirichlet allocation: extracting topics from software engineering data}.
In: \bbtitle{The Art and Science of Analyzing Software Data},
pp. \bfpage{139}--\blpage{159}.
\bpublisher{Elsevier}, \blocation{???}
(\byear{2015})
\end{bchapter}
\endbibitem

\bibitem[\protect\citeauthoryear{Wang et~al.}{2019}]{wang2019does}
\begin{bchapter}
\bauthor{\bsnm{Wang}, \binits{X.}},
\bauthor{\bsnm{Lee}, \binits{M.}},
\bauthor{\bsnm{Pinchbeck}, \binits{A.}},
\bauthor{\bsnm{Fard}, \binits{F.}}:
\bctitle{Where does lda sit for github?}
In: \bbtitle{2019 34th IEEE/ACM International Conference on Automated Software Engineering Workshop (ASEW)},
pp. \bfpage{94}--\blpage{97}
(\byear{2019}).
\bcomment{IEEE}
\end{bchapter}
\endbibitem

\bibitem[\protect\citeauthoryear{Treude and Wagner}{2019}]{treude2019predicting}
\begin{bchapter}
\bauthor{\bsnm{Treude}, \binits{C.}},
\bauthor{\bsnm{Wagner}, \binits{M.}}:
\bctitle{Predicting good configurations for github and stack overflow topic models}.
In: \bbtitle{2019 IEEE/ACM 16th International Conference on Mining Software Repositories (MSR)},
pp. \bfpage{84}--\blpage{95}
(\byear{2019}).
\bcomment{IEEE}
\end{bchapter}
\endbibitem

\bibitem[\protect\citeauthoryear{Nadi et~al.}{2016}]{nadi2016jumping}
\begin{bchapter}
\bauthor{\bsnm{Nadi}, \binits{S.}},
\bauthor{\bsnm{Kr{\"u}ger}, \binits{S.}},
\bauthor{\bsnm{Mezini}, \binits{M.}},
\bauthor{\bsnm{Bodden}, \binits{E.}}:
\bctitle{Jumping through hoops: Why do java developers struggle with cryptography apis?}
In: \bbtitle{Proceedings of the 38th International Conference on Software Engineering},
pp. \bfpage{935}--\blpage{946}
(\byear{2016})
\end{bchapter}
\endbibitem

\bibitem[\protect\citeauthoryear{Han et~al.}{2020}]{han2020programmers}
\begin{barticle}
\bauthor{\bsnm{Han}, \binits{J.}},
\bauthor{\bsnm{Shihab}, \binits{E.}},
\bauthor{\bsnm{Wan}, \binits{Z.}},
\bauthor{\bsnm{Deng}, \binits{S.}},
\bauthor{\bsnm{Xia}, \binits{X.}}:
\batitle{What do programmers discuss about deep learning frameworks}.
\bjtitle{Empirical Software Engineering}
\bvolume{25},
\bfpage{2694}--\blpage{2747}
(\byear{2020})
\end{barticle}
\endbibitem

\bibitem[\protect\citeauthoryear{Barua et~al.}{2014}]{barua2014developers}
\begin{barticle}
\bauthor{\bsnm{Barua}, \binits{A.}},
\bauthor{\bsnm{Thomas}, \binits{S.W.}},
\bauthor{\bsnm{Hassan}, \binits{A.E.}}:
\batitle{What are developers talking about? an analysis of topics and trends in stack overflow}.
\bjtitle{Empirical software engineering}
\bvolume{19},
\bfpage{619}--\blpage{654}
(\byear{2014})
\end{barticle}
\endbibitem

\bibitem[\protect\citeauthoryear{Spjuth et~al.}{2015}]{spjuth2015experiences}
\begin{barticle}
\bauthor{\bsnm{Spjuth}, \binits{O.}},
\bauthor{\bsnm{Bongcam-Rudloff}, \binits{E.}},
\bauthor{\bsnm{Hern{\'a}ndez}, \binits{G.C.}},
\bauthor{\bsnm{Forer}, \binits{L.}},
\bauthor{\bsnm{Giovacchini}, \binits{M.}},
\bauthor{\bsnm{Guimera}, \binits{R.V.}},
\bauthor{\bsnm{Kallio}, \binits{A.}},
\bauthor{\bsnm{Korpelainen}, \binits{E.}},
\bauthor{\bsnm{Ka{\'n}du{\l}a}, \binits{M.M.}},
\bauthor{\bsnm{Krachunov}, \binits{M.}}, \betal:
\batitle{Experiences with workflows for automating data-intensive bioinformatics}.
\bjtitle{Biology direct}
\bvolume{10},
\bfpage{1}--\blpage{12}
(\byear{2015})
\end{barticle}
\endbibitem

\bibitem[\protect\citeauthoryear{Jaeger et~al.}{2005}]{jaeger2005scientific}
\begin{bchapter}
\bauthor{\bsnm{Jaeger}, \binits{E.}},
\bauthor{\bsnm{Altintas}, \binits{I.}},
\bauthor{\bsnm{Zhang}, \binits{J.}},
\bauthor{\bsnm{Lud{\"a}scher}, \binits{B.}},
\bauthor{\bsnm{Pennington}, \binits{D.}},
\bauthor{\bsnm{Michener}, \binits{W.}}:
\bctitle{A scientific workflow approach to distributed geospatial data processing using web services.}
In: \bbtitle{SSDBM},
vol. \bseriesno{3},
pp. \bfpage{87}--\blpage{90}
(\byear{2005})
\end{bchapter}
\endbibitem

\bibitem[\protect\citeauthoryear{Carver et~al.}{2007}]{carver2007software}
\begin{bchapter}
\bauthor{\bsnm{Carver}, \binits{J.C.}},
\bauthor{\bsnm{Kendall}, \binits{R.P.}},
\bauthor{\bsnm{Squires}, \binits{S.E.}},
\bauthor{\bsnm{Post}, \binits{D.E.}}:
\bctitle{Software development environments for scientific and engineering software: A series of case studies}.
In: \bbtitle{29th International Conference on Software Engineering (ICSE'07)},
pp. \bfpage{550}--\blpage{559}
(\byear{2007}).
\bcomment{Ieee}
\end{bchapter}
\endbibitem

\bibitem[\protect\citeauthoryear{Nouri et~al.}{2021}]{nouri2021exploring}
\begin{botherref}
\oauthor{\bsnm{Nouri}, \binits{A.}},
\oauthor{\bsnm{Davis}, \binits{P.E.}},
\oauthor{\bsnm{Subedi}, \binits{P.}},
\oauthor{\bsnm{Parashar}, \binits{M.}}:
Exploring the role of machine learning in scientific workflows: Opportunities and challenges.
arXiv preprint arXiv:2110.13999
(2021)
\end{botherref}
\endbibitem

\bibitem[\protect\citeauthoryear{Yu and Buyya}{2005}]{yu2005taxonomy}
\begin{barticle}
\bauthor{\bsnm{Yu}, \binits{J.}},
\bauthor{\bsnm{Buyya}, \binits{R.}}:
\batitle{A taxonomy of scientific workflow systems for grid computing}.
\bjtitle{ACM Sigmod Record}
\bvolume{34}(\bissue{3}),
\bfpage{44}--\blpage{49}
(\byear{2005})
\end{barticle}
\endbibitem

\bibitem[\protect\citeauthoryear{Rocklin et~al.}{2015}]{rocklin2015dask}
\begin{bchapter}
\bauthor{\bsnm{Rocklin}, \binits{M.}}, \betal:
\bctitle{Dask: Parallel computation with blocked algorithms and task scheduling.}
In: \bbtitle{SciPy},
pp. \bfpage{126}--\blpage{132}
(\byear{2015})
\end{bchapter}
\endbibitem

\bibitem[\protect\citeauthoryear{Wikipedia}{2024}]{swfmswiki}
\begin{botherref}
\oauthor{\bsnm{Wikipedia}}:
Scientific Workflow System.
Online; last accessed January, 2024
(2024).
\url{https://en.wikipedia.org/wiki/Scientific\_workflow\_system}
\end{botherref}
\endbibitem

\bibitem[\protect\citeauthoryear{pditommaso}{2024}]{swfmspipeline}
\begin{botherref}
\oauthor{\bsnm{pditommaso}}:
Workflow Systems.
Online; last accessed January, 2024
(2024).
\url{https://github.com/pditommaso/awesome-pipeline}
\end{botherref}
\endbibitem

\bibitem[\protect\citeauthoryear{cwl}{2024}]{swfmscwl}
\begin{botherref}
\oauthor{\bsnm{cwl}}:
Existing Workflow Systems.
Online; last accessed January, 2024
(2024).
\url{https://github.com/common-workflow-language/common-workflow-language/wiki/Existing-Workflow-systems}
\end{botherref}
\endbibitem

\bibitem[\protect\citeauthoryear{workflow community}{2024}]{wcs}
\begin{botherref}
\oauthor{\bsnm{community}}:
Workflow Systems.
Online; last accessed January, 2024
(2024).
\url{https://workflows.community/systems}
\end{botherref}
\endbibitem

\bibitem[\protect\citeauthoryear{Exchange}{2024}]{stackexchangeforum}
\begin{botherref}
\oauthor{\bsnm{Exchange}, \binits{S.}}:
Stack Exchange Data Explorer.
\url{https://data.stackexchange.com/}.
accessed: January 31, 2024
(2024)
\end{botherref}
\endbibitem

\bibitem[\protect\citeauthoryear{Bird}{2006}]{bird2006nltk}
\begin{bchapter}
\bauthor{\bsnm{Bird}, \binits{S.}}:
\bctitle{Nltk: the natural language toolkit}.
In: \bbtitle{Proceedings of the COLING/ACL 2006 Interactive Presentation Sessions},
pp. \bfpage{69}--\blpage{72}
(\byear{2006})
\end{bchapter}
\endbibitem

\bibitem[\protect\citeauthoryear{Vasiliev}{2020}]{vasiliev2020natural}
\begin{bbook}
\bauthor{\bsnm{Vasiliev}, \binits{Y.}}:
\bbtitle{Natural Language Processing with Python and spaCy: A Practical Introduction}.
\bpublisher{No Starch Press}, \blocation{???}
(\byear{2020})
\end{bbook}
\endbibitem

\bibitem[\protect\citeauthoryear{github}{2024}]{githubrestapi}
\begin{botherref}
\oauthor{\bsnm{github}}:
GitHub REST API.
Online;January 2024
(2024).
\url{https://docs.github.com/en/rest?apiVersion=2022-11-28}
\end{botherref}
\endbibitem

\bibitem[\protect\citeauthoryear{Face}{2024}]{huggingfacemodels}
\begin{botherref}
\oauthor{\bsnm{Face}, \binits{H.}}:
{Hugging Face Model Hub}.
Online; last accessed August, 2024
(2024).
\url{https://huggingface.co/docs/hub/en/models-the-hub}
\end{botherref}
\endbibitem

\bibitem[\protect\citeauthoryear{McInnes et~al.}{2018}]{mcinnes2018umap}
\begin{botherref}
\oauthor{\bsnm{McInnes}, \binits{L.}},
\oauthor{\bsnm{Healy}, \binits{J.}},
\oauthor{\bsnm{Melville}, \binits{J.}}:
Umap: Uniform manifold approximation and projection for dimension reduction.
arXiv preprint arXiv:1802.03426
(2018)
\end{botherref}
\endbibitem

\bibitem[\protect\citeauthoryear{McInnes et~al.}{2017}]{mcinnes2017hdbscan}
\begin{barticle}
\bauthor{\bsnm{McInnes}, \binits{L.}},
\bauthor{\bsnm{Healy}, \binits{J.}},
\bauthor{\bsnm{Astels}, \binits{S.}}, \betal:
\batitle{hdbscan: Hierarchical density based clustering.}
\bjtitle{J. Open Source Softw.}
\bvolume{2}(\bissue{11}),
\bfpage{205}
(\byear{2017})
\end{barticle}
\endbibitem

\bibitem[\protect\citeauthoryear{Openja et~al.}{2020}]{openja2020analysis}
\begin{bchapter}
\bauthor{\bsnm{Openja}, \binits{M.}},
\bauthor{\bsnm{Adams}, \binits{B.}},
\bauthor{\bsnm{Khomh}, \binits{F.}}:
\bctitle{Analysis of modern release engineering topics:--a large-scale study using stackoverflow--}.
In: \bbtitle{2020 IEEE International Conference on Software Maintenance and Evolution (ICSME)},
pp. \bfpage{104}--\blpage{114}
(\byear{2020}).
\bcomment{IEEE}
\end{bchapter}
\endbibitem

\bibitem[\protect\citeauthoryear{Bajaj et~al.}{2014}]{bajaj2014mining}
\begin{bchapter}
\bauthor{\bsnm{Bajaj}, \binits{K.}},
\bauthor{\bsnm{Pattabiraman}, \binits{K.}},
\bauthor{\bsnm{Mesbah}, \binits{A.}}:
\bctitle{Mining questions asked by web developers}.
In: \bbtitle{Proceedings of the 11th Working Conference on Mining Software Repositories},
pp. \bfpage{112}--\blpage{121}
(\byear{2014})
\end{bchapter}
\endbibitem

\bibitem[\protect\citeauthoryear{Hochstein and Moser}{2017}]{hochstein2017ansible}
\begin{botherref}
\oauthor{\bsnm{Hochstein}, \binits{L.}},
\oauthor{\bsnm{Moser}, \binits{R.}}:
Ansible: Up and running: Automating configuration management and deployment the easy way.
" O'Reilly Media, Inc."
(2017)
\end{botherref}
\endbibitem

\bibitem[\protect\citeauthoryear{Blackman and Koval}{2000}]{blackman2000interval}
\begin{barticle}
\bauthor{\bsnm{Blackman}, \binits{N.J.-M.}},
\bauthor{\bsnm{Koval}, \binits{J.J.}}:
\batitle{Interval estimation for cohen's kappa as a measure of agreement}.
\bjtitle{Statistics in medicine}
\bvolume{19}(\bissue{5}),
\bfpage{723}--\blpage{741}
(\byear{2000})
\end{barticle}
\endbibitem

\bibitem[\protect\citeauthoryear{McHugh}{2012}]{mchugh2012interrater}
\begin{barticle}
\bauthor{\bsnm{McHugh}, \binits{M.L.}}:
\batitle{Interrater reliability: the kappa statistic}.
\bjtitle{Biochemia medica}
\bvolume{22}(\bissue{3}),
\bfpage{276}--\blpage{282}
(\byear{2012})
\end{barticle}
\endbibitem

\bibitem[\protect\citeauthoryear{}{2008}]{ref1}
\begin{botherref}
Spearman Rank Correlation Coefficient,
pp. 502--505.
Springer,
New York, NY
(2008).
\doiurl{10.1007/978-0-387-32833-1_379} .
\url{https://doi.org/10.1007/978-0-387-32833-1_379}
\end{botherref}
\endbibitem

\bibitem[\protect\citeauthoryear{Towns et~al.}{2014}]{6866038}
\begin{barticle}
\bauthor{\bsnm{Towns}, \binits{J.}},
\bauthor{\bsnm{Cockerill}, \binits{T.}},
\bauthor{\bsnm{Dahan}, \binits{M.}},
\bauthor{\bsnm{Foster}, \binits{I.}},
\bauthor{\bsnm{Gaither}, \binits{K.}},
\bauthor{\bsnm{Grimshaw}, \binits{A.}},
\bauthor{\bsnm{Hazlewood}, \binits{V.}},
\bauthor{\bsnm{Lathrop}, \binits{S.}},
\bauthor{\bsnm{Lifka}, \binits{D.}},
\bauthor{\bsnm{Peterson}, \binits{G.D.}},
\bauthor{\bsnm{Roskies}, \binits{R.}},
\bauthor{\bsnm{Scott}, \binits{J.R.}},
\bauthor{\bsnm{Wilkins-Diehr}, \binits{N.}}:
\batitle{Xsede: Accelerating scientific discovery}.
\bjtitle{Computing in Science and Engineering}
\bvolume{16}(\bissue{5}),
\bfpage{62}--\blpage{74}
(\byear{2014})
\doiurl{10.1109/MCSE.2014.80}
\end{barticle}
\endbibitem

\bibitem[\protect\citeauthoryear{Nadeem et~al.}{2019}]{nadeem2019using}
\begin{barticle}
\bauthor{\bsnm{Nadeem}, \binits{F.}},
\bauthor{\bsnm{Alghazzawi}, \binits{D.}},
\bauthor{\bsnm{Mashat}, \binits{A.}},
\bauthor{\bsnm{Faqeeh}, \binits{K.}},
\bauthor{\bsnm{Almalaise}, \binits{A.}}:
\batitle{Using machine learning ensemble methods to predict execution time of e-science workflows in heterogeneous distributed systems}.
\bjtitle{IEEE Access}
\bvolume{7},
\bfpage{25138}--\blpage{25149}
(\byear{2019})
\end{barticle}
\endbibitem

\bibitem[\protect\citeauthoryear{Wratten et~al.}{2021}]{wratten2021reproducible}
\begin{barticle}
\bauthor{\bsnm{Wratten}, \binits{L.}},
\bauthor{\bsnm{Wilm}, \binits{A.}},
\bauthor{\bsnm{G{\"o}ke}, \binits{J.}}:
\batitle{Reproducible, scalable, and shareable analysis pipelines with bioinformatics workflow managers}.
\bjtitle{Nature methods}
\bvolume{18}(\bissue{10}),
\bfpage{1161}--\blpage{1168}
(\byear{2021})
\end{barticle}
\endbibitem

\bibitem[\protect\citeauthoryear{de~Souza and Redmiles}{2008}]{de2008empirical}
\begin{bchapter}
\bauthor{\bsnm{Souza}, \binits{C.R.}},
\bauthor{\bsnm{Redmiles}, \binits{D.F.}}:
\bctitle{An empirical study of software developers' management of dependencies and changes}.
In: \bbtitle{Proceedings of the 30th International Conference on Software Engineering},
pp. \bfpage{241}--\blpage{250}
(\byear{2008})
\end{bchapter}
\endbibitem

\bibitem[\protect\citeauthoryear{Cataldo et~al.}{2009}]{cataldo2009software}
\begin{barticle}
\bauthor{\bsnm{Cataldo}, \binits{M.}},
\bauthor{\bsnm{Mockus}, \binits{A.}},
\bauthor{\bsnm{Roberts}, \binits{J.A.}},
\bauthor{\bsnm{Herbsleb}, \binits{J.D.}}:
\batitle{Software dependencies, work dependencies, and their impact on failures}.
\bjtitle{IEEE Transactions on Software Engineering}
\bvolume{35}(\bissue{6}),
\bfpage{864}--\blpage{878}
(\byear{2009})
\end{barticle}
\endbibitem

\bibitem[\protect\citeauthoryear{Hailpern and Santhanam}{2002}]{hailpern2002software}
\begin{barticle}
\bauthor{\bsnm{Hailpern}, \binits{B.}},
\bauthor{\bsnm{Santhanam}, \binits{P.}}:
\batitle{Software debugging, testing, and verification}.
\bjtitle{IBM Systems Journal}
\bvolume{41}(\bissue{1}),
\bfpage{4}--\blpage{12}
(\byear{2002})
\end{barticle}
\endbibitem

\bibitem[\protect\citeauthoryear{Duboc et~al.}{2007}]{duboc2007framework}
\begin{bchapter}
\bauthor{\bsnm{Duboc}, \binits{L.}},
\bauthor{\bsnm{Rosenblum}, \binits{D.}},
\bauthor{\bsnm{Wicks}, \binits{T.}}:
\bctitle{A framework for characterization and analysis of software system scalability}.
In: \bbtitle{Proceedings of the the 6th Joint Meeting of the European Software Engineering Conference and the ACM SIGSOFT Symposium on The Foundations of Software Engineering},
pp. \bfpage{375}--\blpage{384}
(\byear{2007})
\end{bchapter}
\endbibitem

\bibitem[\protect\citeauthoryear{Deelman et~al.}{2015}]{deelman2015pegasus}
\begin{barticle}
\bauthor{\bsnm{Deelman}, \binits{E.}},
\bauthor{\bsnm{Vahi}, \binits{K.}},
\bauthor{\bsnm{Juve}, \binits{G.}},
\bauthor{\bsnm{Rynge}, \binits{M.}},
\bauthor{\bsnm{Callaghan}, \binits{S.}},
\bauthor{\bsnm{Maechling}, \binits{P.J.}},
\bauthor{\bsnm{Mayani}, \binits{R.}},
\bauthor{\bsnm{Chen}, \binits{W.}},
\bauthor{\bsnm{Da~Silva}, \binits{R.F.}},
\bauthor{\bsnm{Livny}, \binits{M.}}, \betal:
\batitle{Pegasus, a workflow management system for science automation}.
\bjtitle{Future Generation Computer Systems}
\bvolume{46},
\bfpage{17}--\blpage{35}
(\byear{2015})
\end{barticle}
\endbibitem

\bibitem[\protect\citeauthoryear{Juve and Deelman}{2010}]{juve2010scientific}
\begin{barticle}
\bauthor{\bsnm{Juve}, \binits{G.}},
\bauthor{\bsnm{Deelman}, \binits{E.}}:
\batitle{Scientific workflows and clouds}.
\bjtitle{XRDS: Crossroads, The ACM Magazine for Students}
\bvolume{16}(\bissue{3}),
\bfpage{14}--\blpage{18}
(\byear{2010})
\end{barticle}
\endbibitem

\bibitem[\protect\citeauthoryear{Berriman et~al.}{2013}]{berriman2013application}
\begin{barticle}
\bauthor{\bsnm{Berriman}, \binits{G.B.}},
\bauthor{\bsnm{Deelman}, \binits{E.}},
\bauthor{\bsnm{Juve}, \binits{G.}},
\bauthor{\bsnm{Rynge}, \binits{M.}},
\bauthor{\bsnm{V{\"o}ckler}, \binits{J.-S.}}:
\batitle{The application of cloud computing to scientific workflows: a study of cost and performance}.
\bjtitle{Philosophical Transactions of the Royal Society A: Mathematical, Physical and Engineering Sciences}
\bvolume{371}(\bissue{1983}),
\bfpage{20120066}
(\byear{2013})
\end{barticle}
\endbibitem

\bibitem[\protect\citeauthoryear{Foster et~al.}{2008}]{foster2008cloud}
\begin{bchapter}
\bauthor{\bsnm{Foster}, \binits{I.}},
\bauthor{\bsnm{Zhao}, \binits{Y.}},
\bauthor{\bsnm{Raicu}, \binits{I.}},
\bauthor{\bsnm{Lu}, \binits{S.}}:
\bctitle{Cloud computing and grid computing 360-degree compared}.
In: \bbtitle{2008 Grid Computing Environments Workshop},
pp. \bfpage{1}--\blpage{10}
(\byear{2008}).
\bcomment{Ieee}
\end{bchapter}
\endbibitem

\bibitem[\protect\citeauthoryear{Boettiger}{2015}]{boettiger2015introduction}
\begin{barticle}
\bauthor{\bsnm{Boettiger}, \binits{C.}}:
\batitle{An introduction to docker for reproducible research}.
\bjtitle{ACM SIGOPS Operating Systems Review}
\bvolume{49}(\bissue{1}),
\bfpage{71}--\blpage{79}
(\byear{2015})
\end{barticle}
\endbibitem

\bibitem[\protect\citeauthoryear{Zaharia et~al.}{2010}]{zaharia2010spark}
\begin{bchapter}
\bauthor{\bsnm{Zaharia}, \binits{M.}},
\bauthor{\bsnm{Chowdhury}, \binits{M.}},
\bauthor{\bsnm{Franklin}, \binits{M.J.}},
\bauthor{\bsnm{Shenker}, \binits{S.}},
\bauthor{\bsnm{Stoica}, \binits{I.}}:
\bctitle{Spark: Cluster computing with working sets}.
In: \bbtitle{2nd USENIX Workshop on Hot Topics in Cloud Computing (HotCloud 10)}
(\byear{2010})
\end{bchapter}
\endbibitem

\bibitem[\protect\citeauthoryear{Rak et~al.}{2012}]{rak2012argo}
\begin{barticle}
\bauthor{\bsnm{Rak}, \binits{R.}},
\bauthor{\bsnm{Rowley}, \binits{A.}},
\bauthor{\bsnm{Black}, \binits{W.}},
\bauthor{\bsnm{Ananiadou}, \binits{S.}}:
\batitle{Argo: an integrative, interactive, text mining-based workbench supporting curation}.
\bjtitle{Database}
\bvolume{2012},
\bfpage{010}
(\byear{2012})
\end{barticle}
\endbibitem

\bibitem[\protect\citeauthoryear{Baldini et~al.}{2017}]{baldini2017serverless}
\begin{botherref}
\oauthor{\bsnm{Baldini}, \binits{I.}},
\oauthor{\bsnm{Castro}, \binits{P.}},
\oauthor{\bsnm{Chang}, \binits{K.}},
\oauthor{\bsnm{Cheng}, \binits{P.}},
\oauthor{\bsnm{Fink}, \binits{S.}},
\oauthor{\bsnm{Ishakian}, \binits{V.}},
\oauthor{\bsnm{Mitchell}, \binits{N.}},
\oauthor{\bsnm{Muthusamy}, \binits{V.}},
\oauthor{\bsnm{Rabbah}, \binits{R.}},
\oauthor{\bsnm{Slominski}, \binits{A.}}, et al.:
Serverless computing: Current trends and open problems.
Research advances in cloud computing,
1--20
(2017)
\end{botherref}
\endbibitem

\bibitem[\protect\citeauthoryear{Adhikari et~al.}{2019}]{adhikari2019survey}
\begin{barticle}
\bauthor{\bsnm{Adhikari}, \binits{M.}},
\bauthor{\bsnm{Amgoth}, \binits{T.}},
\bauthor{\bsnm{Srirama}, \binits{S.N.}}:
\batitle{A survey on scheduling strategies for workflows in cloud environment and emerging trends}.
\bjtitle{ACM Computing Surveys (CSUR)}
\bvolume{52}(\bissue{4}),
\bfpage{1}--\blpage{36}
(\byear{2019})
\end{barticle}
\endbibitem

\bibitem[\protect\citeauthoryear{Bean}{2007}]{bean2007qualitative}
\begin{barticle}
\bauthor{\bsnm{Bean}, \binits{C.J.}}:
\batitle{Qualitative research design: An interactive approach}.
\bjtitle{Organizational Research Methods}
\bvolume{10}(\bissue{2}),
\bfpage{393}
(\byear{2007})
\end{barticle}
\endbibitem

\bibitem[\protect\citeauthoryear{Wohlin et~al.}{2012}]{wohlin2012experimentation}
\begin{botherref}
\oauthor{\bsnm{Wohlin}, \binits{C.}},
\oauthor{\bsnm{Runeson}, \binits{P.}},
\oauthor{\bsnm{H{\"o}st}, \binits{M.}},
\oauthor{\bsnm{Ohlsson}, \binits{M.C.}},
\oauthor{\bsnm{Regnell}, \binits{B.}},
\oauthor{\bsnm{Wessl{\'e}n}, \binits{A.}}:
Experimentation in software engineering.
Springer Science \& Business Media
(2012)
\end{botherref}
\endbibitem

\end{thebibliography}

\end{document}